\documentclass[12pt]{article}
\usepackage{amssymb}
\usepackage{graphicx}
\begin{document}
\newcommand{\beq}{\begin{equation}}
\newcommand{\eeq}{\end{equation}}
\newcommand{\beqa}{\begin{eqnarray}}
\newcommand{\eeqa}{\end{eqnarray}}
\newcommand{\beqar}{\begin{eqnarray*}}
\newcommand{\eeqar}{\end{eqnarray*}}
\newcommand{\al}{\alpha}
\newcommand{\be}{\beta}
\newcommand{\del}{\delta}
\newcommand{\D}{\Delta}
\newcommand{\eps}{\epsilon}
\newcommand{\ga}{\gamma}
\newcommand{\Ga}{\Gamma}
\newcommand{\ka}{\kappa}
\newcommand{\nn}{\nonumber}
\newcommand{\inn}{\!\cdot\!}
\newcommand{\h}{\eta}
\newcommand{\ii}{\iota}
\newcommand{\kk}{\varphi}
\newcommand\F{{}_3F_2}
\newcommand{\la}{\lambda}
\newcommand{\La}{\Lambda}
\newcommand{\na}{\prt}
\newcommand{\Om}{\Omega}
\newcommand{\om}{\omega}
\newcommand{\p}{\Phi}
\newcommand{\sig}{\sigma}
\renewcommand{\t}{\theta}
\newcommand{\z}{\zeta}
\newcommand{\ssc}{\scriptscriptstyle}
\newcommand{\eg}{{\it e.g.,}\ }
\newcommand{\ie}{{\it i.e.,}\ }
\newcommand{\labell}[1]{\label{#1}} 
\newcommand{\reef}[1]{(\ref{#1})}
\newcommand\prt{\partial}
\newcommand\veps{\varepsilon}
\newcommand{\pol}{\varepsilon}
\newcommand\vp{\varphi}
\newcommand\ls{\ell_s}
\newcommand\cF{{\cal F}}
\newcommand\cA{{\cal A}}
\newcommand\cS{{\cal S}}
\newcommand\cT{{\cal T}}
\newcommand\cV{{\cal V}}
\newcommand\cL{{\cal L}}
\newcommand\cM{{\cal M}}
\newcommand\cN{{\cal N}}
\newcommand\cG{{\cal G}}
\newcommand\cK{{\cal K}}
\newcommand\cH{{\cal H}}
\newcommand\cI{{\cal I}}
\newcommand\cJ{{\cal J}}
\newcommand\cl{{\iota}}
\newcommand\cP{{\cal P}}
\newcommand\cQ{{\cal Q}}
\newcommand\cg{{\tilde {{\cal G}}}}
\newcommand\cR{{\cal R}}
\newcommand\cB{{\cal B}}
\newcommand\cO{{\cal O}}
\newcommand\tcO{{\tilde {{\cal O}}}}
\newcommand\bz{\bar{z}}
\newcommand\bb{\bar{b}}
\newcommand\ba{\bar{a}}
\newcommand\bg{\bar{g}}
\newcommand\bc{\bar{c}}
\newcommand\bw{\bar{w}}
\newcommand\bX{\bar{X}}
\newcommand\bK{\bar{K}}
\newcommand\bA{\bar{A}}
\newcommand\bH{\bar{H}}
\newcommand\bF{\bar{F}}
\newcommand\bxi{\bar{\xi}}
\newcommand\bphi{\bar{\phi}}
\newcommand\bpsi{\bar{\psi}}
\newcommand\bprt{\bar{\prt}}
\newcommand\bet{\bar{\eta}}
\newcommand\btau{\bar{\tau}}
\newcommand\hF{\hat{F}}
\newcommand\hA{\hat{A}}
\newcommand\hT{\hat{T}}
\newcommand\htau{\hat{\tau}}
\newcommand\hD{\hat{D}}
\newcommand\hf{\hat{f}}
\newcommand\hK{\hat{K}}
\newcommand\hg{\hat{g}}
\newcommand\hp{\hat{\Phi}}
\newcommand\hi{\hat{i}}
\newcommand\ha{\hat{a}}
\newcommand\hb{\hat{b}}
\newcommand\hQ{\hat{Q}}
\newcommand\hP{\hat{\Phi}}
\newcommand\hS{\hat{S}}
\newcommand\hX{\hat{X}}
\newcommand\tL{\tilde{\cal L}}
\newcommand\hL{\hat{\cal L}}
\newcommand\tG{{\tilde G}}
\newcommand\tg{{\tilde g}}
\newcommand\tphi{{\widetilde \Phi}}
\newcommand\tPhi{{\widetilde \Phi}}
\newcommand\te{{\tilde e}}
\newcommand\tk{{\tilde k}}
\newcommand\tf{{\tilde f}}
\newcommand\tH{{\tilde H}}
\newcommand\ta{{\tilde a}}
\newcommand\tb{{\tilde b}}
\newcommand\tc{{\tilde c}}
\newcommand\td{{\tilde d}}
\newcommand\tm{{\tilde m}}
\newcommand\tmu{{\tilde \mu}}
\newcommand\tnu{{\tilde \nu}}
\newcommand\talpha{{\tilde \alpha}}
\newcommand\tbeta{{\tilde \beta}}
\newcommand\trho{{\tilde \rho}}
 \newcommand\tR{{\tilde R}}
\newcommand\teta{{\tilde \eta}}
\newcommand\tF{{\widetilde F}}
\newcommand\tK{{\tilde K}}
\newcommand\tE{{\widetilde E}}
\newcommand\tpsi{{\tilde \psi}}
\newcommand\tX{{\widetilde X}}
\newcommand\tD{{\widetilde D}}
\newcommand\tO{{\widetilde O}}
\newcommand\tS{{\tilde S}}
\newcommand\tB{{\tilde B}}
\newcommand\tA{{\widetilde A}}
\newcommand\tT{{\widetilde T}}
\newcommand\tC{{\widetilde C}}
\newcommand\tV{{\widetilde V}}
\newcommand\thF{{\widetilde {\hat {F}}}}
\newcommand\Tr{{\rm Tr}}
\newcommand\tr{{\rm tr}}
\newcommand\STr{{\rm STr}}
\newcommand\hR{\hat{R}}
\newcommand\M[2]{M^{#1}{}_{#2}}
\newcommand\MZ{\mathbb{Z}}
\newcommand\MR{\mathbb{R}}
\newcommand\bS{\textbf{ S}}
\newcommand\bI{\textbf{ I}}
\newcommand\bJ{\textbf{ J}}

\begin{titlepage}
\begin{center}

\vskip 0.5 cm
{\LARGE \bf  
Effective action of bosonic string theory\\at order $\alpha'^3$}\\
\vskip 1.25 cm
Mehdi Ameri\footnote{amerimehdi9755@gmail.com}, Alireza Pahlavan\footnote{alirezapahlavan08@gmail.com}, Mohammad R. Garousi \footnote{garousi@um.ac.ir}

\vskip 1 cm
{{\it Department of Physics, Faculty of Science, Ferdowsi University of Mashhad\\}{\it P.O. Box 1436, Mashhad, Iran}\\}
\vskip .1 cm
 \end{center}

\begin{abstract}
In this work, we derive the classical effective action of bosonic string theory at order $\alpha'^{3}$ for the metric, Kalb-Ramond field, and dilaton by imposing a higher-derivative extension of the Buscher rules on the circular reduction of the minimal basis at this order, in the schemes where their corresponding actions at order $\alpha'$ are the Meissner and the Metsaev-Tseytlin schemes. We find that T-duality fixes all coupling constants in terms of the known overall factor at order $\alpha'$ and a single remaining parameter. This final parameter is determined by matching the single-trace term $\Tr(\epsilon \epsilon \epsilon \epsilon)$ in the four-graviton S-matrix element—which lacks a massless pole—with the corresponding string theory amplitude. Our results for the Riemann quartic terms are in full agreement with those obtained from the nonlinear sigma-model approach.

\end{abstract}

\end{titlepage}

\section{Introduction}

Free  closed string theory describes a fundamental  string whose oscillations give rise to a spectrum of fields, including massless modes, a tachyon, and an infinite tower of massive states. These states are characterized by the dimensionful parameter $\alpha'$ and reside in a critical spacetime dimension $D$. The perturbative formulation of the interacting theory provides a systematic framework for computing S-matrix elements involving these modes across all world-sheet genera \cite{Becker:2007zj}.
At low energies and to order $\alpha'^n$, the theory is governed by an effective field theory for the massless fields, whose Feynman amplitudes match the $\alpha'$-expansion of the string S-matrix elements for massless states at that order. The sphere-level S-matrix exhibits poles corresponding to the exchange of tachyon, massless, and massive states. While the tachyon poles reduce to ordinary numbers, the massive poles produce zeta functions—$\zeta(3), \zeta(5), \zeta(7), \dots$—in their residues. As a result, the low-energy expansion of the S-matrix retains only massless poles and contact terms at various orders in $\alpha'$.
For instance, the $\alpha'$-expansion of the 4-point S-matrix element takes the form \cite{Schlotterer:2018zce}:  
\beqa
A^4 &=& \sum_{m,m_1,m_2,\dots=0}^{\infty} c^{m} \alpha'^m \left( \prod_{k=1}^{\infty} \zeta(2k+1)^{m_k} \alpha'^{m_k(2k+1)} \right) f^4_{m,m_1,m_2,\dots}\,,\labell{A4}
\eeqa
where $c \sim 1$ is a numerical factor arising from the expansion of tachyon poles, and $f^4_{m,m_1,m_2,\dots}$ is a function of the Mandelstam variables, momenta, and polarizations of the four external massless states. A similar expansion is expected to hold for any $N$-point S-matrix element, with the only difference being the function $f^N_{m,m_1,m_2,\dots}$, which now depends on the Mandelstam variables, momenta, and polarizations of the $N$ external massless states. 
The effective field theory that  reproduces this expansion of the S-matrix elements should have the following structure in the string frame:
\beqa
\bS&=&-\frac{2}{\kappa^2}\int d^Dx\sqrt{-G}\, e^{-2\Phi}\!\!\!\sum_{m,m_1,m_2,\dots=0}^{\infty} c^{m} \alpha'^m \left( \prod_{k=1}^{\infty} \zeta(2k+1)^{m_k} \alpha'^{m_k(2k+1)} \right)\cL_{m,m_1,m_2,\dots}\labell{S}.  
\eeqa
The Lagrangian density $\mathcal{L}_{m,m_1,m_2,\dots}$ is a function of the Riemann curvature $R_{\mu\nu\alpha\beta}$, the Kalb-Ramond field strength $H_{\mu\nu\alpha}$, $\nabla_\mu\Phi$, and their derivatives at order $\alpha'^{m+3m_1+5m_2+\cdots}$. At the leading order in $\alpha'$, this Lagrangian is known as
\beqa
\cL_{0,0,\cdots,0}=R+4\nabla_\mu\Phi\nabla^\mu\Phi-\frac{1}{12}H^2\,.\labell{L0}
\eeqa 
 A  well-known example at order $\alpha'^3$ is the $\z(3)t_8 t_8 R^4$ coupling in superstring theory, which lacks tachyon poles and was first discovered through this approach \cite{Gross:1986iv,Gross:1986mw}.

Evaluating explicitly the $N$-point function for $N > 4$ and its $\alpha'$-expansion is a significant challenge, requiring substantial effort to derive the corresponding effective action. An alternative approach to constructing the effective action is to exploit the symmetries inherent in string theory, including both worldsheet and spacetime symmetries.
In the first case, the conformal symmetry of the worldsheet theory can be used to impose the vanishing of the nonlinear sigma model $\beta$-functions. This condition yields equations of motion for the background fields, from which the effective action can be systematically derived \cite{Tseytlin:1988rr}. The coupling $\zeta(3)t_8 t_8 R^4$ is reproduced by this method, and the coupling $\zeta(3)\epsilon_{10} \epsilon_{10} R^4$, which cannot be found from the 4-point function, was first obtained this way \cite{Grisaru:1986kw,Grisaru:1986vi}. In this approach, the $\zeta$-functions appear in the spacetime effective action as a consequence of regularizing worldsheet loop-level divergences in the computation of these $\beta$-functions.

In the second case, spacetime symmetries - such as supersymmetry \cite{Ozkan:2024euj} or the $O(d,d,\mathbb{R})$ symmetry of toroidally compactified spacetime effective actions \cite{Sen:1991zi,Hohm:2014sxa} - can be used to constrain the effective action. However, the spacetime symmetries alone cannot  determine the coefficients of multiple $\zeta$-function terms in the effective action. This limitation has two important consequences.
The first consequence is that spacetime symmetries alone cannot fully determine the effective action at all orders in $\alpha'$—they leave undetermined coefficients proportional to $\zeta(3), \zeta(5), \zeta(7), \dots$ that must be fixed by the string S-matrix expansion  \reef{A4}. The expansion \reef{A4} indicates that
at orders $\alpha'$ and $\alpha'^2$, the symmetry should fix the bosonic string effective action up to one  parameter, which the S-matrix determines  to be an ordinary number.
At orders $\alpha'^3$ and $\alpha'^4$, the symmetry should leave another free parameter that the S-matrix fixes to $\zeta(3)$.
At orders $\alpha'^5$ and $\alpha'^6$, a new parameter emerges, which the S-matrix fixes to $\zeta(5)$. This pattern persists for higher orders of $\alpha'$.
The second consequence is that the symmetry transformations themselves must incorporate higher-derivative corrections to account for multiple $\zeta$-function terms. For example, the appearance of $\zeta(3)^2$ at order $\alpha'^6$ implies that 
the leading-order transformation receives a correction $\delta$ at order $\alpha'^3$ which has the factor $\z(3)$.
The second-order variation $\delta\delta$ then generates the $\zeta(3)^2$ coupling at order $\alpha'^6$ when applied to the leading-order action \reef{L0}.

The classical effective action of string theory possesses an \(O(1,1,\mathbb{R})\) symmetry when dimensionally reduced on a circle \cite{Sen:1991zi,Hohm:2014sxa}. The non-geometric component of this transformation, known as the Buscher rules \cite{Buscher:1987sk,Rocek:1991ps} (which form a \(\mathbb{Z}_2\) group), can be utilized to constrain the effective action. This symmetry has been employed in \cite{Kaloper:1997ux,Garousi:2019wgz,Garousi:2019mca} to completely determine the bosonic string effective action at orders \(\alpha'\) and \(\alpha'^2\), up to one coefficient that the S-matrix fixes to be \(1/4\). Similarly, in superstring theory at order \(\alpha'^3\) for the NS-NS sector, this symmetry fixes the effective action up to an overall factor of \(\zeta(3)\), as determined by the S-matrix \cite{Garousi:2020gio}. These couplings are consistent with the partial results that have already been found by the S-matrix method \cite{Gross:1986iv,Gross:1986mw,Liu:2019ses}.
In all these cases, it is essential that the Buscher rules receive higher-derivative corrections \cite{Kaloper:1997ux,Garousi:2019wgz,Garousi:2020gio,Hsia:2024kpi}. While the bosonic theory at orders \(\alpha'\) and \(\alpha'^2\) requires corrections with ordinary coefficients, the superstring theory at order \(\alpha'^3\) demands corrections with coefficient \(\zeta(3)\) \cite{Garousi:2020gio,Hsia:2024kpi}. We expect that imposing the \(\mathbb{Z}_2\) symmetry along with the expansion \(\reef{A4}\) will determine the effective action to all orders in \(\alpha'\). Since the expansion \(\reef{A4}\) only requires fixing residual parameters after applying spacetime symmetry constraints, we focus on the four-point function for specific terms that lack massless poles.
For example, considering terms in the expansion involving the single trace of four graviton polarizations, \(\Tr(\epsilon_1 \epsilon_2 \epsilon_3 \epsilon_4)\), the corresponding S-matrix exhibits no massless poles. This cancellation occurs because the massless pole is exactly compensated by the Mandelstam variables in the numerator, effectively reducing the expression to a contact term.
In this work, we employ this T-duality symmetry in bosonic string theory to derive the effective action at order \(\alpha'^3\). While this effective action has been previously derived for gravity couplings via the non-linear sigma model approach \cite{Jack:1989vp}, our results for the case with vanishing Kalb-Ramond field show complete agreement with those obtained in \cite{Jack:1989vp}.

The paper is organized as follows:  
In Section 2, we review the construction of the effective action using spacetime $\mathbb{Z}_2$-symmetry, which first requires establishing a minimal basis at each order of $\alpha'$. We present the basis at order $\alpha'^3$, as derived in \cite{Garousi:2020mqn}. To determine the coupling constants of this basis, we employ the effective actions at orders $\alpha'$ and $\alpha'^2$. Specifically, we adopt the Meissner scheme for the $\alpha'$-order action and utilize the corresponding $\alpha'^2$-order effective action from \cite{Gholian:2023kjj}. We demonstrate that the $\mathbb{Z}_2$-symmetry fixes all coupling constants up to a single undetermined parameter.  
In Subsection 2.2, we resolve this remaining parameter by comparing the four-graviton S-matrix element of the resulting field theory with the corresponding string theory S-matrix element at order $\alpha'^3$.  
In Subsection 2.3, we apply field redefinitions to express the couplings at order $\alpha'^3$ in a canonical form, eliminating dilaton couplings and terms involving  $\nabla_\mu H^{\mu\alpha\beta}$.  
In Section 3, we repeat the analysis in the Metsaev-Tseytlin scheme, where the $\alpha'$-order couplings are formulated differently. This scheme aligns with the non-linear sigma-model approach used to derive the Riemann quartic couplings \cite{Jack:1989vp,Codina:2021cxh}. Notably, we find that for the case $H=0$, the T-duality-induced couplings match exactly those obtained in \cite{Jack:1989vp,Codina:2021cxh}.  
Section 6 concludes with a brief discussion of our results and their broader implications.  
All calculations were performed using the "xAct" package \cite{Nutma:2013zea} for tensor algebra and perturbation theory.

\section{Effective action in Meissner scheme}

To determine the classical effective action \(\reef{S}\) by imposing spacetime symmetries, we first expand it in powers of \(\alpha'\) as  
\beqa
\bS=\sum_{n=0}^{\infty}\alpha'^n\bS^{(n)}&;& \bS^{(n)}=-\frac{2}{\kappa^2}\int d^Dx\sqrt{-G}\,e^{-2\Phi}\cL_n\,,\labell{Sn}
\eeqa 
and then construct a basis for the couplings in \(\cL_n\), each with undetermined coefficients. These coefficients are fixed by enforcing the symmetry constraints on the effective action.  
For the \(\alpha'^3\) order, the basis of couplings in a particular minimal scheme has been identified in \cite{Garousi:2020mqn} and takes the following form: 
\beqa
\mathcal{L}_3&=&a_1 R_{\alpha \gamma \beta \delta } R^{\alpha \beta \gamma \delta } R_{\epsilon \mu\varepsilon \zeta } R^{\epsilon \varepsilon \mu \zeta } +\cdots +a_{872} H_{\alpha }{}^{\delta \epsilon } H^{\alpha \beta \gamma } H_{\beta }{}^{\varepsilon \mu} H_{\gamma }{}^{\zeta \eta } H_{\delta \varepsilon }{}^{\theta } H_{\epsilon \zeta }{}^{\iota } H_{\mu \iota }{}^{\kappa } H_{\eta \theta \kappa },\labell{L3}
\eeqa
where the ellipsis represents the remaining 870 terms, and \(a_1, a_2, \dots, a_{872}\) are the 872 coupling constants. At the classical level, they are independent of the background, so they can be determined via T-duality for a background with one Killing circle. Once they are found for this background, background independence guarantees that they are valid for any other spacetime as well \cite{Garousi:2022ovo}.

Unlike the leading-order action \(\reef{L0}\), which is scheme-independent, the \(\alpha'\)-corrected effective action depends on the chosen field redefinition scheme \cite{Gross:1986iv,Metsaev:1987zx}. Two widely used schemes at  order $\alpha'$ are the Meissner scheme (where the propagators derived from the leading-order action remain unmodified by \(\alpha'\) corrections) and the Metsaev-Tseytlin scheme (which minimizes the number of couplings at order \(\alpha'\)). Furthermore, it is known from S-matrix calculations that the effective action at order \(\alpha'^2\) and higher inherits scheme dependence from the \(\alpha'\) order \cite{Bento:1990nv}. This is consistent with the speculation that spacetime symmetry transformations must acquire higher-derivative corrections. These higher-derivative corrections depend on the scheme of the effective action \cite{Garousi:2019wgz}.
In this section, we derive the $\alpha'^3$ couplings specifically compatible with the Meissner scheme at order $\alpha'$.

\subsection{T-duality constraint}

In this section, we review how the $\mathbb{Z}_2$-symmetry constrains the coupling constants in the basis at each order of $\alpha'$ \cite{Kaloper:1997ux,Garousi:2023kxw}, and subsequently apply this constraint to the 872 coupling constants in \reef{L3}. To implement this symmetry, we consider a spacetime with one circular dimension (coordinate $y$) where all fields are $y$-independent.  

The Buscher transformations for the NS-NS fields simplify significantly when using the following circular reduction ansatz \cite{Maharana:1992my}:  
\beqa  
G_{\mu\nu} = \left(\matrix{\bg_{ab} + e^{\varphi} g_{a} g_{b} &  e^{\varphi} g_{a} \cr e^{\varphi} g_{b} & e^{\varphi} &}\!\!\!\!\!\right), \quad  
B_{\mu\nu} = \left(\matrix{\bb_{ab} + b_{[a} g_{b]} & b_{a} \cr -b_{b} & 0 &}\!\!\!\!\!\right), \quad  
\Phi = \bar{\phi} + \varphi/4,  
\labell{reduc}  
\eeqa  
where indices $a,b$ denote directions orthogonal to the Killing coordinate $y$.  
Under Buscher transformations, the (D-1)-dimensional base space fields - $\bg_{ab}$ (metric), $\bb_{ab}$ (antisymmetric tensor), and $\bar{\phi}$ (dilaton) - remain invariant. The base space scalar field $\varphi$ acquires a minus sign, while the vector fields $g_a$ and $b_a$ are interchanged.

Unlike in the original $D$-dimensional spacetime where the exact form $H=3dB$ appears explicitly in the action, the $(D-1)$-dimensional base space action does not contain the exact form $d\bb$. Instead, the action features the following field strength \cite{Kaloper:1997ux}:
\beqa
\bar{H}_{abc} &= 3\partial_{[a}\bb_{bc]} - \frac{3}{2}g_{[a}W_{bc]} - \frac{3}{2}b_{[a}V_{bc]}\,,
\eeqa
where $W = 2db$ and $V = 2dg$ are exact forms. Unlike these field strengths, the above field strength satisfies the anomalous Bianchi identity. That is, 
\begin{equation}
dW=0\,,\,\,\,\,\, dV=0\,,\,\,\,\,\, d\bar{H} = -\frac{3}{2}V\wedge W\,. \labell{bian0}
\end{equation}
The Buscher transformation rules for the base space fields are then given by:
\beqa  
\bg_{ab}' &= \bg_{ab}\,, \quad 
\bar{H}_{abc}' = \bar{H}_{abc}\,, \quad 
\bar{\phi}' = \bar{\phi}\,, \nn\\
\varphi' &= -\varphi\,, \quad 
g_a' = b_a\,, \quad 
b_a' = g_a\,.  
\labell{flac}  
\eeqa  
These transformations should be generalized to include higher-derivative corrections.

If $\psi$ denotes any base space field in the Buscher rules above, then the generalized Buscher rules take the form:
\beqa
\psi' = \psi_0 + \sum_{n=1}^{\infty} \alpha'^n \psi_n \,,\labell{psi}
\eeqa
where $\psi_0$ represents the original Buscher rules in \reef{flac}, and $\psi_n$ corresponds to the correction at order $\alpha'^n$ that must be determined through T-duality. 
Due to the Bianchi identity \reef{bian0}, the $\bar{H}$-corrections are related to the $b$- and $g$-corrections as \cite{Garousi:2023kxw}:
\beqa
\sum_{n=1}^{\infty} \alpha'^n \bar{H}_{n} &=& 3\sum_{n=1}^{\infty} \alpha'^n d(\bar{B}_{n}) - 3\sum_{n=1}^{\infty} \alpha'^n W \wedge b_{n} - 3\sum_{n=1}^{\infty} \alpha'^n g_{n} \wedge V \labell{dHbar} \\
&& - 3\sum_{n,m=1}^{\infty} \alpha'^{n+m} \Big[ g_{n} \wedge db_{m} + dg_{n} \wedge b_{m} \Big]\,, \nn
\eeqa
where the 2-form $\bar{B}_{n}$ contains all gauge-invariant terms at order $\alpha'^{n}$, along with additional corrections $\psi_n$. This expression matches the result found in \cite{Garousi:2023kxw} after applying convenient rescalings of the $b$- and $g$-corrections as 
\beqa
g_n = e^{\varphi/2} \Delta g^{(n)}&;&b_n = e^{-\varphi/2} \Delta g^{(n)},
\eeqa
 and absorbing the $1/n!$ factors into the definition of the corrections as in \reef{psi}.
The correction $\bar{H}_{n}$ can be expressed as:
\beqa
\bar{H}_{n} &=& 3d(\bar{B}_{n}) - 3 W \wedge b_{n} - 3 g_{n} \wedge V \labell{dHbar1} \\
&& - 3\sum_{0<k<n} \Big[ g_{k} \wedge db_{n-k} + dg_{k} \wedge b_{n-k} \Big]\,.\nn
\eeqa
Note that while the first three terms contain corrections at order $\alpha'^n$, the last term involves only corrections at orders $\alpha', \alpha'^2, \cdots, \alpha'^{n-1}$.

Since the Buscher transformations receive higher-derivative corrections, T-duality cannot be imposed solely on the reduced effective action at a specific order of $\alpha'$. Instead, it must be imposed on the reduction of the complete effective action. This requirement is expressed as:
\beqa
S(\psi) \sim S(\psi')\,, \labell{SS}
\eeqa
where $S(\psi)$ represents the reduced effective action from \reef{Sn}, and $S(\psi')$ is its transformation under the generalized Buscher rules \reef{psi}. The $\sim$ symbol indicates equality up to total derivative terms and application of the Bianchi identities \reef{bian0}.
Expanding $S(\psi)$ in $\alpha'$ as in \reef{Sn} to obtain $\sum_{n=0}^\infty \alpha'^n S^{(n)}(\psi)$, and performing a Taylor expansion of $S^{(n)}(\psi')$ around the Buscher rules $\psi_0$:
\beqa
S^{(n)}(\psi') = S^{(n)}(\psi_0) + \sum_{m=1}^{\infty} \alpha'^m S^{(n,m)}(\psi_0) \,,\labell{Tay}
\eeqa
we can rewrite the constraint \reef{SS} as:
\beqa
\sum_{n=0}^{\infty} \alpha'^n \left(S^{(n)}(\psi) - S^{(n)}(\psi_0)\right) \sim \sum_{n=0}^{\infty} \sum_{m=1}^{\infty} \alpha'^{n+m} S^{(n,m)}(\psi_0)\,.\labell{TSS}
\eeqa
In the absence of higher-derivative corrections to the Buscher rules, the right-hand side would vanish. This would imply that T-duality constrains all coupling constants at order $\alpha'$ and higher to be zero – a conclusion that clearly contradicts known results from the S-matrix method.

In the Taylor expansion \reef{Tay}, the last term in \( S^{(n,m)} \) contains multiple contributions of
\( \alpha'\psi_1, \alpha'^2\psi_1^2, \alpha'^3\psi_1^3, \cdots \), \( \alpha'^2\psi_2, \alpha'^4\psi_2^2, \alpha'^6\psi_2^3, \cdots \),  \( \alpha'^3\psi_3, \alpha'^6\psi_3^2, \alpha'^9\psi_3^3, \cdots \) and higher, 
all appearing at order \( \alpha'^m \). However, in the T-duality constraint at order \( \alpha'^n \), only the coupling constants in the effective action \( S^{(n)} \) and the parameters of \( \psi_n \) are unknown. All lower-order effective actions \( S^{(0)}, S^{(1)}, \cdots, S^{(n-1)} \) and Buscher rule corrections \( \psi_1, \psi_2, \cdots, \psi_{n-1} \) are already determined from T-duality constraints at lower orders of \( \alpha' \).
To isolate terms containing unknown parameters in the Taylor expansion, we decompose \( S^{(0,n)} \) as:
\beqa
S^{(0,n)}(\psi_0) = S^{(0,n)}_{(n)}(\psi_0) + S^{(0,n)}_{(n')}(\psi_0)\,,
\eeqa
where the first term on the right-hand side is linear in the unknown correction \( \psi_n \), while the second term depends nonlinearly on the known corrections  \( \psi_1, \psi_2, \dots, \psi_{n-1} \).
The T-duality constraint \reef{TSS} at order \( \alpha'^n \) (for \( n > 0 \)) then becomes:
\beqa
S^{(0,n)}_{(n)}(\psi_0) \sim S^{(n)}(\psi) - S^{(n)}(\psi_0) - S^{(0,n)}_{(n')}(\psi_0) - \sum_{0<k<n} S^{(k,n-k)}(\psi_0)\,.\labell{Tn}
\eeqa
Note that the left-hand side of the equation encodes the parameters of the T-duality corrections at order \( \alpha'^n \), while the first two terms on the right-hand side contains the parameters of the effective action at the same order. The last two terms on the right-hand side are parameter-free, as they are already determined by T-duality constraints at lower orders (\( \alpha', \alpha'^2, \dots, \alpha'^{n-1} \)).  
The equation admits both homogeneous solutions (where the right-hand side vanishes) and inhomogeneous solutions (where it does not). Our focus is solely on the inhomogeneous case.  
To simplify the calculation, we assume a flat base space metric. Despite this restriction, the resulting constraints are sufficient to fix the coupling constants in the effective action.  

It is important to note that if one ignores the corrections to the Buscher rules and instead imposes the equations of motion on the transformation of the base space effective action under the Buscher rules (see \reef{flac}), a similar constraint to \reef{Tn} is found. In this case, the third term on the right-hand side is removed, and the last term depends only linearly on the corrections  \( \psi_1, \psi_2, \dots, \psi_{n-1} \). While the resulting constraint on the effective action is identical to \reef{Tn} for the leading-order correction (which involves only linear corrections to the Buscher rules), it differs significantly for the next-to-leading-order action and beyond. These incorrect constraints admit no solution for the coupling constants in the effective action.

The T-duality constraint \reef{Tn} serves two primary purposes:
Determining T-duality corrections for known effective actions (where all parameters are already fixed), and 
simultaneously solving for both the minimal basis parameters in the effective action and their corresponding T-duality corrections. 

\subsubsection{Couplings at order $\alpha'$} 

The T-duality constraint \(\reef{Tn}\) at order \(\alpha'\) is given by  
\beqa  
S^{(0,1)}_{(1)}(\psi_0) \sim S^{(1)}(\psi) - S^{(1)}(\psi_0). \labell{T1}  
\eeqa  
This equation can be used to verify the effective action at order \(\alpha'\) in the Meissner scheme, as derived in \cite{Meissner:1996sa}\footnote{The T-duality constraint \reef{T1} involve the corrections to the Buscher rules only linearly, hence, the constraint is the same as the constraint  that one  imposes the base space equations of motion instead of using corrections to the Buscher rules.}
\beqa
{\bf S}_M^{(1)}&=&-\frac{2}{4\kappa^2}\int d^{26}x \sqrt{-G}e^{-2\Phi}\Big[ \frac{1}{24} H_{\alpha 
}{}^{\delta \epsilon } H^{\alpha \beta \gamma } H_{\beta 
\delta }{}^{\varepsilon } H_{\gamma \epsilon \varepsilon } -  
\frac{1}{8} H_{\alpha \beta }{}^{\delta } H^{\alpha \beta 
\gamma } H_{\gamma }{}^{\epsilon \varepsilon } H_{\delta 
\epsilon \varepsilon } \nn\\&&+ \frac{1}{144} H_{\alpha \beta \gamma 
} H^{\alpha \beta \gamma } H_{\delta \epsilon \varepsilon } H^{
\delta \epsilon \varepsilon } + H_{\alpha }{}^{\gamma \delta } 
H_{\beta \gamma \delta } R^{\alpha \beta } - 4 
R_{\alpha \beta } R^{\alpha \beta } -  
\frac{1}{6} H_{\alpha \beta \gamma } H^{\alpha \beta \gamma } 
R + R^2 \nn\\&&+ R_{\alpha \beta \gamma 
\delta } R^{\alpha \beta \gamma \delta } -  
\frac{1}{2} H_{\alpha }{}^{\delta \epsilon } H^{\alpha \beta 
\gamma } R_{\beta \gamma \delta \epsilon } -  
\frac{2}{3} H_{\beta \gamma \delta } H^{\beta \gamma \delta } 
\nabla_{\alpha }\nabla^{\alpha }\Phi + \frac{2}{3} H_{\beta 
\gamma \delta } H^{\beta \gamma \delta } \nabla_{\alpha }\Phi 
\nabla^{\alpha }\Phi \nn\\&&+ 8 R \nabla_{\alpha }\Phi 
\nabla^{\alpha }\Phi - 16 R_{\alpha \beta } 
\nabla^{\alpha }\Phi \nabla^{\beta }\Phi + 16 \nabla_{\alpha 
}\Phi \nabla^{\alpha }\Phi \nabla_{\beta }\Phi \nabla^{\beta 
}\Phi - 32 \nabla^{\alpha }\Phi \nabla_{\beta }\nabla_{\alpha 
}\Phi \nabla^{\beta }\Phi \nn\\&&+ 2 H_{\alpha }{}^{\gamma \delta } 
H_{\beta \gamma \delta } \nabla^{\beta }\nabla^{\alpha }\Phi \Big]\,.\labell{fourmax}
\eeqa
The action satisfies the constraint \reef{T1} for the following corrections to the Buscher rules \cite{Kaloper:1997ux}:
\beqa
&&\bphi_1= 0\,,
 \bg_{1ab}=0 \,,
\bar{B}_{1ab}= - V_{[b}{}^{c} W_{a]c} \,,g_{1a}=\frac{1}{4}\Big( \bH_{abc} V^{bc}+2{e^{-\vp}} W_{ab}\prt^b\vp\Big)\nn\\
&&\vp_1=\frac{1}{4}\Big( e^{\vp} V_{ab} V^{ab}  + {e^{-\vp}} W_{ab} W^{ab} + 2 \prt_{a}\vp \prt^{a}\vp\Big)\,,\labell{d1}
\eeqa
and $ b_{1a}(\psi)=- g_{1a}(\psi_0)$. These transformations are essential for studying the effective action at order $\alpha'^2$.

\subsubsection{Couplings at order $\alpha'^2$}

The T-duality constraint \(\reef{Tn}\) at order \(\alpha'^2\) is given by  
\beqa
S^{(0,2)}_{(2)}(\psi_0) \sim S^{(2)}(\psi) - S^{(2)}(\psi_0) - S^{(0,2)}_{(2')}(\psi_0) - S^{(1,1)}(\psi_0)\,.\labell{T2}
\eeqa
This constraint has been utilized in \cite{Garousi:2023kxw} to determine all the coupling constants in the minimal basis at order $\alpha'^2$, which consists of 60 terms \cite{Garousi:2019cdn}\footnote{The T-duality constraint \reef{T2} fixes all 60 parameters of the minimal basis. It is clear that if the third term on its right-hand side is ignored—which corresponds to imposing the base space equations of motion instead of including higher-derivative corrections to the Buscher rules—the resulting equation has no solution. Therefore, it is essential to use the full corrections to the Buscher rules rather than relying on the equations of motion.}. The analysis reveals that T-duality generates 27 non-zero couplings. However, these 27 non-zero terms are specific to the particular minimal basis identified in \cite{Garousi:2019cdn}. If a different minimal basis is considered within an alternative scheme, the number of non-zero couplings may increase or decrease. This indicates that field redefinitions can be employed to express the 27 non-zero terms using fewer or more than 27 couplings.
The 27 couplings can be reduced to the following 12 terms through an appropriate field redefinition \cite{Gholian:2023kjj}\footnote{Utilizing $O(d,d)$ symmetry, expressions for the couplings at orders $\alpha'$ and $\alpha'^2$ were derived in \cite{Wulff:2024ips}. The $\alpha'$ couplings coincide with those in the Metsaev-Tseytlin action plus some terms involving equations of motion, while the $\alpha'^2$ couplings are significantly more complex. Field redefinitions were subsequently applied to simplify the latter, resulting in the couplings given in \reef{finalB}. 
}:
\beqa
\bS^{(2)}_{M}&\!\!\!\!\!\!\!\!=\!\!\!\!\!\!\!\!&-\frac{2}{16\kappa^2}\int d^{26} x\sqrt{-G} e^{-2\Phi}\Big[-  
\frac{4}{3}R_{\alpha }{}^{\kappa }{}_{\gamma 
}{}^{\lambda }R^{\alpha \beta \gamma \theta } 
R_{\beta \lambda \theta \kappa } + \frac{4}{3} 
R_{\alpha \beta }{}^{\kappa \lambda } 
R^{\alpha \beta \gamma \theta }R_{\gamma 
\kappa \theta \lambda }\nn\\&& - \frac{1}{12} H_{\alpha }{}^{\theta \kappa } H^{\alpha \beta 
\gamma } H_{\beta \theta }{}^{\lambda } H_{\gamma }{}^{\mu \nu 
} H_{\kappa \mu }{}^{\tau } H_{\lambda \nu \tau } + 
\frac{1}{4} H_{\alpha \beta }{}^{\theta } H^{\alpha \beta 
\gamma } H_{\gamma }{}^{\kappa \lambda } H_{\theta }{}^{\mu 
\nu } H_{\kappa \mu }{}^{\tau } H_{\lambda \nu \tau } \nn\\&&+ 
\frac{1}{48} H_{\alpha \beta }{}^{\theta } H^{\alpha \beta 
\gamma } H_{\gamma }{}^{\kappa \lambda } H_{\theta }{}^{\mu 
\nu } H_{\kappa \lambda }{}^{\tau } H_{\mu \nu \tau } - 2 H_{\alpha }{}^{\theta \kappa } 
H^{\alpha \beta \gamma }R_{\beta \theta }{}^{\lambda 
\mu }R_{\gamma \lambda \kappa \mu } -  H_{\alpha \beta 
}{}^{\theta } H^{\alpha \beta \gamma }R_{\gamma 
}{}^{\kappa \lambda \mu }R_{\theta \lambda \kappa 
\mu } \nn\\&&+ 2 H^{\alpha 
\beta \gamma } H^{\theta \kappa \lambda }R_{\alpha 
\beta \theta }{}^{\mu }R_{\gamma \mu \kappa \lambda 
} - 2 H_{\alpha }{}^{\theta \kappa } H^{\alpha \beta \gamma } 
R_{\beta }{}^{\lambda }{}_{\gamma }{}^{\mu } 
R_{\theta \lambda \kappa \mu }+ \frac{1}{4} H^{\alpha \beta \gamma } H^{\theta \kappa 
\lambda } \nabla_{\gamma }H_{\kappa \lambda \mu } 
\nabla_{\theta }H_{\alpha \beta }{}^{\mu } \nn\\&& + \frac{1}{2} 
H_{\alpha }{}^{\theta \kappa } H^{\alpha \beta \gamma } 
\nabla_{\kappa }H_{\theta \lambda \mu } \nabla^{\mu }H_{\beta 
\gamma }{}^{\lambda } + H_{\alpha }{}^{\theta \kappa } 
H^{\alpha \beta \gamma } \nabla_{\mu }H_{\gamma \kappa 
\lambda } \nabla^{\mu }H_{\beta \theta }{}^{\lambda }\Big]\,.\labell{finalB}
\eeqa
By combining the above fixed action at order \(\alpha'^2\) with the Meissner action (\ref{fourmax}) at order \(\alpha'\) and its corresponding T-duality corrections (\ref{d1}), we can solve the constraint (\ref{T2}) to derive the T-duality corrections at order \(\alpha'^2\). Due to their lengthy expressions, we omit these corrections here, but they are essential for constructing the effective action at order \(\alpha'^3\) in the following section.

\subsubsection{Couplings at order $\alpha'^3$}

The T-duality constraint \(\reef{Tn}\) at order \(\alpha'^3\) is given by  
\beqa
S^{(0,3)}_{(3)}(\psi_0) \sim S^{(3)}(\psi) - S^{(3)}(\psi_0) - S^{(0,3)}_{(3')}(\psi_0) - S^{(1,2)}(\psi_0)- S^{(2,1)}(\psi_0)\,.\labell{T3}
\eeqa
Using the action at order \(\alpha'^3\) in the minimal basis (\ref{L3}), along with the actions (\ref{fourmax}) and (\ref{finalB}) at orders \(\alpha'\) and \(\alpha'^2\) and their corresponding T-duality transformations, we can employ the constraint (\ref{T3}) to determine both the Lagrangian (\ref{L3}) and its associated T-duality corrections. These are expressed in terms of a single unfixed parameter\footnote{As noted, the T-duality constraint \reef{T3} fixes 871 parameters in the minimal basis. Crucially, if the third term and the nonlinear contribution to the fourth term on the right-hand side are omitted—which is equivalent to imposing the base space equations of motion rather than including higher-derivative corrections to the Buscher rules—the resulting equation becomes unsolvable. }. The resulting effective action that we have found,  is:
\beqa
{\bf S}_M^{(3)}&=&-\frac{2 }{\kappa^2}\int d^{26}x \sqrt{-G}e^{-2\Phi}\Big[ \frac{1}{16} R_{\alpha \beta }{}^{\epsilon \varepsilon } 
R^{\alpha \beta \gamma \delta } R_{\gamma }{}^{\zeta 
}{}_{\epsilon }{}^{\mu } R_{\delta \zeta \varepsilon \mu } + 
(\frac{1}{72} -  \frac{a}{9} ) R_{\alpha }{}^{\epsilon 
}{}_{\gamma }{}^{\varepsilon } R^{\alpha \beta \gamma \delta } 
R_{\beta }{}^{\zeta }{}_{\epsilon }{}^{\mu } R_{\delta \mu 
\varepsilon \zeta } \nn\\&&\qquad\qquad\qquad\qquad\quad+ \frac{1}{18} (-1 - a) R_{\alpha 
\beta }{}^{\epsilon \varepsilon } R^{\alpha \beta \gamma 
\delta } R_{\gamma }{}^{\zeta }{}_{\epsilon }{}^{\mu } 
R_{\delta \mu \varepsilon \zeta } + \frac{3}{4} R_{\alpha 
\gamma \beta }{}^{\epsilon } R^{\alpha \beta \gamma \delta } 
R_{\delta }{}^{\varepsilon \zeta \mu } R_{\epsilon \zeta 
\varepsilon \mu } \nn\\&&\qquad\qquad\qquad\qquad\quad -  \frac{1}{16} R_{\alpha \gamma \beta 
\delta } R^{\alpha \beta \gamma \delta } R_{\epsilon \zeta 
\varepsilon \mu } R^{\epsilon \varepsilon \zeta \mu }+\cdots\Big]\,,\labell{SM3}
\eeqa
where the dots represent 754 terms containing the field strength $H$ or $\nabla H$, and the parameter $a$ remains as the sole unfixed parameter among the 872 parameters in (\ref{L3}).
For superstring theory, which lacks effective actions at orders $\alpha'$ and $\alpha'^2$, only terms with coefficient $a$ survive. Comparison with the 4-point S-matrix element fixes this parameter to be proportional to $\zeta(3)$ \cite{Gross:1986iv}.
However, in the bosonic string theory, one cannot simply assume that $a$ must similarly be proportional to $\zeta(3)$. Instead, the parameter $a$ must be determined by explicitly comparing the gravitational couplings with the 4-point S-matrix element in the bosonic string theory. We perform this calculation in the following section.

\subsection{S-matrix constraint}

In this section, we compare the 4-point S-matrix elements derived from the gravity couplings presented in \reef{L0}, \reef{fourmax}, \reef{finalB}, \reef{SM3} with the low-energy expansion of the sphere-level S-matrix element for four-graviton vertex operators in string theory.

As observed in \cite{Kawai:1985xq}, the sphere-level closed string S-matrix element can be expressed in terms of disk-level S-matrix elements of open strings. The disk-level S-matrix element for four gauge boson vertex operators on the disk boundary was calculated in \cite{Schwarz:1982jn}. For the $(s-t)$ channel, this amplitude takes the form
\beqa
\cA (\alpha' s,\alpha' t)&\sim&\frac{\Gamma(-\alpha's)\Gamma(-\alpha' t)}{\Gamma(1+\alpha' u)}K\,,
\eeqa
where the Mandelstam variables are:
\beqa
s=-(k_1+k_2)^2,\quad t=-(k_1+k_4)^2,\quad u=-(k_1+k_3)^2;\quad s+t+u=0\,.
\eeqa
The kinematic factor \( K \), which contains tachyon poles, incorporates the open string momenta \( k_1, k_2, k_3, k_4 \) and polarizations \( \zeta_1, \zeta_2, \zeta_3, \zeta_4 \). In bosonic string theory, it is given by:
\begin{equation}
K^{B}(\alpha') = \left(\frac{\alpha'^2 s t}{1 + \alpha' u}\right) \zeta_1 \cdot \zeta_3 \, \zeta_2 \cdot \zeta_4 + \left(\frac{\alpha'^2 s u}{1 + \alpha' t}\right) \zeta_2 \cdot \zeta_3 \, \zeta_1 \cdot \zeta_4 + \left(\frac{\alpha'^2 t u}{1 + \alpha' s}\right) \zeta_1 \cdot \zeta_2 \, \zeta_3 \cdot \zeta_4 + \cdots,
\end{equation}
where the dots represent terms involving contractions between momenta and polarizations. Since our goal is to compare the field couplings with the string theory S-matrix element to fix the residual parameter \( a \) in \reef{SM3}, we disregard terms where momenta and polarizations are contracted—hence, we ignore the dotted terms in the equation above. Note that these kinematic factors are \( stu \)-symmetric.
If the above amplitude corresponds to right-moving modes, then analogous amplitudes \( \bar{\mathcal{A}} \) exist for the left-moving modes, with momenta \( k_1, k_2, k_3, k_4 \) and polarizations \( \bar{\zeta}_1, \bar{\zeta}_2, \bar{\zeta}_3, \bar{\zeta}_4 \).

The closed string amplitude of four gravitons with momenta $k_i$ and symmetric polarizations $\epsilon_i=\z_i\bar{\z_i}$ for $i=1,2,3,4$ is then given as \cite{Kawai:1985xq}:
\beqa
A&=&-\left(\frac{\kappa^2}{\pi\alpha'}\right)\sin(\frac{\alpha'\pi}{2}k_2\cdot k_3)\cA(\frac{\alpha'}{4}s,\frac{\alpha'}{4}t)\bar{\cA}(\frac{\alpha'}{4}t,\frac{\alpha'}{4}u),\nn\\
&=&-\left(\frac{\kappa^2}{\alpha'}\right)\frac{\Gamma(-\frac{\alpha'}{4}s)\Gamma(-\frac{\alpha'}{4}t)\Gamma(-\frac{\alpha'}{4}u)}{\Gamma(1+\frac{\alpha'}{4}s)\Gamma(1+\frac{\alpha'}{4}t)\Gamma(1+\frac{\alpha'}{4}u)}K^B(\frac{\alpha'}{4})\bar{K}^B(\frac{\alpha'}{4}),\labell{KLT}
\eeqa
where we have normalized the amplitude to be consistent with the leading-order action \reef{L0}. The Gamma functions have massless poles and infinite tower of massive poles.

To derive the $\alpha'$-expansion of the closed string S-matrix element above, one must expand both the massive poles in the Gamma functions and the tachyon poles in the kinematic factor $K^B$. The tachyon pole expansion is the primary contribution, while the massive poles generate the following additional terms:
\beqa
\frac{\Gamma(-\frac{\alpha'}{4}s)\Gamma(-\frac{\alpha'}{4}t)\Gamma(-\frac{\alpha'}{4}u)}{\Gamma(1+\frac{\alpha'}{4}s)\Gamma(1+\frac{\alpha'}{4}t)\Gamma(1+\frac{\alpha'}{4}u)}=-\frac{64}{\alpha'^3stu}-2\z(3)+\cdots\,,
\eeqa
where the dots represent higher-order terms in $\alpha'$. Using these expansions, one finds that the amplitude \reef{KLT}  contains both massless and contact terms. To simplify the analysis further, we restrict our attention to single-trace terms of the form $\Tr(\eps\eps\eps\eps)$ . In this case, the amplitude reduces to purely contact terms, which take the following explicit form:  
\beqa
A&=&f(s,t,u)\Tr(\epsilon_1\epsilon_3\epsilon_2\epsilon_4) +f(u,t,s) \Tr(\epsilon_1\epsilon_2\epsilon_3\epsilon_4)+f(t,u,s)
\Tr(\epsilon_1\epsilon_2\epsilon_4\epsilon_3)\,,\labell{Agggg}
\eeqa
where the function $f(s,t,u)$ is
\beqa
    f(s,t,u)&=&\frac{\kappa^2}{2}\Big[s+\frac{\alpha^{\prime}}{4}s^2+\frac{\alpha^{\prime 2}}{16}(s^3-stu)+\frac{\alpha^{\prime 3}}{64}s^2\Big(s^2+2st+2t^2+\z(3)tu\Big)+\cdots\Big],
    \labell{KLT2}
\eeqa
where the dots represent terms at higher orders of $\alpha'$ in which we are not interested.

We now compute the four-graviton S-matrix element in low-energy field theory up to order \(\alpha'^3\) in the Meissner scheme, which includes the contributions from \(\reef{L0}\), \(\reef{fourmax}\), \(\reef{finalB}\), and \(\reef{SM3}\).  
The metric perturbation is given by:  
\[  
G_{\mu\nu} = \eta_{\mu\nu} + \kappa h_{\mu\nu}.  
\]  
The four-graviton Feynman amplitude consists of both contact terms and massless pole contributions, where the latter involve graviton and dilaton propagation between vertices. All vertices and propagators must be computed in the Einstein frame, where the Einstein-Hilbert term has no overall dilaton factor. Notably, for the single-trace terms, only the graviton propagates between vertices.  
The leading-order action \(\reef{L0}\) yields the following graviton propagator:  
\beqa
(\tG_h)_{\mu\nu, \lambda\rho}&=&\frac{1}{2k^2}\left(\eta_{\mu\lambda}\eta_{\nu\rho}+\eta_{\mu\rho}\eta_{\nu\lambda}-\frac{1}{\frac{D}{2}-1}\eta_{\mu\nu}\eta_{\lambda\rho}\right),
\eeqa
where $k^\mu$ is the momentum of the graviton  in the propagator, and we have used a Euclidean signature spacetime metric in which the partition function is $Z\sim\int e^{-S}$.

Using the gravity couplings in \(\reef{SM3}\), we obtain the following contact terms at order \(\alpha'^3\):
\beqa
	A_{\rm contact}&=&\frac{\alpha'^3}{576}\Big[ \bigl(-18 s^4 - 2 (31 + 4 a) s^3 t -  (89 + 8 a) 
s^2 t^2 - 54 s t^3 - 27 t^4\bigr) \Tr(\epsilon_1\epsilon_3\epsilon_2\epsilon_4) \nn\\&& +  
\bigl(-18 s^4 + 2 (-5 + 4 a) s^3 t + (-11 + 16 a) s^2 t^2 + 2 (-5 
+ 4a) s t^3 - 18 t^4\bigr) \Tr(\epsilon_1\epsilon_2\epsilon_3\epsilon_4) \nn\\&&+ 
 \bigl(-27 s^4 - 54 s^3 t -  (89 + 8 a) s^2 t^2 - 2 
(31 + 4 a) s t^3 - 18 t^4\bigr) \Tr(\epsilon_1\epsilon_2\epsilon_4\epsilon_3)\Big]\,.
\eeqa
The pole amplitude arises from two cubic vertices derived from the actions \(\reef{L0}\), \(\reef{fourmax}\), and \(\reef{finalB}\), connected by a graviton propagator. However, the single-trace terms produce only contact terms. These terms appear at orders \(\alpha'^0\), \(\alpha'\), \(\alpha'^2\), \(\alpha'^3\), and \(\alpha'^4\). The \(\alpha'^3\) terms are:
\beqa
A_{\rm pole}=\widetilde{V} (hhh) \widetilde{G}_h \widetilde{V}(hhh)&=&
\frac{3\alpha'^2}{128}\Big[\left(s^4+4s^3t+6s^2t^2+4st^3 +2t^4\right)\Tr(\epsilon_1\epsilon_3\epsilon_2\epsilon_4)\nn\\&&
+\left(s^4 +t^4\right)\Tr(\epsilon_1\epsilon_2\epsilon_3\epsilon_4)\nn\\&&
+\left(2s^4+4s^3t+6s^2t^2+4st^3+t^4\right)\Tr(\epsilon_1\epsilon_2\epsilon_4\epsilon_3)\Big]\,.
\eeqa
Note that the parameter $a$ does not appear in the amplitude above because the action  \(\reef{SM3}\) lacks a three-field coupling.

The combination $A_{\rm contact}+A_{\rm pole}$ must reproduce the string amplitude \(\reef{Agggg}\). This condition fixes the parameter \(a\) to the value:
\beqa
a = \frac{1}{8} - \frac{9}{8}\zeta(3) \,.\labell{a}
\eeqa
Consequently, the effective action \(\reef{SM3}\) is fully determined
\beqa
{\bf S}_M^{(3)}&\!\!\!=\!\!\!&-\frac{2 }{\kappa^2}\int d^{26}x \sqrt{-G}e^{-2\Phi}\Big[\frac{1}{16} R_{\alpha \beta }{}^{\epsilon \varepsilon } 
R^{\alpha \beta \gamma \delta } R_{\gamma }{}^{\zeta 
}{}_{\epsilon }{}^{\mu } R_{\delta \zeta \varepsilon \mu } + 
\frac{3}{4} R_{\alpha \gamma \beta }{}^{\epsilon } R^{\alpha 
\beta \gamma \delta } R_{\delta }{}^{\varepsilon \zeta \mu } 
R_{\epsilon \zeta \varepsilon \mu }\labell{S32}\\&&\qquad\qquad\qquad\qquad\quad  -  \frac{1}{16} R_{\alpha 
\gamma \beta \delta } R^{\alpha \beta \gamma \delta } 
R_{\epsilon \zeta \varepsilon \mu } R^{\epsilon \varepsilon 
\zeta \mu }- \frac{1}{16} R_{\alpha \beta }{}^{\epsilon 
\varepsilon } R^{\alpha \beta \gamma \delta } R_{\gamma 
}{}^{\zeta }{}_{\epsilon }{}^{\mu } R_{\delta \mu \varepsilon 
\zeta }  \nn\\&&\qquad\qquad\qquad\qquad\quad + \frac{\z(3)}{16} \Big(R_{\alpha \beta }{}^{\epsilon 
\varepsilon } R^{\alpha \beta \gamma \delta } R_{\gamma 
}{}^{\zeta }{}_{\epsilon }{}^{\mu } R_{\delta \mu \varepsilon 
\zeta } +2 R_{\alpha 
}{}^{\epsilon }{}_{\gamma }{}^{\varepsilon } R^{\alpha \beta 
\gamma \delta } R_{\beta }{}^{\zeta }{}_{\epsilon }{}^{\mu } 
R_{\delta \mu \varepsilon \zeta } \Big)+\cdots\Big].\nn
\eeqa
where the dots represent couplings involving the field \( H \) with coefficients that are either \(\zeta(3)\) or fractional numbers. The terms with \(\zeta(3)\) coefficients correspond to couplings that also appear in superstring theory. Among these, there are 735 couplings with fractional coefficients involving \( H \). Some of these terms additionally include \(\nabla\Phi\) or \(\nabla\nabla\Phi\). In the next section, we use field redefinitions to eliminate the dilaton terms and express the couplings in a canonical form.

\subsection{$\alpha'^3$-couplings in a  canonical form}

In superstring theory, the T-duality constraint on the minimal basis \(\reef{L3}\) initially requires 445 non-zero couplings \cite{Garousi:2020gio}. This number reduces to 251 couplings \cite{Garousi:2020lof} after applying an appropriate field redefinition to express the action in canonical form—eliminating all dilaton terms and terms containing \(\nabla_\mu H^{\mu\alpha\beta}\). In this section, we employ field redefinitions for the 739 fractionally-weighted couplings in \(\reef{S32}\) to rewrite them in this canonical form. Since the \(\alpha'\) and \(\alpha'^2\) order couplings are already fixed, our field redefinition involves only perturbations of the leading-order action \(\reef{L0}\) at \(\alpha'^3\). The redefinition takes the form:
\beqa
\bS^{(0,3)}_{(3)} \sim \bS^{(3)} - \bS'^{(3)}\,, \labell{F3}
\eeqa
where \(\!\bS'^{(3)}\) represents the transformed effective action related to \(\!\bS^{(3)}\) through field redefinitions, total derivative terms, and Bianchi identities. Here, the \(\sim\) symbol denotes equality modulo total derivatives and Bianchi identities in \(D\)-dimensional spacetime.
Following \cite{Garousi:2020lof}, we find the fractionally-weighted effective action contains 232 non-zero couplings. These are:
\beqa
{\bf S'}_M^{(3)}&\!\!\!\!=\!\!\!\!&-\frac{2}{\kappa^2}\int d^{26}x \sqrt{-G}e^{-2\Phi}\Big[ \frac{\z(3)}{2^4}\cL_{\rm II}+[R^4]_{3}+[R^2\nabla H^2]_{7}+[\nabla H^4]_{5}+[H^8]_{5}\nn\\&&+[H^6R]_{1}+[H^4R^2]_{15}+[H^2 R^3 ]_{11}+[H^2\nabla H^2 R]_{92}+[H^4\nabla H^2]_{93}\Big]\,,\labell{L22}
\eeqa
where $\cL_{\rm II}$ represents the couplings in superstring theory \cite{Garousi:2020lof}, while the remaining couplings are
\beqa
[R^4]_{3}&=&\frac{3}{8} R_{\alpha \gamma \beta }{}^{\epsilon } R^{\alpha 
\beta \gamma \delta } R_{\delta }{}^{\varepsilon \mu \nu } R_{
\mu \nu \epsilon \varepsilon } -  \frac{1}{64} R^{\alpha 
\beta \gamma \delta } R_{\gamma \delta \alpha \beta } 
R^{\epsilon \varepsilon \mu \nu } R_{\mu \nu \epsilon 
\varepsilon }\nn\\&& + \frac{1}{32} R_{\alpha \beta }{}^{\epsilon 
\varepsilon } R^{\alpha \beta \gamma \delta } R_{\mu \nu 
\delta \varepsilon } R^{\mu \nu }{}_{\gamma \epsilon }\,,
\nn\\
{[}R^2\nabla H^2{]}_{7}&=&\frac{1}{8} R_{\alpha }{}^{\epsilon \varepsilon \mu } 
R^{\alpha \beta \gamma \delta } \nabla_{\epsilon }H_{\beta 
\gamma }{}^{\nu } \nabla_{\mu }H_{\delta \varepsilon \nu } -  
\frac{1}{32} R_{\alpha }{}^{\epsilon \varepsilon \mu } 
R^{\alpha \beta \gamma \delta } \nabla_{\epsilon }H_{\gamma 
\delta \nu } \nabla^{\nu }H_{\beta \varepsilon \mu } \nn\\&&+ 
\frac{1}{32} R_{\alpha }{}^{\epsilon \varepsilon \mu } 
R^{\alpha \beta \gamma \delta } \nabla_{\nu }H_{\gamma \delta 
\epsilon } \nabla^{\nu }H_{\beta \varepsilon \mu } -  
\frac{1}{8} R_{\alpha \gamma }{}^{\epsilon \varepsilon } 
R^{\alpha \beta \gamma \delta } \nabla_{\beta }H_{\varepsilon 
\mu \nu } \nabla^{\nu }H_{\delta \epsilon }{}^{\mu } \nn\\&&+ 
\frac{3}{32} R_{\alpha \gamma \beta }{}^{\epsilon } R^{\alpha 
\beta \gamma \delta } \nabla_{\nu }H_{\epsilon \varepsilon 
\mu } \nabla^{\nu }H_{\delta }{}^{\varepsilon \mu } + 
\frac{5}{64} R^{\alpha \beta \gamma \delta } R_{\gamma \delta 
\alpha }{}^{\epsilon } \nabla_{\beta }H_{\varepsilon \mu \nu } 
\nabla^{\nu }H_{\epsilon }{}^{\varepsilon \mu }\nn\\&& -  
\frac{1}{128} R^{\alpha \beta \gamma \delta } R_{\gamma 
\delta \alpha \beta } \nabla_{\nu }H_{\epsilon \varepsilon 
\mu } \nabla^{\nu }H^{\epsilon \varepsilon \mu }\,,
\nn\\
{[}\nabla H^4{]}_{5}&=&\frac{1}{64} \nabla^{\delta }H^{\alpha \beta \gamma } 
\nabla_{\epsilon }H_{\delta \mu \nu } \nabla^{\varepsilon }H_{
\alpha \beta }{}^{\epsilon } \nabla^{\nu }H_{\gamma 
\varepsilon }{}^{\mu } -  \frac{1}{128} \nabla^{\delta 
}H^{\alpha \beta \gamma } \nabla^{\varepsilon }H_{\alpha \beta 
}{}^{\epsilon } \nabla_{\mu }H_{\delta \epsilon \nu } \nabla^{
\nu }H_{\gamma \varepsilon }{}^{\mu }\nn\\&& -  \frac{1}{128} 
\nabla^{\delta }H^{\alpha \beta \gamma } \nabla^{\varepsilon 
}H_{\alpha \beta }{}^{\epsilon } \nabla_{\nu }H_{\delta 
\epsilon \mu } \nabla^{\nu }H_{\gamma \varepsilon }{}^{\mu } + 
\frac{1}{64} \nabla_{\gamma }H_{\varepsilon \mu \nu } 
\nabla^{\delta }H^{\alpha \beta \gamma } \nabla^{\epsilon }H_{
\alpha \beta \delta } \nabla^{\nu }H_{\epsilon 
}{}^{\varepsilon \mu } \nn\\&&-  \frac{1}{1152} \nabla_{\delta 
}H_{\alpha \beta \gamma } \nabla^{\delta }H^{\alpha \beta 
\gamma } \nabla_{\nu }H_{\epsilon \varepsilon \mu } 
\nabla^{\nu }H^{\epsilon \varepsilon \mu }\,,
\nn\\
{[}H^8{]}_{5}&\!\!\!=\!\!\!&\frac{1}{256} H_{\alpha }{}^{\delta \epsilon } H^{\alpha 
\beta \gamma } H_{\beta \delta }{}^{\varepsilon } H_{\gamma 
}{}^{\mu \zeta } H_{\epsilon \mu }{}^{\eta } H_{\varepsilon 
}{}^{\theta \iota } H_{\zeta \theta }{}^{\kappa } H_{\eta 
\iota \kappa } \nn\\&&+ \frac{1}{2048} H_{\alpha }{}^{\delta \epsilon 
} H^{\alpha \beta \gamma } H_{\beta }{}^{\varepsilon \mu } 
H_{\gamma }{}^{\zeta \eta } H_{\delta \varepsilon }{}^{\theta } 
H_{\epsilon \zeta }{}^{\iota } H_{\theta \iota \kappa } H_{\mu 
\eta }{}^{\kappa }\nn\\&& -  \frac{1}{512} H_{\alpha }{}^{\delta 
\epsilon } H^{\alpha \beta \gamma } H_{\beta \delta 
}{}^{\varepsilon } H_{\gamma \epsilon }{}^{\mu } H_{\varepsilon 
}{}^{\zeta \eta } H_{\zeta \theta }{}^{\kappa } H_{\eta \iota 
\kappa } H_{\mu }{}^{\theta \iota }\nn\\&& + \frac{1}{2048} H_{\alpha 
}{}^{\delta \epsilon } H^{\alpha \beta \gamma } H_{\beta 
}{}^{\varepsilon \mu } H_{\gamma }{}^{\zeta \eta } H_{\delta 
\varepsilon }{}^{\theta } H_{\epsilon \zeta }{}^{\iota } 
H_{\eta \theta \kappa } H_{\mu \iota }{}^{\kappa }\nn\\&& -  
\frac{1}{36864} H_{\alpha }{}^{\delta \epsilon } H^{\alpha 
\beta \gamma } H_{\beta \delta }{}^{\varepsilon } H_{\gamma 
\epsilon \varepsilon } H_{\zeta \theta }{}^{\kappa } H_{\eta 
\iota \kappa } H_{\mu }{}^{\theta \iota } H^{\mu \zeta \eta }\,,
\nn\\
 {[}H^6R{]}_{1}&=&\frac{1}{768} H_{\alpha \beta }{}^{\epsilon } H_{\gamma 
\delta \epsilon } H_{\varepsilon }{}^{\tau \theta } 
H^{\varepsilon \mu \nu } H_{\mu \tau }{}^{\lambda } H_{\nu 
\theta \lambda } R^{\alpha \beta \gamma \delta }\,,
\nn\\
{[}H^4R^2{]}_{15}&=& - \frac{1}{768} H_{\epsilon }{}^{\mu \nu } H^{\epsilon 
\varepsilon \theta } H_{\varepsilon \mu }{}^{\tau } H_{\theta 
\nu \tau } R^{\alpha \beta \gamma \delta } R_{\gamma \delta 
\alpha \beta } -  \frac{1}{512} H_{\epsilon \varepsilon 
}{}^{\mu } H^{\epsilon \varepsilon \theta } H_{\theta }{}^{\nu 
\tau } H_{\mu \nu \tau } R^{\alpha \beta \gamma \delta } 
R_{\gamma \delta \alpha \beta } \nn\\&&+ \frac{3}{128} H_{\alpha 
\epsilon }{}^{\theta } H_{\beta \varepsilon }{}^{\mu } 
H_{\theta }{}^{\nu \tau } H_{\mu \nu \tau } R^{\alpha \beta 
\gamma \delta } R^{\epsilon \varepsilon }{}_{\gamma \delta } -  
\frac{15}{256} H_{\alpha \beta }{}^{\mu } H_{\delta \mu 
}{}^{\nu } H_{\epsilon \varepsilon }{}^{\tau } H_{\theta \nu 
\tau } R^{\alpha \beta \gamma \delta } R^{\epsilon \varepsilon 
}{}_{\gamma }{}^{\theta }\nn\\&& -  \frac{57}{256} H_{\alpha \beta 
}{}^{\mu } H_{\delta \epsilon \mu } H_{\varepsilon }{}^{\nu 
\tau } H_{\theta \nu \tau } R^{\alpha \beta \gamma \delta } 
R^{\epsilon \varepsilon }{}_{\gamma }{}^{\theta } -  
\frac{3}{32} H_{\alpha \epsilon }{}^{\mu } H_{\beta \theta 
}{}^{\nu } H_{\delta \varepsilon }{}^{\tau } H_{\mu \nu \tau } 
R^{\alpha \beta \gamma \delta } R^{\epsilon \varepsilon 
}{}_{\gamma }{}^{\theta } \nn\\&&+ \frac{11}{256} H_{\alpha \beta 
\epsilon } H_{\delta \varepsilon }{}^{\mu } H_{\theta }{}^{\nu 
\tau } H_{\mu \nu \tau } R^{\alpha \beta \gamma \delta } 
R^{\epsilon \varepsilon }{}_{\gamma }{}^{\theta } + 
\frac{3}{32} H_{\alpha \epsilon }{}^{\nu } H_{\beta 
\varepsilon }{}^{\tau } H_{\gamma \theta \nu } H_{\delta \mu 
\tau } R^{\alpha \beta \gamma \delta } R^{\epsilon \varepsilon 
\theta \mu } \nn\\&&-  \frac{1}{128} H_{\alpha \epsilon }{}^{\nu } 
H_{\beta \varepsilon \nu } H_{\gamma \theta }{}^{\tau } 
H_{\delta \mu \tau } R^{\alpha \beta \gamma \delta } 
R^{\epsilon \varepsilon \theta \mu } + \frac{5}{64} H_{\alpha 
\epsilon }{}^{\nu } H_{\beta \varepsilon \nu } H_{\gamma 
\delta }{}^{\tau } H_{\theta \mu \tau } R^{\alpha \beta 
\gamma \delta } R^{\epsilon \varepsilon \theta \mu } \nn\\&&-  
\frac{33}{256} H_{\alpha \beta }{}^{\nu } H_{\gamma \epsilon 
\nu } H_{\delta \varepsilon }{}^{\tau } H_{\theta \mu \tau } 
R^{\alpha \beta \gamma \delta } R^{\epsilon \varepsilon \theta 
\mu } + \frac{1}{128} H_{\alpha \beta }{}^{\nu } H_{\gamma 
\delta }{}^{\tau } H_{\epsilon \varepsilon \nu } H_{\theta \mu 
\tau } R^{\alpha \beta \gamma \delta } R^{\epsilon \varepsilon 
\theta \mu } \nn\\&&-  \frac{1}{64} H_{\alpha \beta }{}^{\nu } 
H_{\gamma \delta \nu } H_{\epsilon \varepsilon }{}^{\tau } 
H_{\theta \mu \tau } R^{\alpha \beta \gamma \delta } 
R^{\epsilon \varepsilon \theta \mu } + \frac{15}{128} 
H_{\alpha \beta \epsilon } H_{\gamma \delta }{}^{\nu } 
H_{\varepsilon \theta }{}^{\tau } H_{\mu \nu \tau } R^{\alpha 
\beta \gamma \delta } R^{\epsilon \varepsilon \theta \mu }\nn\\&& + 
\frac{53}{512} H_{\alpha \beta \epsilon } H_{\gamma \delta 
\theta } H_{\varepsilon }{}^{\nu \tau } H_{\mu \nu \tau } 
R^{\alpha \beta \gamma \delta } R^{\epsilon \varepsilon \theta 
\mu }\,,
\nn\\
{[}H^2 R^3 {]}_{11}&=&- \frac{1}{32} H_{\varepsilon }{}^{\nu \tau } H_{\mu \nu 
\tau } R^{\alpha \beta \gamma \delta } R_{\gamma \delta 
\alpha }{}^{\epsilon } R_{\epsilon }{}^{\varepsilon }{}_{\beta 
}{}^{\mu } + \frac{5}{128} H_{\epsilon \varepsilon }{}^{\tau } 
H_{\mu \nu \tau } R^{\alpha \beta \gamma \delta } R_{\gamma 
\delta \alpha \beta } R^{\epsilon \varepsilon \mu \nu } \nn\\&&+ 
\frac{13}{16} H_{\beta \varepsilon }{}^{\tau } H_{\mu \nu 
\tau } R^{\alpha \beta \gamma \delta } R_{\gamma \delta 
\alpha }{}^{\epsilon } R^{\varepsilon \mu }{}_{\epsilon 
}{}^{\nu } -  \frac{1}{4} H_{\beta \varepsilon \mu } 
H_{\epsilon \nu \tau } R^{\alpha \beta \gamma \delta } 
R_{\gamma }{}^{\epsilon }{}_{\alpha \delta } R^{\varepsilon \mu 
\nu \tau } \nn\\&&-  \frac{1}{8} H_{\beta \mu }{}^{\tau } 
H_{\epsilon \nu \tau } R_{\alpha }{}^{\epsilon }{}_{\gamma 
}{}^{\varepsilon } R^{\alpha \beta \gamma \delta } R^{\mu \nu 
}{}_{\delta \varepsilon } -  \frac{1}{8} H_{\beta \epsilon 
}{}^{\tau } H_{\mu \nu \tau } R_{\alpha }{}^{\epsilon 
}{}_{\gamma }{}^{\varepsilon } R^{\alpha \beta \gamma \delta } 
R^{\mu \nu }{}_{\delta \varepsilon } \nn\\&&+ \frac{1}{8} H_{\beta 
\mu }{}^{\tau } H_{\epsilon \nu \tau } R^{\alpha \beta \gamma 
\delta } R_{\gamma }{}^{\epsilon }{}_{\alpha }{}^{\varepsilon } 
R^{\mu \nu }{}_{\delta \varepsilon } + \frac{1}{8} H_{\beta 
\epsilon }{}^{\tau } H_{\mu \nu \tau } R^{\alpha \beta \gamma 
\delta } R_{\gamma }{}^{\epsilon }{}_{\alpha }{}^{\varepsilon } 
R^{\mu \nu }{}_{\delta \varepsilon } \nn\\&&+ \frac{1}{8} H_{\alpha 
\beta \epsilon } H_{\mu \nu \tau } R^{\alpha \beta \gamma 
\delta } R_{\gamma }{}^{\epsilon \varepsilon \mu } R^{\nu \tau 
}{}_{\delta \varepsilon } + \frac{1}{8} H_{\alpha \beta \nu } 
H_{\delta \epsilon \tau } R^{\alpha \beta \gamma \delta } 
R_{\gamma }{}^{\epsilon \varepsilon \mu } R^{\nu \tau 
}{}_{\varepsilon \mu } \nn\\&&+ \frac{1}{16} H_{\alpha \beta \epsilon 
} H_{\delta \nu \tau } R^{\alpha \beta \gamma \delta } 
R_{\gamma }{}^{\epsilon \varepsilon \mu } R^{\nu \tau 
}{}_{\varepsilon \mu }\,.
\eeqa
The couplings with structures \([H^2\nabla H^2 R]_{92}\) and \([H^4\nabla H^2]_{93}\) appear in the Appendix. When rewriting the four terms in the first and second lines of \(\reef{S32}\) in terms of the three terms in \([R^4]_3\), we have utilized the cyclic symmetry of the Riemann curvature.

To compare the above effective action (for \(H=0\)) with the one generated by the non-linear sigma model \cite{Jack:1989vp}, appropriate field redefinitions must be applied to match the couplings at orders \(\alpha'\) and \(\alpha'^2\) with those in \cite{Jack:1989vp}. The couplings at order \(\alpha'^3\) are identified in \cite{Jack:1989vp} when the corresponding effective action at order \(\alpha'\) is expressed in either the Meissner scheme or the Metsaev-Tseytlin scheme. However, the gravity couplings at order \(\alpha'^2\) differ from those in \(\reef{finalB}\).  
In \cite{Codina:2021cxh}, a suitable field redefinition was applied to the couplings in the Metsaev-Tseytlin scheme, aligning the \(\alpha'^2\)-order couplings with those in \(\reef{finalB}\). The corresponding dilaton and gravity couplings at order \(\alpha'^3\) have been determined, and their cosmological reduction has been studied in \cite{Codina:2021cxh}.  
To compare the T-duality-derived couplings with those found in \cite{Jack:1989vp,Codina:2021cxh}, we identify in the next section the \(\alpha'^3\)-order couplings whose corresponding \(\alpha'\)-order terms are in the Metsaev-Tseytlin scheme and whose \(\alpha'^2\)-order terms match those in \(\reef{finalB}\) for \(H=0\).

\section{Effective action in  Metsaev-Tseytlin scheme}

The effective action at order $\alpha'$ in the Metsaev-Tseytlin scheme is given by \cite{Metsaev:1987zx}
\beqa
{\bf S}_{MT}^{(1)}&=&-\frac{2 }{4\kappa^2}\int d^{10}x \sqrt{-G}e^{-2\Phi}\Big[ \frac{1}{24} H_{\alpha }{}^{\delta \epsilon } H^{\alpha \beta \gamma } H_{\beta \delta }{}^{\varepsilon } H_{\gamma \epsilon \varepsilon } -  \frac{1}{8} H_{\alpha \beta }{}^{\delta } H^{\alpha \beta \gamma } H_{\gamma }{}^{\epsilon \varepsilon } H_{\delta \epsilon \varepsilon }  \nn\\&&\qquad\qquad\qquad\qquad\qquad+ R_{\alpha \beta \gamma \delta } R^{\alpha \beta \gamma \delta }-  \frac{1}{2} H_{\alpha }{}^{\delta \epsilon } H^{\alpha \beta \gamma } R_{\beta \gamma \delta \epsilon }\Big]\,.\labell{fourmin}
\eeqa
Using this effective action as input for the T-duality constraint  \reef{T1} at order $\alpha'$, the corresponding corrections to the Buscher rules were derived in \cite{Garousi:2019wgz}
\beqa
\bphi_1&=&  \frac{1}{4} (e^{\vp} V_{ab} V^{ab}-{ e^{-\vp}}W_{ab} W^{ab})\,,\nn\\
 \bg_{1ab}&=&\frac{1}{2} (e^{\vp} V_{a}{}^{c} V_{bc}-{e^{-\vp}} W_{a}{}^{c} W_{bc})\,,\nn\\
 B_{1ab}&=&- V_{[b}{}^{c} W_{a]c} \,,\nn\\
\vp_1&=& \frac{1}{2} (e^{\vp} V_{ab} V^{ab}+{e^{-\vp}} W_{ab} W^{ab} + \prt_{a}\vp \prt^{a}\vp)\,,\nn\\
 g_{1a}&=&\frac{1}{4}( \bH_{abc} V^{bc}+ 2{e^{-\vp}}\prt_{b}W_{a}{}^{b} - 4 {e^{- \vp}}W_{ab}\prt^{b}\bphi )\,,\labell{minrule}
\eeqa
and $b_{1a}(\psi)=- g_{1a}(\psi_0)$. These transformations are crucial for deriving the corresponding effective action at order \(\alpha'^2\).

Applying the T-duality constraint  \(\reef{T2}\) fixes the coupling constants in the \(\alpha'^2\)-order minimal basis \cite{Garousi:2019cdn}. While this basis initially contains 60 terms, T-duality yields only 27 non-vanishing couplings \cite{Garousi:2019mca}. These can be further simplified via \(\alpha'^2\)-order field redefinitions to the following 17 fundamental terms:
\beqa
\bS^{(2)}_{MT}&\!\!\!\!\!=\!\!\!\!\!&-\frac{2}{16\kappa^2}\int d^{26} x\sqrt{-G} e^{-2\Phi}\Big[ -  \frac{4}{3} R_{\alpha }{}^{\epsilon }{}_{\gamma 
}{}^{\varepsilon } R^{\alpha \beta \gamma \delta } R_{\beta 
\varepsilon \delta \epsilon } + \frac{4}{3} R_{\alpha \beta 
}{}^{\epsilon \varepsilon } R^{\alpha \beta \gamma \delta } R_{
\gamma \epsilon \delta \varepsilon }\nn\\&&- \frac{1}{12} H_{\alpha }{}^{\delta \epsilon } 
H^{\alpha \beta \gamma } H_{\beta \delta }{}^{\varepsilon } H_{
\gamma }{}^{\mu \theta } H_{\epsilon \mu }{}^{\lambda } 
H_{\varepsilon \theta \lambda } + \frac{1}{4} H_{\alpha \beta 
}{}^{\delta } H^{\alpha \beta \gamma } H_{\gamma }{}^{\epsilon 
\varepsilon } H_{\delta }{}^{\mu \theta } H_{\epsilon \mu }{}^{
\lambda } H_{\varepsilon \theta \lambda }\nn\\&& -  \frac{1}{6} 
H_{\alpha \beta }{}^{\delta } H^{\alpha \beta \gamma } 
H_{\gamma }{}^{\epsilon \varepsilon } H_{\delta }{}^{\mu \theta 
} H_{\epsilon \varepsilon }{}^{\lambda } H_{\mu \theta \lambda 
} - 2 H_{\alpha }{}^{\delta 
\epsilon } H^{\alpha \beta \gamma } R_{\beta \delta 
}{}^{\varepsilon \mu } R_{\gamma \varepsilon \epsilon \mu } + 
\frac{1}{4} H_{\alpha \beta }{}^{\delta } H^{\alpha \beta 
\gamma } H_{\epsilon \varepsilon }{}^{\theta } H^{\epsilon 
\varepsilon \mu } R_{\gamma \mu \delta \theta } \nn\\&&+ 2 H^{\alpha 
\beta \gamma } H^{\delta \epsilon \varepsilon } R_{\alpha 
\beta \delta }{}^{\mu } R_{\gamma \mu \epsilon \varepsilon } - 
2 H_{\alpha }{}^{\delta \epsilon } H^{\alpha \beta \gamma } R_{
\beta }{}^{\varepsilon }{}_{\gamma }{}^{\mu } R_{\delta 
\varepsilon \epsilon \mu } + 3 H_{\alpha \beta }{}^{\delta } 
H^{\alpha \beta \gamma } R_{\gamma }{}^{\epsilon \varepsilon 
\mu } R_{\delta \varepsilon \epsilon \mu } \nn\\&&+ 4 H_{\alpha \beta 
}{}^{\delta } H^{\alpha \beta \gamma } H_{\gamma }{}^{\epsilon 
\varepsilon } H_{\epsilon }{}^{\mu \theta } R_{\delta \mu 
\varepsilon \theta } - 2 H_{\alpha \beta }{}^{\delta } 
H^{\alpha \beta \gamma } H_{\gamma }{}^{\epsilon \varepsilon } 
H_{\delta }{}^{\mu \theta } R_{\epsilon \mu \varepsilon \theta 
} -  H_{\alpha }{}^{\delta \epsilon } H^{\alpha \beta \gamma } 
\nabla_{\delta }H_{\beta }{}^{\varepsilon \mu } 
\nabla_{\epsilon }H_{\gamma \varepsilon \mu } \nn\\&&+ \frac{1}{36} 
H^{\alpha \beta \gamma } H^{\delta \epsilon \varepsilon } 
\nabla_{\mu }H_{\delta \epsilon \varepsilon } \nabla^{\mu }H_{
\alpha \beta \gamma } -  H^{\alpha \beta \gamma } H^{\delta 
\epsilon \varepsilon } \nabla_{\varepsilon }H_{\gamma \epsilon 
\mu } \nabla^{\mu }H_{\alpha \beta \delta } -  \frac{3}{4} 
H_{\alpha }{}^{\delta \epsilon } H^{\alpha \beta \gamma } 
\nabla_{\mu }H_{\delta \epsilon \varepsilon } \nabla^{\mu }H_{
\beta \gamma }{}^{\varepsilon } \nn\\&&+ \frac{1}{2} H_{\alpha \beta 
}{}^{\delta } H^{\alpha \beta \gamma } \nabla_{\mu }H_{\delta 
\epsilon \varepsilon } \nabla^{\mu }H_{\gamma }{}^{\epsilon 
\varepsilon }\Big]\,.\labell{SMT2}
\eeqa
The above action contains the minimal number of couplings at order $\alpha'^2$ when the effective action at order $\alpha'$ is given by the Metsaev-Tseytlin action \reef{fourmin}. The T-duality constraint  \reef{T2} at order $\alpha'^2$  determines the corresponding corrections to the Buscher rules at this order. While we omit these lengthy expressions here, both they and the corrections in \reef{minrule} are essential for deriving the effective action at order $\alpha'^3$.

By combining the effective actions in \reef{fourmin} and \reef{SMT2} with the minimal basis \reef{L3}, the T-duality constraint \reef{T3} fixes 871 parameters in \reef{L3}. Our analysis reveals that
\beqa
{\bf S}_{MT}^{(3)}&=&-\frac{2 }{\kappa^2}\int d^{26}x \sqrt{-G}e^{-2\Phi}\Big[ \frac{1}{16} R_{\alpha \beta }{}^{\epsilon \varepsilon } 
R^{\alpha \beta \gamma \delta } R_{\gamma }{}^{\zeta 
}{}_{\epsilon }{}^{\mu } R_{\delta \zeta \varepsilon \mu } + 
(\frac{1}{72} -  \frac{a}{9} ) R_{\alpha }{}^{\epsilon 
}{}_{\gamma }{}^{\varepsilon } R^{\alpha \beta \gamma \delta } 
R_{\beta }{}^{\zeta }{}_{\epsilon }{}^{\mu } R_{\delta \mu 
\varepsilon \zeta } \nn\\&&+ \frac{1}{18} (-1 - a) R_{\alpha 
\beta }{}^{\epsilon \varepsilon } R^{\alpha \beta \gamma 
\delta } R_{\gamma }{}^{\zeta }{}_{\epsilon }{}^{\mu } 
R_{\delta \mu \varepsilon \zeta }- \frac{1}{4} R_{\alpha 
\gamma \beta }{}^{\epsilon } R^{\alpha \beta \gamma \delta } 
R_{\delta }{}^{\varepsilon \zeta \mu } R_{\epsilon \zeta 
\varepsilon \mu }\nn\\&&- \frac{1}{4} H_{\alpha \epsilon }{}^{\mu } R^{\alpha \beta 
\gamma \delta } R_{\gamma }{}^{\epsilon \varepsilon \zeta } 
\nabla_{\beta }\nabla_{\zeta }H_{\delta \varepsilon \mu }+\cdots\Big]\,,\labell{SMT3}
\eeqa
where the dots denote terms containing either $H$ or $\nabla H$. These couplings differ from their counterparts in \reef{SM3} in one key aspect: they include exactly one coupling with second derivatives on $H$ (specifically, $\nabla\nabla H$). Aside from the second term in the second line, all other terms remain invariant under field redefinitions. Consequently, these terms must match the corresponding couplings in the Meissner scheme \reef{S32}, which implies that the parameter $a$ in this action must coincide with that defined in \reef{a}. This completely determines the effective action \reef{SMT3}
\beqa
{\bf S}_{MT}^{(3)}&\!\!\!\!\!=\!\!\!\!\!&-\frac{2 }{\kappa^2}\int d^{26}x \sqrt{-G}e^{-2\Phi}\Big[ \frac{\z(3)}{16} \Big(R_{\alpha \beta }{}^{\epsilon 
\varepsilon } R^{\alpha \beta \gamma \delta } R_{\gamma 
}{}^{\zeta }{}_{\epsilon }{}^{\mu } R_{\delta \mu \varepsilon 
\zeta } +2 R_{\alpha 
}{}^{\epsilon }{}_{\gamma }{}^{\varepsilon } R^{\alpha \beta 
\gamma \delta } R_{\beta }{}^{\zeta }{}_{\epsilon }{}^{\mu } 
R_{\delta \mu \varepsilon \zeta } \Big)\nn\\&&+\frac{1}{16} R_{\alpha \beta }{}^{\epsilon \varepsilon } 
R^{\alpha \beta \gamma \delta } R_{\gamma }{}^{\zeta 
}{}_{\epsilon }{}^{\mu } R_{\delta \zeta \varepsilon \mu } - 
\frac{1}{4} R_{\alpha \gamma \beta }{}^{\epsilon } R^{\alpha 
\beta \gamma \delta } R_{\delta }{}^{\varepsilon \zeta \mu } 
R_{\epsilon \zeta \varepsilon \mu } - \frac{1}{16} R_{\alpha \beta }{}^{\epsilon 
\varepsilon } R^{\alpha \beta \gamma \delta } R_{\gamma 
}{}^{\zeta }{}_{\epsilon }{}^{\mu } R_{\delta \mu \varepsilon 
\zeta } \nn\\&&- \frac{1}{4} H_{\alpha \epsilon }{}^{\mu } R^{\alpha \beta 
\gamma \delta } R_{\gamma }{}^{\epsilon \varepsilon \zeta } 
\nabla_{\beta }\nabla_{\zeta }H_{\delta \varepsilon \mu }+\cdots \Big]\,,\labell{ST32}
\eeqa
where the dots represent couplings involving either $H$ or $\nabla H$, with coefficients that are either $\zeta(3)$ or fractional numbers. The $\zeta(3)$-weighted terms correspond precisely to those appearing in superstring theory. Among these, we identify 748 distinct couplings with fractional coefficients that depend on $H$.

Applying the field redefinition  \reef{F3}, we express all couplings involving $H$ or $\nabla H$ in terms of 259 canonical couplings. This transformation reduces the original set of couplings \reef{ST32} to the following simplified set of 262 terms:
\beqa
{\bf S'}_{MT}^{(3)}&\!\!\!\!=\!\!\!\!&-\frac{2}{\kappa^2}\int d^{26}x \sqrt{-G}e^{-2\Phi}\Big[ \frac{\z(3)}{2^4}\cL_{\rm II}+[R^4]_{2}+[R^2\nabla H^2]_{17}+[\nabla H^4]_{6}+[H\nabla\nabla H R^2]_1+[H^8]_{4}\nn\\&&+[H^6R]_{1}+[H^4R^2]_{12}+[H^2 R^3 ]_{20}+[H^2\nabla H^2 R]_{100}+[H^4\nabla H^2]_{99}\Big]\,,\labell{L221}
\eeqa
where
\beqa
[R^4]_{2}&=&\frac{1}{32} R_{\alpha \beta }{}^{\epsilon \varepsilon } 
R^{\alpha \beta \gamma \delta } R_{\gamma \epsilon }{}^{\mu 
\nu } R_{\delta \varepsilon \mu \nu } -  \frac{1}{16} 
R_{\alpha \beta \gamma }{}^{\epsilon } R^{\alpha \beta \gamma 
\delta } R_{\delta }{}^{\varepsilon \mu \nu } R_{\epsilon 
\varepsilon \mu \nu }\,,
\nn\\
{[}R^2\nabla H^2{]}_{17}&=&\frac{17}{16} R^{\alpha \beta \gamma \delta } R_{\gamma }{}^{
\epsilon \varepsilon \mu } \nabla_{\beta }H_{\delta \mu \nu } 
\nabla_{\varepsilon }H_{\alpha \epsilon }{}^{\nu } -  
\frac{15}{8} R^{\alpha \beta \gamma \delta } R^{\epsilon 
\varepsilon }{}_{\gamma }{}^{\mu } \nabla_{\epsilon }H_{\alpha 
\delta }{}^{\nu } \nabla_{\varepsilon }H_{\beta \mu \nu }\nn\\&&-  
\frac{1}{16} R^{\alpha \beta \gamma \delta } R^{\epsilon 
\varepsilon \mu \nu } \nabla_{\gamma }H_{\alpha \beta 
\epsilon } \nabla_{\varepsilon }H_{\delta \mu \nu } + 
\frac{7}{16} R_{\alpha }{}^{\epsilon }{}_{\gamma 
}{}^{\varepsilon } R^{\alpha \beta \gamma \delta } 
\nabla_{\epsilon }H_{\beta }{}^{\mu \nu } \nabla_{\varepsilon 
}H_{\delta \mu \nu } \nn\\&&-  \frac{3}{16} R^{\alpha \beta \gamma 
\delta } R_{\gamma }{}^{\epsilon }{}_{\alpha }{}^{\varepsilon } 
\nabla_{\epsilon }H_{\beta }{}^{\mu \nu } \nabla_{\varepsilon 
}H_{\delta \mu \nu } + \frac{15}{16} R^{\alpha \beta \gamma 
\delta } R^{\epsilon \varepsilon }{}_{\gamma }{}^{\mu } 
\nabla_{\beta }H_{\alpha \delta }{}^{\nu } \nabla_{\varepsilon 
}H_{\epsilon \mu \nu }\nn\\&& -  \frac{1}{4} R^{\alpha \beta \gamma 
\delta } R_{\gamma }{}^{\epsilon }{}_{\alpha }{}^{\varepsilon } 
\nabla_{\delta }H_{\beta }{}^{\mu \nu } \nabla_{\varepsilon 
}H_{\epsilon \mu \nu } -  \frac{1}{8} R^{\alpha \beta \gamma 
\delta } R^{\epsilon \varepsilon \mu \nu } \nabla_{\varepsilon 
}H_{\beta \delta \nu } \nabla_{\mu }H_{\alpha \gamma \epsilon 
} \nn\\&&+ \frac{13}{16} R^{\alpha \beta \gamma \delta } R_{\gamma 
}{}^{\epsilon \varepsilon \mu } \nabla_{\epsilon }H_{\alpha 
\beta }{}^{\nu } \nabla_{\mu }H_{\delta \varepsilon \nu } + 
\frac{3}{16} R^{\alpha \beta \gamma \delta } R^{\epsilon 
\varepsilon \mu \nu } \nabla_{\mu }H_{\alpha \gamma \epsilon 
} \nabla_{\nu }H_{\beta \delta \varepsilon }\nn\\&& -  \frac{17}{32} 
R^{\alpha \beta \gamma \delta } R_{\gamma }{}^{\epsilon 
\varepsilon \mu } \nabla_{\beta }H_{\varepsilon \mu \nu } 
\nabla^{\nu }H_{\alpha \delta \epsilon } -  \frac{3}{8} 
R^{\alpha \beta \gamma \delta } R^{\epsilon \varepsilon 
}{}_{\gamma }{}^{\mu } \nabla_{\varepsilon }H_{\beta \mu \nu } 
\nabla^{\nu }H_{\alpha \delta \epsilon } \nn\\&&-  \frac{13}{16} 
R^{\alpha \beta \gamma \delta } R_{\gamma }{}^{\epsilon 
\varepsilon \mu } \nabla_{\mu }H_{\beta \varepsilon \nu } 
\nabla^{\nu }H_{\alpha \delta \epsilon } + \frac{15}{8} 
R_{\alpha }{}^{\epsilon }{}_{\gamma }{}^{\varepsilon } R^{\alpha 
\beta \gamma \delta } \nabla_{\varepsilon }H_{\epsilon \mu 
\nu } \nabla^{\nu }H_{\beta \delta }{}^{\mu } \nn\\&&+ \frac{1}{4} 
R_{\alpha }{}^{\epsilon }{}_{\gamma }{}^{\varepsilon } R^{\alpha 
\beta \gamma \delta } \nabla_{\varepsilon }H_{\delta \mu \nu 
} \nabla^{\nu }H_{\beta \epsilon }{}^{\mu } -  \frac{15}{8} 
R^{\alpha \beta \gamma \delta } R_{\gamma }{}^{\epsilon 
}{}_{\alpha }{}^{\varepsilon } \nabla_{\varepsilon }H_{\delta 
\mu \nu } \nabla^{\nu }H_{\beta \epsilon }{}^{\mu }\nn\\&& -  
\frac{1}{8} R^{\alpha \beta \gamma \delta } R_{\gamma 
}{}^{\epsilon }{}_{\alpha \delta } \nabla_{\epsilon 
}H_{\varepsilon \mu \nu } \nabla^{\nu }H_{\beta 
}{}^{\varepsilon \mu }\,,
\nn\\
{[}\nabla H^4{]}_{6}&=&\frac{1}{64} \nabla^{\delta }H^{\alpha \beta \gamma } 
\nabla_{\epsilon }H_{\delta \mu \nu } \nabla^{\varepsilon }H_{
\alpha \beta }{}^{\epsilon } \nabla^{\nu }H_{\gamma 
\varepsilon }{}^{\mu } -  \frac{1}{128} \nabla^{\delta 
}H^{\alpha \beta \gamma } \nabla^{\varepsilon }H_{\alpha \beta 
}{}^{\epsilon } \nabla_{\mu }H_{\delta \epsilon \nu } \nabla^{
\nu }H_{\gamma \varepsilon }{}^{\mu } \nn\\&&-  \frac{1}{128} 
\nabla^{\delta }H^{\alpha \beta \gamma } \nabla^{\varepsilon 
}H_{\alpha \beta }{}^{\epsilon } \nabla_{\nu }H_{\delta 
\epsilon \mu } \nabla^{\nu }H_{\gamma \varepsilon }{}^{\mu } - 
 \frac{9}{64} \nabla_{\delta }H_{\alpha \beta }{}^{\epsilon } 
\nabla^{\delta }H^{\alpha \beta \gamma } \nabla_{\nu 
}H_{\epsilon \varepsilon \mu } \nabla^{\nu }H_{\gamma 
}{}^{\varepsilon \mu }\nn\\&& -  \frac{1}{32} \nabla^{\delta 
}H^{\alpha \beta \gamma } \nabla^{\epsilon }H_{\alpha \beta 
\gamma } \nabla_{\nu }H_{\epsilon \varepsilon \mu } 
\nabla^{\nu }H_{\delta }{}^{\varepsilon \mu } + \frac{7}{1152} 
\nabla_{\delta }H_{\alpha \beta \gamma } \nabla^{\delta 
}H^{\alpha \beta \gamma } \nabla_{\nu }H_{\epsilon \varepsilon 
\mu } \nabla^{\nu }H^{\epsilon \varepsilon \mu }\,,
\nn\\
{[}H\nabla\nabla HR^2{]}_{1}&=& - \frac{1}{4} H_{\alpha \epsilon }{}^{\mu } R^{\alpha \beta 
\gamma \delta } R_{\gamma }{}^{\epsilon \varepsilon \zeta } 
\nabla_{\beta }\nabla_{\zeta }H_{\delta \varepsilon \mu }\,,
\nn\\
{[}H^8{]}_{4}&\!\!\!=\!\!\!& \frac{1}{2048} H_{\alpha }{}^{\delta \epsilon } H^{\alpha 
\beta \gamma } H_{\beta }{}^{\varepsilon \zeta } H_{\gamma 
}{}^{\eta \theta } H_{\delta \varepsilon }{}^{\iota } 
H_{\epsilon \eta }{}^{\kappa } H_{\zeta \kappa }{}^{\mu } 
H_{\theta \iota \mu } \nn\\&&+ \frac{1}{256} H_{\alpha }{}^{\delta 
\epsilon } H^{\alpha \beta \gamma } H_{\beta \delta 
}{}^{\varepsilon } H_{\gamma }{}^{\zeta \eta } H_{\epsilon 
\zeta }{}^{\theta } H_{\varepsilon }{}^{\iota \kappa } H_{\eta 
\iota }{}^{\mu } H_{\theta \kappa \mu }  \nn\\&&-  \frac{1}{512} 
H_{\alpha }{}^{\delta \epsilon } H^{\alpha \beta \gamma } 
H_{\beta \delta }{}^{\varepsilon } H_{\gamma \epsilon 
}{}^{\zeta } H_{\varepsilon }{}^{\eta \theta } H_{\zeta 
}{}^{\iota \kappa } H_{\eta \iota }{}^{\mu } H_{\theta \kappa 
\mu }  \nn\\&&+ \frac{1}{2048} H_{\alpha }{}^{\delta \epsilon } 
H^{\alpha \beta \gamma } H_{\beta }{}^{\varepsilon \zeta } 
H_{\gamma }{}^{\eta \theta } H_{\delta \varepsilon }{}^{\iota } 
H_{\epsilon \eta }{}^{\kappa } H_{\zeta \theta }{}^{\mu } 
H_{\iota \kappa \mu }
\,,
\nn\\
 {[}H^6R{]}_{1}&=&\frac{1}{32} H_{\alpha \gamma }{}^{\epsilon } H_{\beta \delta 
}{}^{\varepsilon } H_{\epsilon }{}^{\theta \lambda } 
H_{\varepsilon }{}^{\mu \nu } H_{\theta \mu }{}^{\tau } 
H_{\lambda \nu \tau } R^{\alpha \beta \gamma \delta }
\,,
\nn\\
{[}H^4R^2{]}_{12}&=&- \frac{1}{64} H_{\beta }{}^{\theta \mu } H_{\delta \theta 
\mu } H_{\epsilon }{}^{\nu \tau } H_{\varepsilon \nu \tau } 
R_{\alpha }{}^{\epsilon }{}_{\gamma }{}^{\varepsilon } R^{\alpha 
\beta \gamma \delta } -  \frac{81}{64} H_{\beta \delta 
}{}^{\theta } H_{\epsilon }{}^{\mu \nu } H_{\varepsilon \mu 
}{}^{\tau } H_{\theta \nu \tau } R_{\alpha }{}^{\epsilon 
}{}_{\gamma }{}^{\varepsilon } R^{\alpha \beta \gamma \delta } \nn\\&&
-  \frac{1}{32} H_{\beta \delta }{}^{\theta } H_{\epsilon 
\varepsilon }{}^{\mu } H_{\theta }{}^{\nu \tau } H_{\mu \nu 
\tau } R_{\alpha }{}^{\epsilon }{}_{\gamma }{}^{\varepsilon } 
R^{\alpha \beta \gamma \delta } + \frac{73}{128} H_{\beta 
}{}^{\varepsilon \theta } H_{\epsilon }{}^{\mu \nu } 
H_{\varepsilon \mu }{}^{\tau } H_{\theta \nu \tau } R^{\alpha 
\beta \gamma \delta } R_{\gamma }{}^{\epsilon }{}_{\alpha 
\delta } \nn\\&&-  \frac{19}{256} H_{\beta }{}^{\varepsilon \theta } 
H_{\epsilon }{}^{\mu \nu } H_{\varepsilon \theta }{}^{\tau } 
H_{\mu \nu \tau } R^{\alpha \beta \gamma \delta } R_{\gamma 
}{}^{\epsilon }{}_{\alpha \delta } -  \frac{75}{64} H_{\beta 
}{}^{\theta \mu } H_{\delta \varepsilon }{}^{\nu } H_{\epsilon 
\theta }{}^{\tau } H_{\mu \nu \tau } R^{\alpha \beta \gamma 
\delta } R_{\gamma }{}^{\epsilon }{}_{\alpha }{}^{\varepsilon } \nn\\&&
-  \frac{17}{64} H_{\alpha \epsilon }{}^{\mu } H_{\beta 
}{}^{\nu \tau } H_{\delta \nu \tau } H_{\varepsilon \theta 
\mu } R^{\alpha \beta \gamma \delta } R_{\gamma }{}^{\epsilon 
\varepsilon \theta } -  \frac{1}{64} H_{\alpha \epsilon 
}{}^{\mu } H_{\beta \mu }{}^{\nu } H_{\delta \nu }{}^{\tau } 
H_{\varepsilon \theta \tau } R^{\alpha \beta \gamma \delta } 
R_{\gamma }{}^{\epsilon \varepsilon \theta }\nn\\&& + \frac{277}{384} 
H_{\alpha \beta }{}^{\mu } H_{\delta \epsilon }{}^{\nu } 
H_{\varepsilon \theta }{}^{\tau } H_{\mu \nu \tau } R^{\alpha 
\beta \gamma \delta } R_{\gamma }{}^{\epsilon \varepsilon 
\theta } -  \frac{3}{32} H_{\alpha \gamma }{}^{\nu } H_{\beta 
\delta }{}^{\tau } H_{\epsilon \theta \nu } H_{\varepsilon \mu 
\tau } R^{\alpha \beta \gamma \delta } R^{\epsilon \varepsilon 
\theta \mu }\nn\\&& + \frac{1}{16} H_{\alpha \gamma \epsilon } 
H_{\beta \delta }{}^{\nu } H_{\varepsilon \nu }{}^{\tau } 
H_{\theta \mu \tau } R^{\alpha \beta \gamma \delta } 
R^{\epsilon \varepsilon \theta \mu } -  \frac{11}{32} 
H_{\alpha \gamma \epsilon } H_{\beta \delta \theta } 
H_{\varepsilon }{}^{\nu \tau } H_{\mu \nu \tau } R^{\alpha 
\beta \gamma \delta } R^{\epsilon \varepsilon \theta \mu }\,,
\nn\\
{[}H^2 R^3 {]}_{20}&=&
\frac{1}{2} H_{\epsilon \mu }{}^{\tau } H_{\varepsilon \nu 
\tau } R_{\alpha }{}^{\epsilon }{}_{\gamma }{}^{\varepsilon } 
R^{\alpha \beta \gamma \delta } R_{\beta }{}^{\mu }{}_{\delta 
}{}^{\nu } + \frac{15}{16} H_{\epsilon \varepsilon }{}^{\tau } 
H_{\mu \nu \tau } R_{\alpha }{}^{\epsilon }{}_{\gamma 
}{}^{\varepsilon } R^{\alpha \beta \gamma \delta } R_{\beta 
}{}^{\mu }{}_{\delta }{}^{\nu }\nn\\&& -  \frac{1}{32} H_{\beta 
}{}^{\nu \tau } H_{\epsilon \nu \tau } R_{\alpha }{}^{\epsilon 
\varepsilon \mu } R^{\alpha \beta \gamma \delta } R_{\gamma 
\varepsilon \delta \mu } -  \frac{5}{16} H_{\beta \delta 
\epsilon } H_{\mu \nu \tau } R_{\alpha }{}^{\epsilon 
\varepsilon \mu } R^{\alpha \beta \gamma \delta } R_{\gamma 
}{}^{\nu }{}_{\varepsilon }{}^{\tau } \nn\\&&+ \frac{3}{16} 
H_{\epsilon }{}^{\nu \tau } H_{\mu \nu \tau } R_{\alpha 
}{}^{\epsilon }{}_{\gamma }{}^{\varepsilon } R^{\alpha \beta 
\gamma \delta } R_{\delta }{}^{\mu }{}_{\beta \varepsilon } + 
\frac{9}{16} H_{\epsilon \mu }{}^{\tau } H_{\varepsilon \nu 
\tau } R_{\alpha }{}^{\epsilon }{}_{\gamma }{}^{\varepsilon } 
R^{\alpha \beta \gamma \delta } R_{\delta }{}^{\mu }{}_{\beta 
}{}^{\nu } \nn\\&&-  \frac{15}{4} H_{\beta \mu }{}^{\tau } 
H_{\epsilon \nu \tau } R_{\alpha }{}^{\epsilon }{}_{\gamma 
}{}^{\varepsilon } R^{\alpha \beta \gamma \delta } R_{\delta 
}{}^{\mu }{}_{\varepsilon }{}^{\nu } -  \frac{1}{8} H_{\beta 
\epsilon }{}^{\tau } H_{\mu \nu \tau } R_{\alpha }{}^{\epsilon 
}{}_{\gamma }{}^{\varepsilon } R^{\alpha \beta \gamma \delta } 
R_{\delta }{}^{\mu }{}_{\varepsilon }{}^{\nu } \nn\\&&+ \frac{13}{8} 
H_{\beta \mu }{}^{\tau } H_{\epsilon \nu \tau } R^{\alpha 
\beta \gamma \delta } R_{\gamma }{}^{\epsilon }{}_{\alpha }{}^{
\varepsilon } R_{\delta }{}^{\mu }{}_{\varepsilon }{}^{\nu } -  
\frac{15}{8} H_{\beta \epsilon }{}^{\tau } H_{\mu \nu \tau } 
R^{\alpha \beta \gamma \delta } R_{\gamma }{}^{\epsilon 
}{}_{\alpha }{}^{\varepsilon } R_{\delta }{}^{\mu 
}{}_{\varepsilon }{}^{\nu }\nn\\&& + \frac{1}{4} H_{\beta \epsilon 
\nu } H_{\delta \mu \tau } R^{\alpha \beta \gamma \delta } 
R_{\gamma }{}^{\epsilon \varepsilon \mu } R_{\varepsilon 
}{}^{\nu }{}_{\alpha }{}^{\tau } + \frac{19}{16} H_{\beta 
\delta \mu } H_{\epsilon \nu \tau } R^{\alpha \beta \gamma 
\delta } R_{\gamma }{}^{\epsilon \varepsilon \mu } 
R_{\varepsilon }{}^{\nu }{}_{\alpha }{}^{\tau } \nn\\&&+ \frac{21}{16} 
H_{\alpha \epsilon \nu } H_{\beta \mu \tau } R^{\alpha \beta 
\gamma \delta } R_{\gamma }{}^{\epsilon \varepsilon \mu } 
R_{\varepsilon }{}^{\nu }{}_{\delta }{}^{\tau } + \frac{1}{4} 
H_{\epsilon \varepsilon }{}^{\tau } H_{\mu \nu \tau } 
R^{\alpha \beta \gamma \delta } R_{\gamma }{}^{\epsilon 
}{}_{\alpha \delta } R^{\varepsilon \mu }{}_{\beta }{}^{\nu }\nn\\&& + 
\frac{1}{8} H_{\beta \varepsilon \nu } H_{\epsilon \mu \tau } 
R^{\alpha \beta \gamma \delta } R_{\gamma }{}^{\epsilon 
}{}_{\alpha \delta } R^{\varepsilon \mu \nu \tau } + 
\frac{15}{8} H_{\beta \mu }{}^{\tau } H_{\epsilon \nu \tau } 
R_{\alpha }{}^{\epsilon }{}_{\gamma }{}^{\varepsilon } R^{\alpha 
\beta \gamma \delta } R^{\mu \nu }{}_{\delta \varepsilon }\nn\\&& -  
\frac{21}{16} H_{\beta \mu }{}^{\tau } H_{\epsilon \nu \tau } 
R^{\alpha \beta \gamma \delta } R_{\gamma }{}^{\epsilon 
}{}_{\alpha }{}^{\varepsilon } R^{\mu \nu }{}_{\delta 
\varepsilon } -  \frac{29}{8} H_{\beta \varepsilon \mu } 
H_{\epsilon \nu \tau } R_{\alpha }{}^{\epsilon }{}_{\gamma 
}{}^{\varepsilon } R^{\alpha \beta \gamma \delta } R^{\mu \nu 
}{}_{\delta }{}^{\tau } \nn\\&&+ \frac{15}{8} H_{\beta \varepsilon 
\mu } H_{\epsilon \nu \tau } R^{\alpha \beta \gamma \delta } 
R_{\gamma }{}^{\epsilon }{}_{\alpha }{}^{\varepsilon } R^{\mu 
\nu }{}_{\delta }{}^{\tau } -  \frac{7}{32} H_{\alpha \delta 
\epsilon } H_{\beta \nu \tau } R^{\alpha \beta \gamma \delta 
} R_{\gamma }{}^{\epsilon \varepsilon \mu } R^{\nu \tau 
}{}_{\varepsilon \mu }\,.
\eeqa
The couplings with structures \([H^2\nabla H^2 R]_{100}\) and \([H^4\nabla H^2]_{99}\) are presented in the Appendix. By exploiting the cyclic symmetry of the Riemann curvature, we have expressed the three Riemann curvature terms in \(\reef{S32}\) as two terms in \([R^4]_2\). These \([R^4]_2\) couplings precisely match those derived from the non-linear sigma model \cite{Jack:1989vp} after appropriate field redefinitions \cite{Codina:2021cxh}. At order \(\alpha'^3\), the couplings in the Meissner scheme \(\reef{L22}\) are equivalent to those in \(\reef{L221}\) up to field redefinitions. Consequently, the  couplings \(\reef{L22}\) also maintain full consistency with the gravity results obtained from the non-linear sigma model \cite{Jack:1989vp}.

\section{Discussion}

In this work, we have utilized the $\MZ_2$-symmetry of the circularly reduced effective action in bosonic string theory to determine the effective action at order $\alpha'^3$. This $\MZ_2$-symmetry fixes 871 of the 872 coupling constants in the minimal basis at this order. The remaining parameter is determined by matching the four-graviton S-matrix element derived from the effective action with the corresponding string theory amplitude.
We have derived this effective action in two distinct schemes: one where the $\alpha'$-order action follows the Meissner scheme (given in \reef{L22}), and another following the Metsaev-Tseytlin scheme \reef{L221}. For the $H=0$ case, both formulations agree with the well-known results previously obtained from the non-linear sigma model.

We have used the field redefinition relation \reef{F3} to express the T-duality-derived couplings in their respective canonical forms: those from \reef{SM3} in the Meissner scheme are written as in \reef{L22}, and those from \reef{ST32} in the Metsaev-Tseytlin scheme are written as in \reef{L221}. For these redefinitions, we assume the corresponding effective actions at the four- and six-derivative levels are fixed. Specifically, we use the actions in \reef{fourmax} and \reef{finalB} for the Meissner scheme, and those in \reef{fourmin} and \reef{SMT2} for the Metsaev-Tseytlin scheme.
If $\Psi$ represents all massless fields, the field redefinition is given by
\beqa
\Psi'=\Psi+\alpha'^3\Psi_3\,,\labell{Psi}
\eeqa
where $\Psi_3$ involves all six-derivative contractions of the massless fields with unknown coefficients. The field redefinition relation \reef{F3} then produces a system of linear algebraic equations relating these coefficients to the parameters in the new eight-derivative effective action. Solving this system fixes the parameters of the effective action to match those in \reef{L22} for the Meissner scheme, as well as the parameters of the field redefinition $\Psi_3$, in which we are not interested.

 One may use field redefinitions in which the four- and six-derivative couplings are altered along with the eight-derivative couplings. In this case, the appropriate field redefinition is:
\beqa
\Psi'=\Psi+\alpha'\Psi_1+\alpha'^2\Psi_2+\alpha'^3\Psi_3\,,\labell{Psi1}
\eeqa
where $\Psi_1$ involves all two-derivative contractions of the massless fields, and $\Psi_2$ involves all four-derivative contractions, all with unknown coefficients.
Extending the relation \reef{F3} for \reef{Psi} to the field redefinition \reef{Psi1} yields a structure similar to the T-duality relation \reef{T3}. This structure involves terms that depend nonlinearly on the parameters in $\Psi_1$ and $\Psi_2$, as well as terms that depend linearly on the parameters in $\Psi_1$, $\Psi_2$, $\Psi_3$, and the parameters in the eight-derivative effective action. Consequently, the algebraic equations that must be solved to change the scheme for the four-, six-, and eight-derivative couplings depend nonlinearly on the parameters of $\Psi_1$ and $\Psi_2$, and linearly on the parameters of the eight-derivative effective action and of $\Psi_1$, $\Psi_2$, and $\Psi_3$. While solving these nonlinear equations would yield the parameters for both the field redefinitions and the new actions, they are too complex to be solved.

To address this, we redefined the multiples of the parameters in $\Psi_1$ and $\Psi_2$ that appear in the algebraic equations as new linear parameters. This converts the nonlinear algebraic system into a linear one, which can then be solved to find the coefficients of the new eight-derivative effective action. We found a scheme in which only 204 coupling constants are non-zero.
However, there is no guarantee that these non-zero parameters, when substituted back into the original nonlinear equations, yield a solution for the parameters in $\Psi_1$, $\Psi_2$, and $\Psi_3$. Even if a solution exists, it would be hard to solve for the parameters of the field redefinitions $\Psi_1$ and $\Psi_2$, which are needed to determine the new four- and six-derivative actions. The eight-derivative action has limited utility without the corresponding four- and six-derivative terms. For example, it cannot be used independently to compute S-matrix elements. For this reason, the 204 eight-derivative couplings are not listed here.
Moreover, as argued in Section 2.1, the nonlinear terms cannot be ignored because imposing field redefinitions at order $\alpha'^2$ and higher is not equivalent to using the equations of motion—a procedure which is effectively the same as ignoring the nonlinear terms—even though they are equivalent at  order $\alpha'$.

Determining the NS-NS couplings in chiral heterotic string theories at order $\alpha'^3$ follows essentially the same procedure as developed in this paper, with two key distinctions. First, the chiral anomaly cancellation requires extending the H-field as \cite{Green:1984sg}
\beqa 
H_{\alpha\beta\gamma} &\rightarrow& H_{\alpha\beta\gamma} + \frac{3}{2}\alpha' \Omega_{\alpha\beta\gamma}\,, \labell{Hreplace} 
\eeqa
where $\Omega$  represents the gravitational Chern-Simons term. This modification, mandated by the Green-Schwarz mechanism \cite{Green:1984sg}, introduces additional terms involving $\Omega$ in the effective action at each $\alpha'$ order. The coefficients of these terms are already fixed through T-duality constraints at lower $\alpha'$ orders.
The second difference involves the structure of the effective action, which must now include both parity-even terms (containing even powers of $H$) and
parity-odd terms (containing odd powers of $H$).
While such couplings at order $\alpha'^2$ have been determined through T-duality \cite{Garousi:2024avb,Garousi:2023kxw,Garousi:2024imy}, their $\alpha'^3$ counterparts remain to be found. For the $H=0$ case, these couplings have been established through both S-matrix calculations \cite{Gross:1986mw} and T-duality arguments \cite{Razaghian:2018svg}. The extension to general $H$-field configurations presents an interesting challenge for future work.

In principle, we expect that the $\MZ_2$-symmetry of the classical effective action, combined with consistency requirements from the four-point string theory S-matrix element, should uniquely determine the effective action to all orders in $\alpha'$. This field theory framework could then be used to systematically construct $n$-point sphere-level S-matrix elements in string theory. Since the four-point function involves only $\z(2k+1)$ terms, we anticipate that closed string $n$-point S-matrix elements (for $n>4$) should similarly depend exclusively on  $\z(2k+1)$ functions. This expectation aligns with existing literature suggesting that while disk-level amplitudes contain both  $\z(2k)$ and  $\z(2k+1)$ terms, the KLT relations preserve only  $\z(2k+1)$ terms in sphere-level amplitudes \cite{Stieberger:2013wea,Schlotterer:2018zce,Alday:2025bjp}. This prediction has been explicitly verified in superstring theory through KLT calculations: as demonstrated in \cite{Stieberger:2009rr}, the 5- and 6-point graviton amplitudes indeed contain no  $\z(2k)$ terms, in perfect agreement with the constraints imposed by the $\MZ_2$-symmetry.

The  $\z(2k+1)$ terms in field theory arise from massive poles in string theory S-matrix elements. If one could consistently truncate string theory to retain only tachyon and massless states while removing massive states, the resulting S-matrix elements would produce the following tower of couplings:
\beqa
\bS&=&-\frac{2}{\kappa^2}\int d^Dx\sqrt{-G}\, e^{-2\Phi}\sum_{m=0}^{\infty}  \alpha'^m \cL_{m}\labell{S1}.  
\eeqa
Here, the Lagrangian $\cL_0$ is given in  \reef{L0}, while $\cL_1$, $\cL_2$, and $\cL_3$ correspond to the couplings presented in this paper after removing all $\z(3)$ terms. T-duality constraints require that all terms in this action contain only one undetermined parameter: the overall factor of the effective action at order $\alpha'$.
A similar structure appears in truncated heterotic string theory, though with an important modification: the replacement   \reef{Hreplace} imposes that T-duality completely fixes the effective action in terms of the leading-order overall factor \cite{Garousi:2025xqn}. Since these actions have uniquely determined couplings, they might be derivable from a Double Field Theory (DFT) like the one in  \cite{Hohm:2013jaa}, where T-duality remains exact but coordinate transformations acquire higher-derivative corrections. In fact, as demonstrated in \cite{Hronek:2020xxi}, DFT cannot reproduce the $\z(3)$ terms in the effective action. This aligns with the speculation that spacetime symmetries alone may be insufficient to fully determine the effective action of string theory.

\vskip 0.5 cm

{\bf Acknowledgments}: 

A.P. and M.A. would like to sincerely thank S.F. Moosavian and H. Erbin for insightful discussions.
M.A. is grateful to G. Moore for a helpful conversation on T-duality and extends appreciation to the organizers of the String-Math 2024 conference for the opportunity to contribute.
A.P. thanks U. Moitra for valuable feedback during the Spring School on Superstring Theory and Related Topics 2025.
Both A.P. and M.A. acknowledge with gratitude the financial support provided by the Abdus Salam International Centre for Theoretical Physics (ICTP) during their stay.   This work is supported by Ferdowsi University of Mashhad under grant  371(1403/06/28).

\vskip 0.5 cm
{\Large \bf Appendix: }{\large\bf $H^2\nabla H^2 R$ and $H^4\nabla H^2$ couplings}
\vskip 0.5 cm
The complete forms of the $H^2\nabla H^2 R$ and $H^4\nabla H^2$ couplings are too extensive for the main text and are therefore provided in this appendix. Below we list the $H^2\nabla H^2 R$ couplings derived in the Meissner scheme:
\beqa
[H^2\nabla H^2 R]_{92}&=&\frac{1}{64} H_{\epsilon }{}^{\nu \tau } H^{\epsilon 
\varepsilon \mu } R^{\alpha \beta \gamma \delta } 
\nabla_{\gamma }H_{\alpha \varepsilon \nu } \nabla_{\delta 
}H_{\beta \mu \tau } -  \frac{1}{64} H_{\epsilon }{}^{\nu 
\tau } H^{\epsilon \varepsilon \mu } R^{\alpha \beta \gamma 
\delta } \nabla_{\gamma }H_{\alpha \varepsilon \mu } 
\nabla_{\delta }H_{\beta \nu \tau } \nn\\&&+ \frac{3}{64} 
H_{\epsilon \varepsilon }{}^{\nu } H^{\epsilon \varepsilon \mu 
} R^{\alpha \beta \gamma \delta } \nabla_{\gamma }H_{\alpha 
\mu }{}^{\tau } \nabla_{\delta }H_{\beta \nu \tau } -  
\frac{1}{32} H_{\alpha }{}^{\epsilon \varepsilon } H^{\mu \nu 
\tau } R^{\alpha \beta \gamma \delta } \nabla_{\gamma 
}H_{\beta \epsilon \mu } \nabla_{\varepsilon }H_{\delta \nu 
\tau } \nn\\&&+ \frac{1}{8} H_{\alpha }{}^{\epsilon \varepsilon } 
H_{\epsilon }{}^{\mu \nu } R^{\alpha \beta \gamma \delta } 
\nabla_{\gamma }H_{\beta \mu }{}^{\tau } \nabla_{\varepsilon 
}H_{\delta \nu \tau } + \frac{1}{64} H_{\alpha }{}^{\epsilon 
\varepsilon } H_{\epsilon \varepsilon }{}^{\mu } R^{\alpha 
\beta \gamma \delta } \nabla_{\gamma }H_{\beta }{}^{\nu \tau 
} \nabla_{\mu }H_{\delta \nu \tau }\nn\\&& + \frac{1}{128} H_{\alpha 
\beta }{}^{\epsilon } H_{\epsilon }{}^{\varepsilon \mu } 
R^{\alpha \beta \gamma \delta } \nabla_{\varepsilon }H_{\gamma 
}{}^{\nu \tau } \nabla_{\mu }H_{\delta \nu \tau } + 
\frac{1}{64} H_{\alpha }{}^{\epsilon \varepsilon } H_{\gamma 
\epsilon }{}^{\mu } R^{\alpha \beta \gamma \delta } 
\nabla_{\beta }H_{\varepsilon \nu \tau } \nabla_{\mu 
}H_{\delta }{}^{\nu \tau }\nn\\&& -  \frac{1}{64} H_{\alpha 
}{}^{\epsilon \varepsilon } H_{\gamma \epsilon }{}^{\mu } 
R^{\alpha \beta \gamma \delta } \nabla_{\delta }H_{\beta }{}^{
\nu \tau } \nabla_{\mu }H_{\varepsilon \nu \tau } -  
\frac{1}{16} H_{\alpha }{}^{\epsilon \varepsilon } H_{\epsilon 
}{}^{\mu \nu } R^{\alpha \beta \gamma \delta } 
\nabla_{\varepsilon }H_{\beta \mu }{}^{\tau } \nabla_{\nu }H_{
\gamma \delta \tau }\nn\\&& -  \frac{1}{32} H_{\alpha }{}^{\epsilon 
\varepsilon } H_{\epsilon }{}^{\mu \nu } R^{\alpha \beta 
\gamma \delta } \nabla_{\gamma }H_{\beta \mu }{}^{\tau } 
\nabla_{\nu }H_{\delta \varepsilon \tau } -  \frac{3}{16} 
H_{\alpha }{}^{\epsilon \varepsilon } H_{\gamma }{}^{\mu \nu } 
R^{\alpha \beta \gamma \delta } \nabla_{\epsilon }H_{\beta 
\mu }{}^{\tau } \nabla_{\nu }H_{\delta \varepsilon \tau }\nn\\&& -  
\frac{7}{64} H_{\alpha }{}^{\epsilon \varepsilon } H_{\epsilon 
}{}^{\mu \nu } R^{\alpha \beta \gamma \delta } \nabla_{\gamma 
}H_{\beta \varepsilon }{}^{\tau } \nabla_{\nu }H_{\delta \mu 
\tau } + \frac{23}{256} H_{\alpha \beta }{}^{\epsilon } 
H^{\varepsilon \mu \nu } R^{\alpha \beta \gamma \delta } 
\nabla_{\epsilon }H_{\gamma \varepsilon }{}^{\tau } 
\nabla_{\nu }H_{\delta \mu \tau }\nn\\&& + \frac{5}{32} H_{\alpha 
}{}^{\epsilon \varepsilon } H_{\beta }{}^{\mu \nu } R^{\alpha 
\beta \gamma \delta } \nabla_{\varepsilon }H_{\gamma \epsilon 
}{}^{\tau } \nabla_{\nu }H_{\delta \mu \tau } -  
\frac{11}{64} H_{\alpha \beta }{}^{\epsilon } H^{\varepsilon 
\mu \nu } R^{\alpha \beta \gamma \delta } \nabla_{\varepsilon 
}H_{\gamma \epsilon }{}^{\tau } \nabla_{\nu }H_{\delta \mu 
\tau }\nn\\&& + \frac{5}{128} H_{\alpha }{}^{\epsilon \varepsilon } 
H_{\gamma }{}^{\mu \nu } R^{\alpha \beta \gamma \delta } 
\nabla_{\beta }H_{\epsilon \varepsilon \tau } \nabla_{\nu }H_{
\delta \mu }{}^{\tau } + \frac{65}{512} H_{\alpha \beta 
}{}^{\epsilon } H^{\varepsilon \mu \nu } R^{\alpha \beta 
\gamma \delta } \nabla_{\varepsilon }H_{\gamma \delta 
}{}^{\tau } \nabla_{\nu }H_{\epsilon \mu \tau }\nn\\&& -  
\frac{7}{32} H_{\alpha }{}^{\epsilon \varepsilon } H_{\epsilon 
}{}^{\mu \nu } R^{\alpha \beta \gamma \delta } \nabla_{\delta 
}H_{\beta \gamma }{}^{\tau } \nabla_{\nu }H_{\varepsilon \mu 
\tau } -  \frac{7}{64} H_{\alpha }{}^{\epsilon \varepsilon } 
H_{\gamma }{}^{\mu \nu } R^{\alpha \beta \gamma \delta } 
\nabla_{\delta }H_{\beta \epsilon }{}^{\tau } \nabla_{\nu }H_{
\varepsilon \mu \tau } \nn\\&&+ \frac{7}{128} H_{\alpha }{}^{\epsilon 
\varepsilon } H_{\beta }{}^{\mu \nu } R^{\alpha \beta \gamma 
\delta } \nabla_{\epsilon }H_{\gamma \delta }{}^{\tau } 
\nabla_{\nu }H_{\varepsilon \mu \tau } -  \frac{41}{256} 
H_{\alpha }{}^{\epsilon \varepsilon } H^{\mu \nu \tau } 
R^{\alpha \beta \gamma \delta } \nabla_{\epsilon }H_{\beta 
\mu \nu } \nabla_{\tau }H_{\gamma \delta \varepsilon } \nn\\&&+ 
\frac{5}{64} H_{\epsilon }{}^{\nu \tau } H^{\epsilon 
\varepsilon \mu } R^{\alpha \beta \gamma \delta } 
\nabla_{\beta }H_{\alpha \varepsilon \nu } \nabla_{\tau 
}H_{\gamma \delta \mu } + \frac{3}{256} H_{\epsilon }{}^{\nu 
\tau } H^{\epsilon \varepsilon \mu } R^{\alpha \beta \gamma 
\delta } \nabla_{\nu }H_{\alpha \beta \varepsilon } 
\nabla_{\tau }H_{\gamma \delta \mu }\nn\\&& + \frac{17}{128} 
H_{\alpha }{}^{\epsilon \varepsilon } H^{\mu \nu \tau } 
R^{\alpha \beta \gamma \delta } \nabla_{\varepsilon }H_{\beta 
\epsilon \mu } \nabla_{\tau }H_{\gamma \delta \nu } + 
\frac{5}{128} H_{\epsilon }{}^{\nu \tau } H^{\epsilon 
\varepsilon \mu } R^{\alpha \beta \gamma \delta } \nabla_{\mu 
}H_{\alpha \beta \varepsilon } \nabla_{\tau }H_{\gamma \delta 
\nu }\nn\\&& + \frac{1}{64} H_{\alpha }{}^{\epsilon \varepsilon } 
H^{\mu \nu \tau } R^{\alpha \beta \gamma \delta } 
\nabla_{\gamma }H_{\beta \mu \nu } \nabla_{\tau }H_{\delta 
\epsilon \varepsilon } -  \frac{1}{64} H_{\alpha }{}^{\epsilon 
\varepsilon } H^{\mu \nu \tau } R^{\alpha \beta \gamma \delta 
} \nabla_{\nu }H_{\beta \gamma \mu } \nabla_{\tau }H_{\delta 
\epsilon \varepsilon }\nn\\&& -  \frac{5}{64} H_{\alpha }{}^{\epsilon 
\varepsilon } H^{\mu \nu \tau } R^{\alpha \beta \gamma \delta 
} \nabla_{\gamma }H_{\beta \epsilon \mu } \nabla_{\tau 
}H_{\delta \varepsilon \nu } + \frac{3}{32} H_{\alpha 
}{}^{\epsilon \varepsilon } H^{\mu \nu \tau } R^{\alpha \beta 
\gamma \delta } \nabla_{\epsilon }H_{\beta \gamma \mu } 
\nabla_{\tau }H_{\delta \varepsilon \nu } \nn\\&&-  \frac{13}{32} H_{
\alpha }{}^{\epsilon \varepsilon } H^{\mu \nu \tau } R^{\alpha 
\beta \gamma \delta } \nabla_{\mu }H_{\beta \gamma \epsilon } 
\nabla_{\tau }H_{\delta \varepsilon \nu } -  \frac{1}{32} 
H_{\epsilon }{}^{\nu \tau } H^{\epsilon \varepsilon \mu } 
R^{\alpha \beta \gamma \delta } \nabla_{\gamma }H_{\alpha 
\beta \varepsilon } \nabla_{\tau }H_{\delta \mu \nu }\nn\\&& + 
\frac{1}{16} H_{\alpha }{}^{\epsilon \varepsilon } H^{\mu \nu 
\tau } R^{\alpha \beta \gamma \delta } \nabla_{\varepsilon 
}H_{\beta \gamma \epsilon } \nabla_{\tau }H_{\delta \mu \nu } 
+ \frac{1}{32} H_{\alpha }{}^{\epsilon \varepsilon } H^{\mu 
\nu \tau } R^{\alpha \beta \gamma \delta } \nabla_{\delta 
}H_{\beta \gamma \mu } \nabla_{\tau }H_{\epsilon \varepsilon 
\nu }\nn\\&& -  \frac{145}{1024} H_{\alpha \beta }{}^{\epsilon } 
H^{\varepsilon \mu \nu } R^{\alpha \beta \gamma \delta } 
\nabla_{\varepsilon }H_{\gamma \delta }{}^{\tau } \nabla_{\tau 
}H_{\epsilon \mu \nu } + \frac{41}{128} H_{\alpha 
}{}^{\epsilon \varepsilon } H^{\mu \nu \tau } R^{\alpha \beta 
\gamma \delta } \nabla_{\delta }H_{\beta \gamma \epsilon } 
\nabla_{\tau }H_{\varepsilon \mu \nu } \nn\\&&-  \frac{3}{256} 
H_{\alpha }{}^{\epsilon \varepsilon } H_{\epsilon }{}^{\mu \nu 
} R^{\alpha \beta \gamma \delta } \nabla_{\delta }H_{\beta 
\gamma }{}^{\tau } \nabla_{\tau }H_{\varepsilon \mu \nu } + 
\frac{35}{128} H_{\alpha }{}^{\epsilon \varepsilon } H_{\gamma 
}{}^{\mu \nu } R^{\alpha \beta \gamma \delta } \nabla_{\delta 
}H_{\beta \epsilon }{}^{\tau } \nabla_{\tau }H_{\varepsilon 
\mu \nu } \nn\\&&+ \frac{11}{512} H_{\alpha }{}^{\epsilon \varepsilon 
} H_{\beta }{}^{\mu \nu } R^{\alpha \beta \gamma \delta } 
\nabla_{\epsilon }H_{\gamma \delta }{}^{\tau } \nabla_{\tau 
}H_{\varepsilon \mu \nu } -  \frac{45}{128} H_{\gamma 
}{}^{\epsilon \varepsilon } H_{\epsilon }{}^{\mu \nu } 
R^{\alpha \beta \gamma \delta } \nabla_{\beta }H_{\delta 
\varepsilon \tau } \nabla^{\tau }H_{\alpha \mu \nu }\nn\\&& -  
\frac{7}{64} H_{\alpha }{}^{\epsilon \varepsilon } H_{\epsilon 
}{}^{\mu \nu } R^{\alpha \beta \gamma \delta } \nabla_{\nu 
}H_{\delta \varepsilon \tau } \nabla^{\tau }H_{\beta \gamma 
\mu } -  \frac{1}{8} H_{\alpha }{}^{\epsilon \varepsilon } 
H_{\gamma }{}^{\mu \nu } R^{\alpha \beta \gamma \delta } 
\nabla_{\tau }H_{\delta \varepsilon \nu } \nabla^{\tau 
}H_{\beta \epsilon \mu }\nn\\&& -  \frac{35}{128} H_{\alpha 
}{}^{\epsilon \varepsilon } H_{\epsilon }{}^{\mu \nu } 
R^{\alpha \beta \gamma \delta } \nabla_{\nu }H_{\gamma \delta 
\tau } \nabla^{\tau }H_{\beta \varepsilon \mu } + 
\frac{5}{128} H_{\alpha }{}^{\epsilon \varepsilon } H_{\epsilon 
}{}^{\mu \nu } R^{\alpha \beta \gamma \delta } \nabla_{\tau 
}H_{\gamma \delta \nu } \nabla^{\tau }H_{\beta \varepsilon 
\mu }\nn\\&& -  \frac{25}{512} H_{\alpha }{}^{\epsilon \varepsilon } 
H_{\epsilon }{}^{\mu \nu } R^{\alpha \beta \gamma \delta } 
\nabla_{\varepsilon }H_{\gamma \delta \tau } \nabla^{\tau }H_{
\beta \mu \nu } -  \frac{11}{256} H_{\alpha }{}^{\epsilon 
\varepsilon } H_{\epsilon }{}^{\mu \nu } R^{\alpha \beta 
\gamma \delta } \nabla_{\tau }H_{\gamma \delta \varepsilon } 
\nabla^{\tau }H_{\beta \mu \nu } \nn\\&&+ \frac{3}{256} H_{\alpha 
}{}^{\epsilon \varepsilon } H_{\gamma }{}^{\mu \nu } R^{\alpha 
\beta \gamma \delta } \nabla_{\tau }H_{\delta \epsilon 
\varepsilon } \nabla^{\tau }H_{\beta \mu \nu } -  
\frac{79}{512} H_{\alpha }{}^{\epsilon \varepsilon } 
H_{\epsilon \varepsilon }{}^{\mu } R^{\alpha \beta \gamma 
\delta } \nabla_{\nu }H_{\gamma \delta \tau } \nabla^{\tau 
}H_{\beta \mu }{}^{\nu } \nn\\&&-  \frac{3}{16} H_{\alpha 
}{}^{\epsilon \varepsilon } H_{\gamma \epsilon }{}^{\mu } 
R^{\alpha \beta \gamma \delta } \nabla_{\nu }H_{\delta 
\varepsilon \tau } \nabla^{\tau }H_{\beta \mu }{}^{\nu } + 
\frac{41}{256} H_{\alpha }{}^{\epsilon \varepsilon } 
H_{\epsilon \varepsilon }{}^{\mu } R^{\alpha \beta \gamma 
\delta } \nabla_{\tau }H_{\gamma \delta \nu } \nabla^{\tau 
}H_{\beta \mu }{}^{\nu } \nn\\&&+ \frac{7}{32} H_{\alpha 
}{}^{\epsilon \varepsilon } H_{\gamma \epsilon }{}^{\mu } 
R^{\alpha \beta \gamma \delta } \nabla_{\tau }H_{\delta 
\varepsilon \nu } \nabla^{\tau }H_{\beta \mu }{}^{\nu } -  
\frac{3}{128} H_{\alpha }{}^{\epsilon \varepsilon } H_{\gamma 
\epsilon \varepsilon } R^{\alpha \beta \gamma \delta } 
\nabla_{\tau }H_{\delta \mu \nu } \nabla^{\tau }H_{\beta 
}{}^{\mu \nu } \nn\\&&+ \frac{13}{128} H_{\alpha }{}^{\epsilon 
\varepsilon } H_{\beta }{}^{\mu \nu } R^{\alpha \beta \gamma 
\delta } \nabla_{\nu }H_{\varepsilon \mu \tau } \nabla^{\tau 
}H_{\gamma \delta \epsilon } -  \frac{3}{128} H_{\alpha 
}{}^{\epsilon \varepsilon } H_{\beta }{}^{\mu \nu } R^{\alpha 
\beta \gamma \delta } \nabla_{\tau }H_{\varepsilon \mu \nu } 
\nabla^{\tau }H_{\gamma \delta \epsilon } \nn\\&&+ \frac{29}{1536} 
H_{\alpha \beta }{}^{\epsilon } H^{\varepsilon \mu \nu } 
R^{\alpha \beta \gamma \delta } \nabla_{\tau }H_{\varepsilon 
\mu \nu } \nabla^{\tau }H_{\gamma \delta \epsilon } -  
\frac{13}{128} H_{\alpha }{}^{\epsilon \varepsilon } 
H_{\epsilon }{}^{\mu \nu } R^{\alpha \beta \gamma \delta } 
\nabla_{\beta }H_{\mu \nu \tau } \nabla^{\tau }H_{\gamma 
\delta \varepsilon } \nn\\&&+ \frac{107}{512} H_{\alpha \beta 
}{}^{\epsilon } H^{\varepsilon \mu \nu } R^{\alpha \beta 
\gamma \delta } \nabla_{\nu }H_{\epsilon \mu \tau } 
\nabla^{\tau }H_{\gamma \delta \varepsilon } -  
\frac{99}{1024} H_{\alpha \beta }{}^{\epsilon } H^{\varepsilon 
\mu \nu } R^{\alpha \beta \gamma \delta } \nabla_{\tau 
}H_{\epsilon \mu \nu } \nabla^{\tau }H_{\gamma \delta 
\varepsilon } \nn\\&&+ \frac{1}{128} H_{\alpha }{}^{\epsilon 
\varepsilon } H_{\epsilon }{}^{\mu \nu } R^{\alpha \beta 
\gamma \delta } \nabla_{\beta }H_{\varepsilon \nu \tau } 
\nabla^{\tau }H_{\gamma \delta \mu } -  \frac{33}{256} 
H_{\alpha }{}^{\epsilon \varepsilon } H_{\epsilon \varepsilon 
}{}^{\mu } R^{\alpha \beta \gamma \delta } \nabla_{\beta 
}H_{\mu \nu \tau } \nabla^{\tau }H_{\gamma \delta }{}^{\nu } 
\nn\\&&+ \frac{41}{256} H_{\alpha }{}^{\epsilon \varepsilon } H_{\beta 
\epsilon }{}^{\mu } R^{\alpha \beta \gamma \delta } 
\nabla_{\nu }H_{\varepsilon \mu \tau } \nabla^{\tau 
}H_{\gamma \delta }{}^{\nu } -  \frac{29}{256} H_{\alpha \beta 
}{}^{\epsilon } H_{\epsilon }{}^{\varepsilon \mu } R^{\alpha 
\beta \gamma \delta } \nabla_{\nu }H_{\varepsilon \mu \tau } 
\nabla^{\tau }H_{\gamma \delta }{}^{\nu }\nn\\&& -  \frac{5}{128} H_{
\alpha }{}^{\epsilon \varepsilon } H_{\beta \epsilon }{}^{\mu } 
R^{\alpha \beta \gamma \delta } \nabla_{\tau }H_{\varepsilon 
\mu \nu } \nabla^{\tau }H_{\gamma \delta }{}^{\nu } + 
\frac{5}{256} H_{\alpha \beta }{}^{\epsilon } H_{\epsilon }{}^{
\varepsilon \mu } R^{\alpha \beta \gamma \delta } 
\nabla_{\tau }H_{\varepsilon \mu \nu } \nabla^{\tau 
}H_{\gamma \delta }{}^{\nu } \nn\\&&-  \frac{7}{256} H_{\alpha 
}{}^{\epsilon \varepsilon } H_{\beta }{}^{\mu \nu } R^{\alpha 
\beta \gamma \delta } \nabla_{\tau }H_{\delta \mu \nu } 
\nabla^{\tau }H_{\gamma \epsilon \varepsilon } + 
\frac{31}{128} H_{\alpha \beta }{}^{\epsilon } H^{\varepsilon 
\mu \nu } R^{\alpha \beta \gamma \delta } \nabla_{\tau 
}H_{\delta \mu \nu } \nabla^{\tau }H_{\gamma \epsilon 
\varepsilon } \nn\\&&-  \frac{1}{16} H_{\alpha }{}^{\epsilon 
\varepsilon } H_{\epsilon }{}^{\mu \nu } R^{\alpha \beta 
\gamma \delta } \nabla_{\beta }H_{\delta \nu \tau } 
\nabla^{\tau }H_{\gamma \varepsilon \mu } -  \frac{175}{512} 
H_{\alpha \beta }{}^{\epsilon } H^{\varepsilon \mu \nu } 
R^{\alpha \beta \gamma \delta } \nabla_{\epsilon }H_{\delta 
\nu \tau } \nabla^{\tau }H_{\gamma \varepsilon \mu }\nn\\&& -  
\frac{3}{64} H_{\alpha }{}^{\epsilon \varepsilon } H_{\beta 
\epsilon }{}^{\mu } R^{\alpha \beta \gamma \delta } 
\nabla_{\nu }H_{\delta \mu \tau } \nabla^{\tau }H_{\gamma 
\varepsilon }{}^{\nu } -  \frac{15}{128} H_{\alpha \beta 
}{}^{\epsilon } H_{\epsilon }{}^{\varepsilon \mu } R^{\alpha 
\beta \gamma \delta } \nabla_{\nu }H_{\delta \mu \tau } 
\nabla^{\tau }H_{\gamma \varepsilon }{}^{\nu } \nn\\&&+ \frac{11}{64} 
H_{\alpha }{}^{\epsilon \varepsilon } H_{\beta \epsilon 
}{}^{\mu } R^{\alpha \beta \gamma \delta } \nabla_{\tau 
}H_{\delta \mu \nu } \nabla^{\tau }H_{\gamma \varepsilon }{}^{
\nu } + \frac{1}{8} H_{\alpha \beta }{}^{\epsilon } 
H_{\epsilon }{}^{\varepsilon \mu } R^{\alpha \beta \gamma 
\delta } \nabla_{\tau }H_{\delta \mu \nu } \nabla^{\tau 
}H_{\gamma \varepsilon }{}^{\nu }\nn\\&& -  \frac{45}{128} H_{\alpha 
}{}^{\epsilon \varepsilon } H_{\epsilon \varepsilon }{}^{\mu } 
R^{\alpha \beta \gamma \delta } \nabla_{\beta }H_{\delta \nu 
\tau } \nabla^{\tau }H_{\gamma \mu }{}^{\nu } -  
\frac{5}{128} H_{\alpha }{}^{\epsilon \varepsilon } H_{\gamma 
}{}^{\mu \nu } R^{\alpha \beta \gamma \delta } \nabla_{\beta 
}H_{\mu \nu \tau } \nabla^{\tau }H_{\delta \epsilon 
\varepsilon } \nn\\&&-  \frac{1}{8} H_{\alpha \gamma }{}^{\epsilon } 
H^{\varepsilon \mu \nu } R^{\alpha \beta \gamma \delta } 
\nabla_{\beta }H_{\mu \nu \tau } \nabla^{\tau }H_{\delta 
\epsilon \varepsilon } + \frac{1}{8} H_{\alpha }{}^{\epsilon 
\varepsilon } H_{\gamma }{}^{\mu \nu } R^{\alpha \beta \gamma 
\delta } \nabla_{\beta }H_{\varepsilon \nu \tau } 
\nabla^{\tau }H_{\delta \epsilon \mu } \nn\\&&-  \frac{27}{256} 
H_{\alpha \beta }{}^{\epsilon } H_{\gamma }{}^{\varepsilon \mu 
} R^{\alpha \beta \gamma \delta } \nabla_{\nu }H_{\varepsilon 
\mu \tau } \nabla^{\tau }H_{\delta \epsilon }{}^{\nu } + 
\frac{29}{256} H_{\alpha \beta }{}^{\epsilon } H_{\gamma 
}{}^{\varepsilon \mu } R^{\alpha \beta \gamma \delta } 
\nabla_{\tau }H_{\varepsilon \mu \nu } \nabla^{\tau 
}H_{\delta \epsilon }{}^{\nu }\nn\\&& -  \frac{39}{64} H_{\alpha 
\gamma }{}^{\epsilon } H^{\varepsilon \mu \nu } R^{\alpha 
\beta \gamma \delta } \nabla_{\beta }H_{\epsilon \nu \tau } 
\nabla^{\tau }H_{\delta \varepsilon \mu } + \frac{5}{16} 
H_{\alpha \gamma }{}^{\epsilon } H_{\epsilon }{}^{\varepsilon 
\mu } R^{\alpha \beta \gamma \delta } \nabla_{\beta }H_{\mu 
\nu \tau } \nabla^{\tau }H_{\delta \varepsilon }{}^{\nu }\nn\\&& + 
\frac{23}{128} H_{\alpha \beta }{}^{\epsilon } H_{\gamma 
}{}^{\varepsilon \mu } R^{\alpha \beta \gamma \delta } 
\nabla_{\mu }H_{\epsilon \nu \tau } \nabla^{\tau }H_{\delta 
\varepsilon }{}^{\nu } -  \frac{5}{128} H_{\alpha \beta 
}{}^{\epsilon } H_{\gamma }{}^{\varepsilon \mu } R^{\alpha 
\beta \gamma \delta } \nabla_{\nu }H_{\epsilon \mu \tau } 
\nabla^{\tau }H_{\delta \varepsilon }{}^{\nu }\nn\\&& + \frac{7}{128} 
H_{\alpha \beta }{}^{\epsilon } H_{\gamma }{}^{\varepsilon \mu 
} R^{\alpha \beta \gamma \delta } \nabla_{\tau }H_{\epsilon 
\mu \nu } \nabla^{\tau }H_{\delta \varepsilon }{}^{\nu } + 
\frac{7}{256} H_{\alpha }{}^{\epsilon \varepsilon } H_{\gamma 
}{}^{\mu \nu } R^{\alpha \beta \gamma \delta } \nabla_{\beta 
}H_{\epsilon \varepsilon \tau } \nabla^{\tau }H_{\delta \mu 
\nu } \nn\\&&-  \frac{7}{32} H_{\alpha }{}^{\epsilon \varepsilon } H_{
\gamma \epsilon }{}^{\mu } R^{\alpha \beta \gamma \delta } 
\nabla_{\beta }H_{\varepsilon \nu \tau } \nabla^{\tau 
}H_{\delta \mu }{}^{\nu } + \frac{1}{128} H_{\alpha 
}{}^{\epsilon \varepsilon } H_{\gamma \epsilon \varepsilon } R^{
\alpha \beta \gamma \delta } \nabla_{\beta }H_{\mu \nu \tau 
} \nabla^{\tau }H_{\delta }{}^{\mu \nu }\nn\\&& -  \frac{7}{256} 
H_{\alpha \beta }{}^{\epsilon } H_{\gamma \epsilon 
}{}^{\varepsilon } R^{\alpha \beta \gamma \delta } 
\nabla_{\tau }H_{\varepsilon \mu \nu } \nabla^{\tau 
}H_{\delta }{}^{\mu \nu } -  \frac{7}{256} H_{\alpha 
}{}^{\epsilon \varepsilon } H_{\gamma \delta }{}^{\mu } 
R^{\alpha \beta \gamma \delta } \nabla_{\beta }H_{\mu \nu 
\tau } \nabla^{\tau }H_{\epsilon \varepsilon }{}^{\nu } \nn\\&&+ 
\frac{1}{16} H_{\alpha }{}^{\epsilon \varepsilon } H_{\gamma 
\delta }{}^{\mu } R^{\alpha \beta \gamma \delta } 
\nabla_{\beta }H_{\varepsilon \nu \tau } \nabla^{\tau 
}H_{\epsilon \mu }{}^{\nu } -  \frac{3}{64} H_{\alpha \beta 
}{}^{\epsilon } H_{\gamma \epsilon }{}^{\varepsilon } R^{\alpha 
\beta \gamma \delta } \nabla_{\delta }H_{\mu \nu \tau } 
\nabla^{\tau }H_{\varepsilon }{}^{\mu \nu } \nn\\&&-  \frac{23}{512} 
H_{\alpha \beta }{}^{\epsilon } H_{\gamma \delta 
}{}^{\varepsilon } R^{\alpha \beta \gamma \delta } 
\nabla_{\epsilon }H_{\mu \nu \tau } \nabla^{\tau 
}H_{\varepsilon }{}^{\mu \nu } + \frac{5}{768} H_{\alpha \beta 
}{}^{\epsilon } H_{\gamma \delta \epsilon } R^{\alpha \beta 
\gamma \delta } \nabla_{\tau }H_{\varepsilon \mu \nu } 
\nabla^{\tau }H^{\varepsilon \mu \nu }.\nn
\eeqa
The couplings with the structure \(H^4\nabla H^2 R\) in the Meissner scheme are as follows:
\beqa
&&[H^4\nabla H^2]_{93}=- \frac{3}{256} H_{\alpha \beta }{}^{\delta } H^{\alpha \beta 
\gamma } H_{\epsilon }{}^{\mu \nu } H^{\epsilon \varepsilon 
\theta } \nabla_{\theta }H_{\delta \nu \tau } \nabla_{\mu 
}H_{\gamma \varepsilon }{}^{\tau } \\&&+ \frac{211}{128} H_{\alpha 
}{}^{\delta \epsilon } H^{\alpha \beta \gamma } H_{\beta 
}{}^{\varepsilon \theta } H_{\delta }{}^{\mu \nu } 
\nabla_{\theta }H_{\epsilon \nu \tau } \nabla_{\mu }H_{\gamma 
\varepsilon }{}^{\tau } -  \frac{1}{256} H_{\alpha }{}^{\delta 
\epsilon } H^{\alpha \beta \gamma } H_{\beta }{}^{\varepsilon 
\theta } H_{\delta \varepsilon }{}^{\mu } \nabla_{\theta 
}H_{\epsilon \nu \tau } \nabla_{\mu }H_{\gamma }{}^{\nu \tau 
}\nn\\&&+ \frac{93}{512} H_{\alpha }{}^{\delta \epsilon } H^{\alpha 
\beta \gamma } H_{\varepsilon }{}^{\nu \tau } H^{\varepsilon 
\theta \mu } \nabla_{\theta }H_{\beta \gamma \delta } 
\nabla_{\mu }H_{\epsilon \nu \tau } + \frac{59}{512} 
H_{\alpha \beta }{}^{\delta } H^{\alpha \beta \gamma } 
H_{\gamma }{}^{\epsilon \varepsilon } H_{\epsilon }{}^{\theta 
\mu } \nabla_{\theta }H_{\delta }{}^{\nu \tau } \nabla_{\mu 
}H_{\varepsilon \nu \tau }\nn\\&& -  \frac{103}{128} H_{\alpha 
}{}^{\delta \epsilon } H^{\alpha \beta \gamma } H_{\beta 
\delta }{}^{\varepsilon } H_{\gamma }{}^{\theta \mu } 
\nabla_{\theta }H_{\epsilon }{}^{\nu \tau } \nabla_{\mu 
}H_{\varepsilon \nu \tau } -  \frac{193}{1024} H_{\alpha \beta 
}{}^{\delta } H^{\alpha \beta \gamma } H_{\gamma }{}^{\epsilon 
\varepsilon } H_{\delta }{}^{\theta \mu } \nabla_{\theta 
}H_{\epsilon }{}^{\nu \tau } \nabla_{\mu }H_{\varepsilon \nu 
\tau }\nn\\&& + \frac{1}{256} H_{\alpha }{}^{\delta \epsilon } 
H^{\alpha \beta \gamma } H_{\beta }{}^{\varepsilon \theta } H_{
\delta \varepsilon }{}^{\mu } \nabla_{\epsilon }H_{\gamma }{}^{
\nu \tau } \nabla_{\mu }H_{\theta \nu \tau } + \frac{53}{64} 
H_{\alpha }{}^{\delta \epsilon } H^{\alpha \beta \gamma } 
H_{\varepsilon }{}^{\nu \tau } H^{\varepsilon \theta \mu } 
\nabla_{\mu }H_{\gamma \epsilon \tau } \nabla_{\nu }H_{\beta 
\delta \theta }\nn\\&& -  \frac{237}{128} H_{\alpha }{}^{\delta 
\epsilon } H^{\alpha \beta \gamma } H_{\beta }{}^{\varepsilon 
\theta } H^{\mu \nu \tau } \nabla_{\theta }H_{\epsilon 
\varepsilon \tau } \nabla_{\nu }H_{\gamma \delta \mu } + 
\frac{83}{128} H_{\alpha }{}^{\delta \epsilon } H^{\alpha 
\beta \gamma } H_{\beta }{}^{\varepsilon \theta } H_{\delta 
}{}^{\mu \nu } \nabla_{\theta }H_{\epsilon \varepsilon \tau } 
\nabla_{\nu }H_{\gamma \mu }{}^{\tau } \nn\\&&+ \frac{75}{256} 
H_{\alpha \beta }{}^{\delta } H^{\alpha \beta \gamma } 
H_{\epsilon }{}^{\mu \nu } H^{\epsilon \varepsilon \theta } 
\nabla_{\mu }H_{\gamma \varepsilon }{}^{\tau } \nabla_{\nu 
}H_{\delta \theta \tau } + \frac{21}{128} H_{\alpha \beta 
}{}^{\delta } H^{\alpha \beta \gamma } H_{\epsilon }{}^{\mu 
\nu } H^{\epsilon \varepsilon \theta } \nabla_{\theta 
}H_{\gamma \varepsilon }{}^{\tau } \nabla_{\nu }H_{\delta \mu 
\tau } \nn\\&&-  \frac{109}{1024} H_{\alpha \beta }{}^{\delta } 
H^{\alpha \beta \gamma } H_{\gamma }{}^{\epsilon \varepsilon } 
H^{\theta \mu \nu } \nabla_{\mu }H_{\delta \theta }{}^{\tau } 
\nabla_{\nu }H_{\epsilon \varepsilon \tau } -  \frac{55}{32} 
H_{\alpha }{}^{\delta \epsilon } H^{\alpha \beta \gamma } 
H_{\beta }{}^{\varepsilon \theta } H_{\delta }{}^{\mu \nu } 
\nabla_{\mu }H_{\gamma \varepsilon }{}^{\tau } \nabla_{\nu 
}H_{\epsilon \theta \tau } \nn\\&&+ \frac{139}{256} H_{\alpha 
}{}^{\delta \epsilon } H^{\alpha \beta \gamma } H_{\beta 
}{}^{\varepsilon \theta } H_{\delta }{}^{\mu \nu } 
\nabla_{\theta }H_{\gamma \varepsilon }{}^{\tau } \nabla_{\nu 
}H_{\epsilon \mu \tau } + \frac{3}{32} H_{\alpha \beta 
}{}^{\delta } H^{\alpha \beta \gamma } H_{\gamma }{}^{\epsilon 
\varepsilon } H^{\theta \mu \nu } \nabla_{\epsilon }H_{\delta 
\theta }{}^{\tau } \nabla_{\nu }H_{\varepsilon \mu \tau }\nn\\&& -  
\frac{1}{32} H_{\alpha }{}^{\delta \epsilon } H^{\alpha \beta 
\gamma } H_{\beta \delta }{}^{\varepsilon } H^{\theta \mu \nu 
} \nabla_{\theta }H_{\gamma \epsilon }{}^{\tau } \nabla_{\nu 
}H_{\varepsilon \mu \tau } -  \frac{15}{512} H_{\alpha \beta 
}{}^{\delta } H^{\alpha \beta \gamma } H_{\gamma }{}^{\epsilon 
\varepsilon } H^{\theta \mu \nu } \nabla_{\theta }H_{\delta 
\epsilon }{}^{\tau } \nabla_{\nu }H_{\varepsilon \mu \tau } \nn\\&&-  
\frac{135}{512} H_{\alpha \beta }{}^{\delta } H^{\alpha \beta 
\gamma } H_{\epsilon }{}^{\mu \nu } H^{\epsilon \varepsilon 
\theta } \nabla_{\delta }H_{\gamma \varepsilon }{}^{\tau } 
\nabla_{\nu }H_{\theta \mu \tau } + \frac{75}{128} H_{\alpha 
}{}^{\delta \epsilon } H^{\alpha \beta \gamma } H_{\beta 
}{}^{\varepsilon \theta } H_{\delta }{}^{\mu \nu } 
\nabla_{\epsilon }H_{\gamma \varepsilon }{}^{\tau } 
\nabla_{\nu }H_{\theta \mu \tau } \nn\\&&+ \frac{63}{128} H_{\alpha 
}{}^{\delta \epsilon } H^{\alpha \beta \gamma } H_{\beta 
}{}^{\varepsilon \theta } H_{\delta }{}^{\mu \nu } 
\nabla_{\varepsilon }H_{\gamma \epsilon }{}^{\tau } 
\nabla_{\nu }H_{\theta \mu \tau } + \frac{49}{256} H_{\alpha 
}{}^{\delta \epsilon } H^{\alpha \beta \gamma } H_{\beta 
}{}^{\varepsilon \theta } H_{\gamma }{}^{\mu \nu } 
\nabla_{\varepsilon }H_{\delta \epsilon }{}^{\tau } 
\nabla_{\nu }H_{\theta \mu \tau } \nn\\&&+ \frac{1}{32} H_{\alpha 
}{}^{\delta \epsilon } H^{\alpha \beta \gamma } H_{\varepsilon 
}{}^{\nu \tau } H^{\varepsilon \theta \mu } \nabla_{\nu 
}H_{\beta \delta \theta } \nabla_{\tau }H_{\gamma \epsilon 
\mu } + \frac{39}{64} H_{\alpha }{}^{\delta \epsilon } 
H^{\alpha \beta \gamma } H_{\varepsilon }{}^{\nu \tau } 
H^{\varepsilon \theta \mu } \nabla_{\mu }H_{\beta \delta 
\theta } \nabla_{\tau }H_{\gamma \epsilon \nu } \nn\\&&+ 
\frac{33}{1024} H_{\alpha }{}^{\delta \epsilon } H^{\alpha 
\beta \gamma } H_{\beta }{}^{\varepsilon \theta } H^{\mu \nu 
\tau } \nabla_{\theta }H_{\delta \epsilon \varepsilon } 
\nabla_{\tau }H_{\gamma \mu \nu } + \frac{9}{128} H_{\alpha 
}{}^{\delta \epsilon } H^{\alpha \beta \gamma } H_{\varepsilon 
}{}^{\nu \tau } H^{\varepsilon \theta \mu } \nabla_{\nu 
}H_{\beta \gamma \theta } \nabla_{\tau }H_{\delta \epsilon 
\mu }\nn\\&& -  \frac{153}{1024} H_{\alpha }{}^{\delta \epsilon } 
H^{\alpha \beta \gamma } H_{\varepsilon }{}^{\nu \tau } 
H^{\varepsilon \theta \mu } \nabla_{\mu }H_{\beta \gamma 
\theta } \nabla_{\tau }H_{\delta \epsilon \nu } -  
\frac{41}{256} H_{\alpha \beta }{}^{\delta } H^{\alpha \beta 
\gamma } H^{\epsilon \varepsilon \theta } H^{\mu \nu \tau } 
\nabla_{\varepsilon }H_{\gamma \epsilon \mu } \nabla_{\tau 
}H_{\delta \theta \nu }\nn\\&& + \frac{93}{512} H_{\alpha \beta }{}^{
\delta } H^{\alpha \beta \gamma } H^{\epsilon \varepsilon 
\theta } H^{\mu \nu \tau } \nabla_{\mu }H_{\gamma \epsilon 
\varepsilon } \nabla_{\tau }H_{\delta \theta \nu } -  
\frac{5}{32} H_{\alpha }{}^{\delta \epsilon } H^{\alpha \beta 
\gamma } H_{\beta }{}^{\varepsilon \theta } H^{\mu \nu \tau } 
\nabla_{\delta }H_{\gamma \mu \nu } \nabla_{\tau }H_{\epsilon 
\varepsilon \theta } \nn\\&&+ \frac{195}{256} H_{\alpha }{}^{\delta 
\epsilon } H^{\alpha \beta \gamma } H_{\beta }{}^{\varepsilon 
\theta } H^{\mu \nu \tau } \nabla_{\nu }H_{\gamma \delta \mu 
} \nabla_{\tau }H_{\epsilon \varepsilon \theta } -  
\frac{69}{128} H_{\alpha }{}^{\delta \epsilon } H^{\alpha 
\beta \gamma } H_{\beta }{}^{\varepsilon \theta } H^{\mu \nu 
\tau } \nabla_{\varepsilon }H_{\gamma \delta \mu } 
\nabla_{\tau }H_{\epsilon \theta \nu } \nn\\&&-  \frac{75}{128} 
H_{\alpha }{}^{\delta \epsilon } H^{\alpha \beta \gamma } 
H_{\varepsilon }{}^{\nu \tau } H^{\varepsilon \theta \mu } 
\nabla_{\delta }H_{\beta \gamma \theta } \nabla_{\tau 
}H_{\epsilon \mu \nu } -  \frac{1}{128} H_{\alpha }{}^{\delta 
\epsilon } H^{\alpha \beta \gamma } H_{\varepsilon }{}^{\nu 
\tau } H^{\varepsilon \theta \mu } \nabla_{\theta }H_{\beta 
\gamma \delta } \nabla_{\tau }H_{\epsilon \mu \nu }\nn\\&& + 
\frac{127}{1024} H_{\alpha \beta }{}^{\delta } H^{\alpha \beta 
\gamma } H^{\epsilon \varepsilon \theta } H^{\mu \nu \tau } 
\nabla_{\delta }H_{\gamma \epsilon \mu } \nabla_{\tau 
}H_{\varepsilon \theta \nu } + \frac{273}{256} H_{\alpha 
}{}^{\delta \epsilon } H^{\alpha \beta \gamma } H_{\beta 
}{}^{\varepsilon \theta } H^{\mu \nu \tau } \nabla_{\epsilon 
}H_{\gamma \delta \mu } \nabla_{\tau }H_{\varepsilon \theta 
\nu } \nn\\&&+ \frac{33}{1024} H_{\alpha }{}^{\delta \epsilon } 
H^{\alpha \beta \gamma } H_{\beta }{}^{\varepsilon \theta } H^{
\mu \nu \tau } \nabla_{\mu }H_{\gamma \delta \epsilon } 
\nabla_{\tau }H_{\varepsilon \theta \nu } -  \frac{27}{1024} 
H_{\alpha \beta }{}^{\delta } H^{\alpha \beta \gamma } 
H_{\gamma }{}^{\epsilon \varepsilon } H^{\theta \mu \nu } 
\nabla_{\theta }H_{\delta \epsilon }{}^{\tau } \nabla_{\tau 
}H_{\varepsilon \mu \nu } \nn\\&&-  \frac{41}{512} H_{\alpha 
}{}^{\delta \epsilon } H^{\alpha \beta \gamma } H_{\beta 
}{}^{\varepsilon \theta } H^{\mu \nu \tau } \nabla_{\epsilon 
}H_{\gamma \delta \varepsilon } \nabla_{\tau }H_{\theta \mu 
\nu } + \frac{39}{128} H_{\alpha }{}^{\delta \epsilon } 
H^{\alpha \beta \gamma } H_{\beta }{}^{\varepsilon \theta } H_{
\delta }{}^{\mu \nu } \nabla_{\epsilon }H_{\gamma \varepsilon 
}{}^{\tau } \nabla_{\tau }H_{\theta \mu \nu }\nn\\&& -  
\frac{47}{512} H_{\alpha }{}^{\delta \epsilon } H^{\alpha 
\beta \gamma } H_{\beta }{}^{\varepsilon \theta } H^{\mu \nu 
\tau } \nabla_{\varepsilon }H_{\gamma \delta \epsilon } 
\nabla_{\tau }H_{\theta \mu \nu } -  \frac{53}{64} H_{\alpha 
}{}^{\delta \epsilon } H^{\alpha \beta \gamma } H_{\beta 
}{}^{\varepsilon \theta } H_{\delta }{}^{\mu \nu } 
\nabla_{\varepsilon }H_{\gamma \epsilon }{}^{\tau } 
\nabla_{\tau }H_{\theta \mu \nu } \nn\\&&+ \frac{1}{256} H_{\alpha 
\beta }{}^{\delta } H^{\alpha \beta \gamma } H_{\gamma 
}{}^{\epsilon \varepsilon } H^{\theta \mu \nu } 
\nabla_{\varepsilon }H_{\delta \epsilon }{}^{\tau } 
\nabla_{\tau }H_{\theta \mu \nu } -  \frac{77}{1024} 
H_{\alpha \beta }{}^{\delta } H^{\alpha \beta \gamma } 
H_{\gamma }{}^{\epsilon \varepsilon } H_{\epsilon }{}^{\theta 
\mu } \nabla_{\varepsilon }H_{\delta }{}^{\nu \tau } 
\nabla_{\tau }H_{\theta \mu \nu } \nn\\&&-  \frac{109}{64} H_{\alpha 
}{}^{\delta \epsilon } H^{\alpha \beta \gamma } H_{\beta 
}{}^{\varepsilon \theta } H_{\delta }{}^{\mu \nu } \nabla_{\nu 
}H_{\theta \mu \tau } \nabla^{\tau }H_{\gamma \epsilon 
\varepsilon } + \frac{27}{32} H_{\alpha }{}^{\delta \epsilon } 
H^{\alpha \beta \gamma } H_{\beta }{}^{\varepsilon \theta } H_{
\delta }{}^{\mu \nu } \nabla_{\tau }H_{\theta \mu \nu } 
\nabla^{\tau }H_{\gamma \epsilon \varepsilon }\nn\\&& + 
\frac{1}{1152} H_{\alpha }{}^{\delta \epsilon } H^{\alpha 
\beta \gamma } H_{\beta \delta }{}^{\varepsilon } H^{\theta 
\mu \nu } \nabla_{\tau }H_{\theta \mu \nu } \nabla^{\tau 
}H_{\gamma \epsilon \varepsilon } + \frac{1}{32} H_{\alpha 
}{}^{\delta \epsilon } H^{\alpha \beta \gamma } H_{\beta 
\delta }{}^{\varepsilon } H^{\theta \mu \nu } \nabla_{\nu }H_{
\varepsilon \mu \tau } \nabla^{\tau }H_{\gamma \epsilon 
\theta } \nn\\&&-  \frac{1}{256} H_{\alpha }{}^{\delta \epsilon } 
H^{\alpha \beta \gamma } H_{\beta }{}^{\varepsilon \theta } H_{
\delta \varepsilon }{}^{\mu } \nabla_{\nu }H_{\theta \mu \tau 
} \nabla^{\tau }H_{\gamma \epsilon }{}^{\nu } + \frac{1}{256} 
H_{\alpha }{}^{\delta \epsilon } H^{\alpha \beta \gamma } 
H_{\beta }{}^{\varepsilon \theta } H_{\delta \varepsilon 
}{}^{\mu } \nabla_{\tau }H_{\theta \mu \nu } \nabla^{\tau 
}H_{\gamma \epsilon }{}^{\nu } \nn\\&&+ \frac{21}{512} H_{\alpha 
\beta }{}^{\delta } H^{\alpha \beta \gamma } H_{\epsilon 
}{}^{\mu \nu } H^{\epsilon \varepsilon \theta } \nabla_{\nu 
}H_{\delta \mu \tau } \nabla^{\tau }H_{\gamma \varepsilon 
\theta } -  \frac{143}{512} H_{\alpha }{}^{\delta \epsilon } 
H^{\alpha \beta \gamma } H_{\beta }{}^{\varepsilon \theta } H_{
\delta }{}^{\mu \nu } \nabla_{\nu }H_{\epsilon \mu \tau } 
\nabla^{\tau }H_{\gamma \varepsilon \theta }\nn\\&& -  \frac{105}{64} 
H_{\alpha }{}^{\delta \epsilon } H^{\alpha \beta \gamma } 
H_{\beta }{}^{\varepsilon \theta } H_{\delta }{}^{\mu \nu } 
\nabla_{\theta }H_{\epsilon \nu \tau } \nabla^{\tau 
}H_{\gamma \varepsilon \mu } -  \frac{149}{128} H_{\alpha 
\beta }{}^{\delta } H^{\alpha \beta \gamma } H_{\epsilon 
}{}^{\mu \nu } H^{\epsilon \varepsilon \theta } \nabla_{\nu 
}H_{\delta \theta \tau } \nabla^{\tau }H_{\gamma \varepsilon 
\mu } \nn\\&&+ \frac{111}{256} H_{\alpha \beta }{}^{\delta } 
H^{\alpha \beta \gamma } H_{\epsilon }{}^{\mu \nu } 
H^{\epsilon \varepsilon \theta } \nabla_{\tau }H_{\delta 
\theta \nu } \nabla^{\tau }H_{\gamma \varepsilon \mu } + 
\frac{1}{128} H_{\alpha }{}^{\delta \epsilon } H^{\alpha \beta 
\gamma } H_{\beta }{}^{\varepsilon \theta } H_{\delta }{}^{\mu 
\nu } \nabla_{\tau }H_{\epsilon \theta \nu } \nabla^{\tau 
}H_{\gamma \varepsilon \mu } \nn\\&&+ \frac{159}{256} H_{\alpha 
}{}^{\delta \epsilon } H^{\alpha \beta \gamma } H_{\beta 
\delta }{}^{\varepsilon } H^{\theta \mu \nu } 
\nabla_{\varepsilon }H_{\epsilon \nu \tau } \nabla^{\tau 
}H_{\gamma \theta \mu } -  \frac{93}{1024} H_{\alpha \beta 
}{}^{\delta } H^{\alpha \beta \gamma } H_{\epsilon \varepsilon 
}{}^{\mu } H^{\epsilon \varepsilon \theta } \nabla_{\nu 
}H_{\delta \mu \tau } \nabla^{\tau }H_{\gamma \theta }{}^{\nu 
} \nn\\&&+ \frac{93}{1024} H_{\alpha \beta }{}^{\delta } H^{\alpha 
\beta \gamma } H_{\epsilon \varepsilon }{}^{\mu } H^{\epsilon 
\varepsilon \theta } \nabla_{\tau }H_{\delta \mu \nu } 
\nabla^{\tau }H_{\gamma \theta }{}^{\nu } -  \frac{3}{4} 
H_{\alpha }{}^{\delta \epsilon } H^{\alpha \beta \gamma } 
H_{\beta }{}^{\varepsilon \theta } H_{\delta }{}^{\mu \nu } 
\nabla_{\theta }H_{\epsilon \varepsilon \tau } \nabla^{\tau 
}H_{\gamma \mu \nu }\nn\\&& + \frac{55}{256} H_{\alpha }{}^{\delta 
\epsilon } H^{\alpha \beta \gamma } H_{\beta }{}^{\varepsilon 
\theta } H_{\delta }{}^{\mu \nu } \nabla_{\tau }H_{\epsilon 
\varepsilon \theta } \nabla^{\tau }H_{\gamma \mu \nu } + 
\frac{23}{256} H_{\alpha }{}^{\delta \epsilon } H^{\alpha 
\beta \gamma } H_{\beta }{}^{\varepsilon \theta } H_{\gamma 
}{}^{\mu \nu } \nabla_{\nu }H_{\theta \mu \tau } 
\nabla^{\tau }H_{\delta \epsilon \varepsilon }\nn\\&& -  \frac{3}{64} 
H_{\alpha }{}^{\delta \epsilon } H^{\alpha \beta \gamma } 
H_{\beta }{}^{\varepsilon \theta } H_{\gamma }{}^{\mu \nu } 
\nabla_{\tau }H_{\theta \mu \nu } \nabla^{\tau }H_{\delta 
\epsilon \varepsilon } + \frac{1}{512} H_{\alpha \beta 
}{}^{\delta } H^{\alpha \beta \gamma } H_{\gamma }{}^{\epsilon 
\varepsilon } H^{\theta \mu \nu } \nabla_{\nu }H_{\varepsilon 
\mu \tau } \nabla^{\tau }H_{\delta \epsilon \theta }\nn\\&& + 
\frac{59}{1024} H_{\alpha \beta }{}^{\delta } H^{\alpha \beta 
\gamma } H_{\gamma }{}^{\epsilon \varepsilon } H^{\theta \mu 
\nu } \nabla_{\tau }H_{\varepsilon \mu \nu } \nabla^{\tau 
}H_{\delta \epsilon \theta } + \frac{29}{1024} H_{\alpha \beta 
}{}^{\delta } H^{\alpha \beta \gamma } H_{\gamma }{}^{\epsilon 
\varepsilon } H_{\epsilon }{}^{\theta \mu } \nabla_{\nu 
}H_{\theta \mu \tau } \nabla^{\tau }H_{\delta \varepsilon 
}{}^{\nu }\nn\\&& -  \frac{61}{1024} H_{\alpha \beta }{}^{\delta } H^{
\alpha \beta \gamma } H_{\gamma }{}^{\epsilon \varepsilon } H_{
\epsilon }{}^{\theta \mu } \nabla_{\tau }H_{\theta \mu \nu } 
\nabla^{\tau }H_{\delta \varepsilon }{}^{\nu } + \frac{5}{256} 
H_{\alpha \beta }{}^{\delta } H^{\alpha \beta \gamma } 
H_{\gamma }{}^{\epsilon \varepsilon } H^{\theta \mu \nu } 
\nabla_{\nu }H_{\epsilon \varepsilon \tau } \nabla^{\tau 
}H_{\delta \theta \mu }\nn\\&& -  \frac{39}{2048} H_{\alpha \beta 
}{}^{\delta } H^{\alpha \beta \gamma } H_{\gamma }{}^{\epsilon 
\varepsilon } H^{\theta \mu \nu } \nabla_{\tau }H_{\epsilon 
\varepsilon \nu } \nabla^{\tau }H_{\delta \theta \mu } + 
\frac{55}{256} H_{\alpha \beta }{}^{\delta } H^{\alpha \beta 
\gamma } H_{\gamma }{}^{\epsilon \varepsilon } H_{\epsilon }{}^{
\theta \mu } \nabla_{\mu }H_{\varepsilon \nu \tau } 
\nabla^{\tau }H_{\delta \theta }{}^{\nu } \nn\\&&+ \frac{1}{128} 
H_{\alpha \beta }{}^{\delta } H^{\alpha \beta \gamma } 
H_{\gamma }{}^{\epsilon \varepsilon } H_{\epsilon }{}^{\theta 
\mu } \nabla_{\nu }H_{\varepsilon \mu \tau } \nabla^{\tau 
}H_{\delta \theta }{}^{\nu } -  \frac{1}{128} H_{\alpha \beta 
}{}^{\delta } H^{\alpha \beta \gamma } H_{\gamma }{}^{\epsilon 
\varepsilon } H_{\epsilon }{}^{\theta \mu } \nabla_{\tau 
}H_{\varepsilon \mu \nu } \nabla^{\tau }H_{\delta \theta }{}^{
\nu } \nn\\&&-  \frac{9}{256} H_{\alpha \beta }{}^{\delta } H^{\alpha 
\beta \gamma } H_{\gamma }{}^{\epsilon \varepsilon } 
H_{\epsilon \varepsilon }{}^{\theta } \nabla_{\nu }H_{\theta 
\mu \tau } \nabla^{\tau }H_{\delta }{}^{\mu \nu } -  
\frac{1}{256} H_{\alpha \beta }{}^{\delta } H^{\alpha \beta 
\gamma } H_{\gamma }{}^{\epsilon \varepsilon } H_{\epsilon 
\varepsilon }{}^{\theta } \nabla_{\tau }H_{\theta \mu \nu } 
\nabla^{\tau }H_{\delta }{}^{\mu \nu }\nn\\&& -  \frac{9}{2048} 
H_{\alpha \beta }{}^{\delta } H^{\alpha \beta \gamma } 
H_{\gamma }{}^{\epsilon \varepsilon } H^{\theta \mu \nu } 
\nabla_{\delta }H_{\mu \nu \tau } \nabla^{\tau }H_{\epsilon 
\varepsilon \theta } + \frac{147}{512} H_{\alpha }{}^{\delta 
\epsilon } H^{\alpha \beta \gamma } H_{\beta \delta 
}{}^{\varepsilon } H_{\gamma }{}^{\theta \mu } \nabla_{\nu }H_{
\theta \mu \tau } \nabla^{\tau }H_{\epsilon \varepsilon 
}{}^{\nu } \nn\\&&+ \frac{193}{2048} H_{\alpha \beta }{}^{\delta } H^{
\alpha \beta \gamma } H_{\gamma }{}^{\epsilon \varepsilon } H_{
\delta }{}^{\theta \mu } \nabla_{\nu }H_{\theta \mu \tau } 
\nabla^{\tau }H_{\epsilon \varepsilon }{}^{\nu } -  
\frac{147}{512} H_{\alpha }{}^{\delta \epsilon } H^{\alpha 
\beta \gamma } H_{\beta \delta }{}^{\varepsilon } H_{\gamma 
}{}^{\theta \mu } \nabla_{\tau }H_{\theta \mu \nu } 
\nabla^{\tau }H_{\epsilon \varepsilon }{}^{\nu }\nn\\&& -  
\frac{185}{2048} H_{\alpha \beta }{}^{\delta } H^{\alpha \beta 
\gamma } H_{\gamma }{}^{\epsilon \varepsilon } H_{\delta 
}{}^{\theta \mu } \nabla_{\tau }H_{\theta \mu \nu } 
\nabla^{\tau }H_{\epsilon \varepsilon }{}^{\nu } + 
\frac{3}{1024} H_{\alpha \beta }{}^{\delta } H^{\alpha \beta 
\gamma } H_{\gamma }{}^{\epsilon \varepsilon } H^{\theta \mu 
\nu } \nabla_{\delta }H_{\varepsilon \nu \tau } \nabla^{\tau 
}H_{\epsilon \theta \mu }\nn\\&& -  \frac{13}{8} H_{\alpha 
}{}^{\delta \epsilon } H^{\alpha \beta \gamma } H_{\beta 
\delta }{}^{\varepsilon } H_{\gamma }{}^{\theta \mu } 
\nabla_{\mu }H_{\varepsilon \nu \tau } \nabla^{\tau 
}H_{\epsilon \theta }{}^{\nu } -  \frac{1}{64} H_{\alpha 
}{}^{\delta \epsilon } H^{\alpha \beta \gamma } H_{\beta 
\delta }{}^{\varepsilon } H_{\gamma }{}^{\theta \mu } 
\nabla_{\nu }H_{\varepsilon \mu \tau } \nabla^{\tau 
}H_{\epsilon \theta }{}^{\nu }\nn\\&& -  \frac{81}{512} H_{\alpha 
\beta }{}^{\delta } H^{\alpha \beta \gamma } H_{\gamma 
}{}^{\epsilon \varepsilon } H_{\delta }{}^{\theta \mu } 
\nabla_{\nu }H_{\varepsilon \mu \tau } \nabla^{\tau 
}H_{\epsilon \theta }{}^{\nu } + \frac{1}{64} H_{\alpha 
}{}^{\delta \epsilon } H^{\alpha \beta \gamma } H_{\beta 
\delta }{}^{\varepsilon } H_{\gamma }{}^{\theta \mu } 
\nabla_{\tau }H_{\varepsilon \mu \nu } \nabla^{\tau 
}H_{\epsilon \theta }{}^{\nu } \nn\\&&+ \frac{81}{512} H_{\alpha 
\beta }{}^{\delta } H^{\alpha \beta \gamma } H_{\gamma 
}{}^{\epsilon \varepsilon } H_{\delta }{}^{\theta \mu } 
\nabla_{\tau }H_{\varepsilon \mu \nu } \nabla^{\tau 
}H_{\epsilon \theta }{}^{\nu } + \frac{161}{256} H_{\alpha 
\beta }{}^{\delta } H^{\alpha \beta \gamma } H_{\gamma 
}{}^{\epsilon \varepsilon } H_{\epsilon }{}^{\theta \mu } 
\nabla_{\delta }H_{\mu \nu \tau } \nabla^{\tau 
}H_{\varepsilon \theta }{}^{\nu } \nn\\&&+ \frac{55}{64} H_{\alpha 
}{}^{\delta \epsilon } H^{\alpha \beta \gamma } H_{\beta 
\delta }{}^{\varepsilon } H_{\gamma \epsilon }{}^{\theta } 
\nabla_{\nu }H_{\theta \mu \tau } \nabla^{\tau 
}H_{\varepsilon }{}^{\mu \nu } + \frac{65}{256} H_{\alpha 
\beta }{}^{\delta } H^{\alpha \beta \gamma } H_{\gamma 
}{}^{\epsilon \varepsilon } H_{\delta \epsilon }{}^{\theta } 
\nabla_{\nu }H_{\theta \mu \tau } \nabla^{\tau 
}H_{\varepsilon }{}^{\mu \nu }\nn\\&& -  \frac{55}{128} H_{\alpha 
}{}^{\delta \epsilon } H^{\alpha \beta \gamma } H_{\beta 
\delta }{}^{\varepsilon } H_{\gamma \epsilon }{}^{\theta } 
\nabla_{\tau }H_{\theta \mu \nu } \nabla^{\tau 
}H_{\varepsilon }{}^{\mu \nu } -  \frac{65}{512} H_{\alpha 
\beta }{}^{\delta } H^{\alpha \beta \gamma } H_{\gamma 
}{}^{\epsilon \varepsilon } H_{\delta \epsilon }{}^{\theta } 
\nabla_{\tau }H_{\theta \mu \nu } \nabla^{\tau 
}H_{\varepsilon }{}^{\mu \nu }\nn\\&& -  \frac{1}{4608} H_{\alpha 
}{}^{\delta \epsilon } H^{\alpha \beta \gamma } H_{\beta 
\delta }{}^{\varepsilon } H_{\gamma \epsilon \varepsilon } 
\nabla_{\tau }H_{\theta \mu \nu } \nabla^{\tau }H^{\theta 
\mu \nu } -  \frac{1}{3072} H_{\alpha \beta }{}^{\delta } 
H^{\alpha \beta \gamma } H_{\gamma }{}^{\epsilon \varepsilon } 
H_{\delta \epsilon \varepsilon } \nabla_{\tau }H_{\theta \mu 
\nu } \nabla^{\tau }H^{\theta \mu \nu }.\nn
\eeqa

The couplings with the structure \(H^2\nabla H^2 R\) in the Metsaev-Tseytlin scheme are as follows:
 \beqa
[H^2\nabla H^2 R]_{100}&=&\frac{77}{256} H_{\epsilon }{}^{\nu \tau } H^{\epsilon 
\varepsilon \mu } R^{\alpha \beta \gamma \delta } 
\nabla_{\alpha }H_{\gamma \delta \varepsilon } \nabla_{\beta 
}H_{\mu \nu \tau } + \frac{397}{768} H_{\epsilon }{}^{\nu 
\tau } H^{\epsilon \varepsilon \mu } R^{\alpha \beta \gamma 
\delta } \nabla_{\beta }H_{\delta \nu \tau } \nabla_{\gamma 
}H_{\alpha \varepsilon \mu } \nn\\&&+ \frac{1}{6} H_{\epsilon 
}{}^{\nu \tau } H^{\epsilon \varepsilon \mu } R^{\alpha \beta 
\gamma \delta } \nabla_{\beta }H_{\delta \mu \tau } 
\nabla_{\gamma }H_{\alpha \varepsilon \nu } -  \frac{113}{96} 
H_{\epsilon \varepsilon }{}^{\nu } H^{\epsilon \varepsilon \mu 
} R^{\alpha \beta \gamma \delta } \nabla_{\beta }H_{\delta 
\nu \tau } \nabla_{\gamma }H_{\alpha \mu }{}^{\tau } \nn\\&& -  
\frac{301}{384} H_{\epsilon }{}^{\nu \tau } H^{\epsilon 
\varepsilon \mu } R^{\alpha \beta \gamma \delta } 
\nabla_{\gamma }H_{\alpha \varepsilon \nu } \nabla_{\delta 
}H_{\beta \mu \tau } -  \frac{55}{192} H_{\epsilon }{}^{\nu 
\tau } H^{\epsilon \varepsilon \mu } R^{\alpha \beta \gamma 
\delta } \nabla_{\gamma }H_{\alpha \varepsilon \mu } 
\nabla_{\delta }H_{\beta \nu \tau }  \nn\\&&+ \frac{29}{24} 
H_{\epsilon \varepsilon }{}^{\nu } H^{\epsilon \varepsilon \mu 
} R^{\alpha \beta \gamma \delta } \nabla_{\gamma }H_{\alpha 
\mu }{}^{\tau } \nabla_{\delta }H_{\beta \nu \tau } -  
\frac{229}{256} H_{\alpha }{}^{\epsilon \varepsilon } H^{\mu 
\nu \tau } R^{\alpha \beta \gamma \delta } \nabla_{\beta }H_{
\gamma \mu \nu } \nabla_{\delta }H_{\epsilon \varepsilon \tau 
}  \nn\\&&+ \frac{137}{128} H_{\alpha }{}^{\epsilon \varepsilon } 
H^{\mu \nu \tau } R^{\alpha \beta \gamma \delta } 
\nabla_{\gamma }H_{\beta \mu \nu } \nabla_{\delta 
}H_{\epsilon \varepsilon \tau } + \frac{3}{32} H_{\alpha 
}{}^{\epsilon \varepsilon } H_{\gamma }{}^{\mu \nu } R^{\alpha 
\beta \gamma \delta } \nabla_{\beta }H_{\mu \nu \tau } 
\nabla_{\delta }H_{\epsilon \varepsilon }{}^{\tau } \nn\\&& -  
\frac{5}{192} H_{\alpha \gamma }{}^{\epsilon } H^{\varepsilon 
\mu \nu } R^{\alpha \beta \gamma \delta } \nabla_{\beta 
}H_{\mu \nu \tau } \nabla_{\delta }H_{\epsilon \varepsilon 
}{}^{\tau } -  \frac{61}{128} H_{\alpha }{}^{\epsilon 
\varepsilon } H^{\mu \nu \tau } R^{\alpha \beta \gamma \delta 
} \nabla_{\beta }H_{\gamma \epsilon \mu } \nabla_{\delta 
}H_{\varepsilon \nu \tau } \nn\\&& -  \frac{239}{96} H_{\alpha 
}{}^{\epsilon \varepsilon } H_{\epsilon }{}^{\mu \nu } 
R^{\alpha \beta \gamma \delta } \nabla_{\beta }H_{\gamma \mu 
}{}^{\tau } \nabla_{\delta }H_{\varepsilon \nu \tau } + 
\frac{3}{16} H_{\alpha }{}^{\epsilon \varepsilon } H^{\mu \nu 
\tau } R^{\alpha \beta \gamma \delta } \nabla_{\gamma 
}H_{\beta \epsilon \mu } \nabla_{\delta }H_{\varepsilon \nu 
\tau }  \nn\\&&+ \frac{241}{192} H_{\alpha }{}^{\epsilon \varepsilon } 
H_{\epsilon }{}^{\mu \nu } R^{\alpha \beta \gamma \delta } 
\nabla_{\gamma }H_{\beta \mu }{}^{\tau } \nabla_{\delta 
}H_{\varepsilon \nu \tau } -  \frac{75}{128} H_{\alpha 
}{}^{\epsilon \varepsilon } H_{\gamma \epsilon }{}^{\mu } 
R^{\alpha \beta \gamma \delta } \nabla_{\beta }H_{\mu \nu 
\tau } \nabla_{\delta }H_{\varepsilon }{}^{\nu \tau } \nn\\&& -  
\frac{1}{32} H_{\alpha }{}^{\epsilon \varepsilon } H^{\mu \nu 
\tau } R^{\alpha \beta \gamma \delta } \nabla_{\beta 
}H_{\gamma \epsilon \varepsilon } \nabla_{\delta }H_{\mu \nu 
\tau } -  \frac{61}{384} H_{\alpha }{}^{\epsilon \varepsilon } 
H_{\epsilon }{}^{\mu \nu } R^{\alpha \beta \gamma \delta } 
\nabla_{\beta }H_{\gamma \varepsilon }{}^{\tau } 
\nabla_{\delta }H_{\mu \nu \tau }  \nn\\&&-  \frac{23}{48} H_{\alpha 
}{}^{\epsilon \varepsilon } H_{\epsilon \varepsilon }{}^{\mu } 
R^{\alpha \beta \gamma \delta } \nabla_{\beta }H_{\gamma }{}^{
\nu \tau } \nabla_{\delta }H_{\mu \nu \tau } -  \frac{5}{64} 
H_{\alpha }{}^{\epsilon \varepsilon } H_{\gamma }{}^{\mu \nu } 
R^{\alpha \beta \gamma \delta } \nabla_{\beta }H_{\epsilon 
\varepsilon }{}^{\tau } \nabla_{\delta }H_{\mu \nu \tau } \nn\\&& -  
\frac{3}{128} H_{\alpha }{}^{\epsilon \varepsilon } H_{\gamma 
\epsilon }{}^{\mu } R^{\alpha \beta \gamma \delta } 
\nabla_{\beta }H_{\varepsilon }{}^{\nu \tau } \nabla_{\delta 
}H_{\mu \nu \tau } -  \frac{45}{64} H_{\alpha \gamma 
}{}^{\epsilon } H_{\epsilon }{}^{\varepsilon \mu } R^{\alpha 
\beta \gamma \delta } \nabla_{\beta }H_{\varepsilon }{}^{\nu 
\tau } \nabla_{\delta }H_{\mu \nu \tau }  \nn\\&&+ \frac{13}{192} H_{
\alpha }{}^{\epsilon \varepsilon } H^{\mu \nu \tau } R^{\alpha 
\beta \gamma \delta } \nabla_{\gamma }H_{\beta \epsilon 
\varepsilon } \nabla_{\delta }H_{\mu \nu \tau } + 
\frac{89}{192} H_{\alpha }{}^{\epsilon \varepsilon } 
H_{\epsilon }{}^{\mu \nu } R^{\alpha \beta \gamma \delta } 
\nabla_{\gamma }H_{\beta \varepsilon }{}^{\tau } 
\nabla_{\delta }H_{\mu \nu \tau }  \nn\\&&+ \frac{239}{768} H_{\alpha 
}{}^{\epsilon \varepsilon } H_{\epsilon \varepsilon }{}^{\mu } 
R^{\alpha \beta \gamma \delta } \nabla_{\gamma }H_{\beta }{}^{
\nu \tau } \nabla_{\delta }H_{\mu \nu \tau } -  
\frac{21}{32} H_{\alpha }{}^{\epsilon \varepsilon } H^{\mu \nu 
\tau } R^{\alpha \beta \gamma \delta } \nabla_{\delta 
}H_{\varepsilon \nu \tau } \nabla_{\epsilon }H_{\beta \gamma 
\mu } \nn\\&& -  \frac{125}{192} H_{\alpha \gamma }{}^{\epsilon } 
H^{\varepsilon \mu \nu } R^{\alpha \beta \gamma \delta } 
\nabla_{\beta }H_{\mu \nu \tau } \nabla_{\epsilon }H_{\delta 
\varepsilon }{}^{\tau } -  \frac{109}{128} H_{\alpha \gamma 
}{}^{\epsilon } H^{\varepsilon \mu \nu } R^{\alpha \beta 
\gamma \delta } \nabla_{\delta }H_{\beta \varepsilon }{}^{\tau 
} \nabla_{\epsilon }H_{\mu \nu \tau } \nn\\&& + \frac{827}{384} 
H_{\alpha }{}^{\epsilon \varepsilon } H^{\mu \nu \tau } 
R^{\alpha \beta \gamma \delta } \nabla_{\beta }H_{\gamma 
\epsilon \mu } \nabla_{\varepsilon }H_{\delta \nu \tau } -  
\frac{1085}{384} H_{\alpha }{}^{\epsilon \varepsilon } H^{\mu 
\nu \tau } R^{\alpha \beta \gamma \delta } \nabla_{\gamma 
}H_{\beta \epsilon \mu } \nabla_{\varepsilon }H_{\delta \nu 
\tau }  \nn\\&&+ \frac{57}{32} H_{\alpha \gamma }{}^{\epsilon } 
H_{\beta }{}^{\varepsilon \mu } R^{\alpha \beta \gamma \delta 
} \nabla_{\delta }H_{\mu \nu \tau } \nabla_{\varepsilon 
}H_{\epsilon }{}^{\nu \tau } + \frac{79}{288} H_{\alpha 
}{}^{\epsilon \varepsilon } H^{\mu \nu \tau } R^{\alpha \beta 
\gamma \delta } \nabla_{\beta }H_{\gamma \delta \epsilon } 
\nabla_{\varepsilon }H_{\mu \nu \tau }  \nn\\&&+ \frac{539}{768} 
H_{\alpha }{}^{\epsilon \varepsilon } H_{\epsilon }{}^{\mu \nu 
} R^{\alpha \beta \gamma \delta } \nabla_{\beta }H_{\gamma 
\delta }{}^{\tau } \nabla_{\varepsilon }H_{\mu \nu \tau } -  
\frac{107}{48} H_{\alpha }{}^{\epsilon \varepsilon } H_{\gamma 
}{}^{\mu \nu } R^{\alpha \beta \gamma \delta } \nabla_{\delta 
}H_{\varepsilon \nu \tau } \nabla_{\mu }H_{\beta \epsilon 
}{}^{\tau } \nn\\&& -  \frac{253}{768} H_{\alpha }{}^{\epsilon 
\varepsilon } H^{\mu \nu \tau } R^{\alpha \beta \gamma \delta 
} \nabla_{\beta }H_{\delta \nu \tau } \nabla_{\mu }H_{\gamma 
\epsilon \varepsilon } + \frac{97}{96} H_{\alpha }{}^{\epsilon 
\varepsilon } H_{\epsilon }{}^{\mu \nu } R^{\alpha \beta 
\gamma \delta } \nabla_{\beta }H_{\delta \nu \tau } 
\nabla_{\mu }H_{\gamma \varepsilon }{}^{\tau }  \nn\\&&+ \frac{39}{64} 
H_{\alpha }{}^{\epsilon \varepsilon } H_{\gamma \epsilon 
}{}^{\mu } R^{\alpha \beta \gamma \delta } \nabla_{\beta 
}H_{\delta }{}^{\nu \tau } \nabla_{\mu }H_{\varepsilon \nu 
\tau } + \frac{425}{768} H_{\epsilon }{}^{\nu \tau } 
H^{\epsilon \varepsilon \mu } R^{\alpha \beta \gamma \delta } 
\nabla_{\beta }H_{\gamma \delta \tau } \nabla_{\nu }H_{\alpha 
\varepsilon \mu } \nn\\&& -  \frac{107}{192} H_{\epsilon \varepsilon 
}{}^{\nu } H^{\epsilon \varepsilon \mu } R^{\alpha \beta 
\gamma \delta } \nabla_{\beta }H_{\gamma \delta \tau } 
\nabla_{\nu }H_{\alpha \mu }{}^{\tau } -  \frac{97}{96} 
H_{\alpha }{}^{\epsilon \varepsilon } H_{\epsilon }{}^{\mu \nu 
} R^{\alpha \beta \gamma \delta } \nabla_{\gamma }H_{\beta 
\mu }{}^{\tau } \nabla_{\nu }H_{\delta \varepsilon \tau } \nn\\&& + 
\frac{499}{384} H_{\alpha }{}^{\epsilon \varepsilon } H_{\gamma 
}{}^{\mu \nu } R^{\alpha \beta \gamma \delta } \nabla_{\beta 
}H_{\delta \mu }{}^{\tau } \nabla_{\nu }H_{\epsilon 
\varepsilon \tau } -  \frac{47}{384} H_{\alpha }{}^{\epsilon 
\varepsilon } H_{\gamma }{}^{\mu \nu } R^{\alpha \beta \gamma 
\delta } \nabla_{\delta }H_{\beta \mu }{}^{\tau } \nabla_{\nu 
}H_{\epsilon \varepsilon \tau } \nn\\&& -  \frac{349}{768} H_{\alpha 
}{}^{\epsilon \varepsilon } H^{\mu \nu \tau } R^{\alpha \beta 
\gamma \delta } \nabla_{\gamma }H_{\beta \mu \nu } 
\nabla_{\tau }H_{\delta \epsilon \varepsilon } + 
\frac{263}{96} H_{\alpha }{}^{\epsilon \varepsilon } 
H_{\epsilon }{}^{\mu \nu } R^{\alpha \beta \gamma \delta } 
\nabla_{\gamma }H_{\beta \mu }{}^{\tau } \nabla_{\tau 
}H_{\delta \varepsilon \nu } \nn\\&& -  \frac{295}{384} H_{\alpha }{}^{
\epsilon \varepsilon } H_{\epsilon }{}^{\mu \nu } R^{\alpha 
\beta \gamma \delta } \nabla_{\beta }H_{\gamma \varepsilon 
}{}^{\tau } \nabla_{\tau }H_{\delta \mu \nu } -  
\frac{257}{128} H_{\alpha }{}^{\epsilon \varepsilon } 
H_{\epsilon }{}^{\mu \nu } R^{\alpha \beta \gamma \delta } 
\nabla_{\gamma }H_{\beta \varepsilon }{}^{\tau } \nabla_{\tau 
}H_{\delta \mu \nu } \nn\\&& -  \frac{289}{96} H_{\alpha }{}^{\epsilon 
\varepsilon } H_{\epsilon \varepsilon }{}^{\mu } R^{\alpha 
\beta \gamma \delta } \nabla_{\gamma }H_{\beta }{}^{\nu \tau 
} \nabla_{\tau }H_{\delta \mu \nu } + \frac{37}{32} H_{\alpha 
}{}^{\epsilon \varepsilon } H^{\mu \nu \tau } R^{\alpha \beta 
\gamma \delta } \nabla_{\beta }H_{\gamma \delta \mu } 
\nabla_{\tau }H_{\epsilon \varepsilon \nu } \nn\\&& -  \frac{613}{384} 
H_{\alpha }{}^{\epsilon \varepsilon } H_{\gamma }{}^{\mu \nu } 
R^{\alpha \beta \gamma \delta } \nabla_{\beta }H_{\delta \mu 
}{}^{\tau } \nabla_{\tau }H_{\epsilon \varepsilon \nu } -  
\frac{893}{384} H_{\alpha }{}^{\epsilon \varepsilon } H_{\gamma 
}{}^{\mu \nu } R^{\alpha \beta \gamma \delta } \nabla_{\delta 
}H_{\beta \mu }{}^{\tau } \nabla_{\tau }H_{\epsilon 
\varepsilon \nu } \nn\\&& + \frac{2035}{384} H_{\alpha \gamma 
}{}^{\epsilon } H^{\varepsilon \mu \nu } R^{\alpha \beta 
\gamma \delta } \nabla_{\delta }H_{\beta \varepsilon }{}^{\tau 
} \nabla_{\tau }H_{\epsilon \mu \nu } -  \frac{491}{768} 
H_{\alpha }{}^{\epsilon \varepsilon } H_{\epsilon }{}^{\mu \nu 
} R^{\alpha \beta \gamma \delta } \nabla_{\beta }H_{\gamma 
\delta }{}^{\tau } \nabla_{\tau }H_{\varepsilon \mu \nu } \nn\\&& + 
\frac{1}{96} H_{\alpha \gamma }{}^{\epsilon } H^{\varepsilon 
\mu \nu } R^{\alpha \beta \gamma \delta } \nabla_{\delta }H_{
\beta \epsilon }{}^{\tau } \nabla_{\tau }H_{\varepsilon \mu 
\nu } -  \frac{251}{96} H_{\alpha }{}^{\epsilon \varepsilon } 
H_{\gamma \epsilon }{}^{\mu } R^{\alpha \beta \gamma \delta } 
\nabla_{\delta }H_{\beta }{}^{\nu \tau } \nabla_{\tau 
}H_{\varepsilon \mu \nu }  \nn\\&&+ \frac{449}{384} H_{\alpha \gamma 
}{}^{\epsilon } H_{\epsilon }{}^{\varepsilon \mu } R^{\alpha 
\beta \gamma \delta } \nabla_{\delta }H_{\beta }{}^{\nu \tau 
} \nabla_{\tau }H_{\varepsilon \mu \nu } + \frac{9}{2} 
H_{\epsilon \varepsilon }{}^{\nu } H^{\epsilon \varepsilon \mu 
} R^{\alpha \beta \gamma \delta } \nabla_{\delta }H_{\beta 
\nu \tau } \nabla^{\tau }H_{\alpha \gamma \mu } \nn\\&& + 
\frac{319}{192} H_{\epsilon \varepsilon }{}^{\nu } H^{\epsilon 
\varepsilon \mu } R^{\alpha \beta \gamma \delta } 
\nabla_{\tau }H_{\beta \delta \nu } \nabla^{\tau }H_{\alpha 
\gamma \mu } -  \frac{431}{192} H_{\alpha }{}^{\epsilon 
\varepsilon } H_{\epsilon }{}^{\mu \nu } R^{\alpha \beta 
\gamma \delta } \nabla_{\delta }H_{\mu \nu \tau } 
\nabla^{\tau }H_{\beta \gamma \varepsilon } \nn\\&& + \frac{397}{192} 
H_{\alpha }{}^{\epsilon \varepsilon } H_{\epsilon }{}^{\mu \nu 
} R^{\alpha \beta \gamma \delta } \nabla_{\tau }H_{\delta \mu 
\nu } \nabla^{\tau }H_{\beta \gamma \varepsilon } + 
\frac{277}{96} H_{\alpha }{}^{\epsilon \varepsilon } 
H_{\epsilon }{}^{\mu \nu } R^{\alpha \beta \gamma \delta } 
\nabla_{\delta }H_{\varepsilon \nu \tau } \nabla^{\tau 
}H_{\beta \gamma \mu } \nn\\&& -  \frac{131}{96} H_{\alpha 
}{}^{\epsilon \varepsilon } H_{\epsilon \varepsilon }{}^{\mu } 
R^{\alpha \beta \gamma \delta } \nabla_{\tau }H_{\delta \mu 
\nu } \nabla^{\tau }H_{\beta \gamma }{}^{\nu } + 
\frac{73}{48} H_{\alpha \gamma }{}^{\epsilon } H^{\varepsilon 
\mu \nu } R^{\alpha \beta \gamma \delta } \nabla_{\tau 
}H_{\epsilon \mu \nu } \nabla^{\tau }H_{\beta \delta 
\varepsilon } \nn\\&& -  \frac{277}{192} H_{\alpha }{}^{\epsilon 
\varepsilon } H_{\gamma }{}^{\mu \nu } R^{\alpha \beta \gamma 
\delta } \nabla_{\nu }H_{\epsilon \varepsilon \tau } 
\nabla^{\tau }H_{\beta \delta \mu } -  \frac{491}{192} 
H_{\alpha }{}^{\epsilon \varepsilon } H_{\gamma \epsilon 
}{}^{\mu } R^{\alpha \beta \gamma \delta } \nabla_{\tau 
}H_{\varepsilon \mu \nu } \nabla^{\tau }H_{\beta \delta 
}{}^{\nu }  \nn\\&&+ \frac{43}{48} H_{\alpha \gamma }{}^{\epsilon } H_{
\epsilon }{}^{\varepsilon \mu } R^{\alpha \beta \gamma \delta 
} \nabla_{\tau }H_{\varepsilon \mu \nu } \nabla^{\tau 
}H_{\beta \delta }{}^{\nu } -  \frac{49}{128} H_{\alpha 
}{}^{\epsilon \varepsilon } H_{\gamma }{}^{\mu \nu } R^{\alpha 
\beta \gamma \delta } \nabla_{\delta }H_{\mu \nu \tau } 
\nabla^{\tau }H_{\beta \epsilon \varepsilon } \nn\\&& + \frac{73}{32} 
H_{\alpha \gamma }{}^{\epsilon } H^{\varepsilon \mu \nu } 
R^{\alpha \beta \gamma \delta } \nabla_{\nu }H_{\delta \mu 
\tau } \nabla^{\tau }H_{\beta \epsilon \varepsilon } + 
\frac{47}{128} H_{\alpha }{}^{\epsilon \varepsilon } H_{\gamma 
}{}^{\mu \nu } R^{\alpha \beta \gamma \delta } \nabla_{\tau 
}H_{\delta \mu \nu } \nabla^{\tau }H_{\beta \epsilon 
\varepsilon }  \nn\\&&+ \frac{97}{96} H_{\alpha }{}^{\epsilon 
\varepsilon } H_{\gamma }{}^{\mu \nu } R^{\alpha \beta \gamma 
\delta } \nabla_{\delta }H_{\varepsilon \nu \tau } 
\nabla^{\tau }H_{\beta \epsilon \mu } -  \frac{97}{96} 
H_{\alpha }{}^{\epsilon \varepsilon } H_{\gamma }{}^{\mu \nu } 
R^{\alpha \beta \gamma \delta } \nabla_{\tau }H_{\delta 
\varepsilon \nu } \nabla^{\tau }H_{\beta \epsilon \mu } \nn\\&& -  
\frac{97}{96} H_{\alpha }{}^{\epsilon \varepsilon } H_{\gamma 
\delta }{}^{\mu } R^{\alpha \beta \gamma \delta } \nabla_{\nu 
}H_{\varepsilon \mu \tau } \nabla^{\tau }H_{\beta \epsilon 
}{}^{\nu } + \frac{33}{16} H_{\alpha }{}^{\epsilon \varepsilon 
} H_{\gamma \delta }{}^{\mu } R^{\alpha \beta \gamma \delta } 
\nabla_{\tau }H_{\varepsilon \mu \nu } \nabla^{\tau }H_{\beta 
\epsilon }{}^{\nu } \nn\\&& -  \frac{169}{192} H_{\alpha \gamma 
}{}^{\epsilon } H^{\varepsilon \mu \nu } R^{\alpha \beta 
\gamma \delta } \nabla_{\delta }H_{\epsilon \nu \tau } 
\nabla^{\tau }H_{\beta \varepsilon \mu } + \frac{179}{64} 
H_{\alpha \gamma }{}^{\epsilon } H^{\varepsilon \mu \nu } 
R^{\alpha \beta \gamma \delta } \nabla_{\epsilon }H_{\delta 
\nu \tau } \nabla^{\tau }H_{\beta \varepsilon \mu } \nn\\&& + 
\frac{33}{32} H_{\alpha \gamma }{}^{\epsilon } H^{\varepsilon 
\mu \nu } R^{\alpha \beta \gamma \delta } \nabla_{\tau 
}H_{\delta \epsilon \nu } \nabla^{\tau }H_{\beta \varepsilon 
\mu } -  \frac{47}{24} H_{\alpha }{}^{\epsilon \varepsilon } 
H_{\gamma \epsilon }{}^{\mu } R^{\alpha \beta \gamma \delta } 
\nabla_{\delta }H_{\mu \nu \tau } \nabla^{\tau }H_{\beta 
\varepsilon }{}^{\nu }  \nn\\&&+ \frac{41}{12} H_{\alpha \gamma 
}{}^{\epsilon } H_{\epsilon }{}^{\varepsilon \mu } R^{\alpha 
\beta \gamma \delta } \nabla_{\delta }H_{\mu \nu \tau } 
\nabla^{\tau }H_{\beta \varepsilon }{}^{\nu } + \frac{1}{32} 
H_{\alpha }{}^{\epsilon \varepsilon } H_{\gamma \epsilon 
}{}^{\mu } R^{\alpha \beta \gamma \delta } \nabla_{\nu 
}H_{\delta \mu \tau } \nabla^{\tau }H_{\beta \varepsilon }{}^{
\nu }  \nn\\&&+ \frac{235}{64} H_{\alpha \gamma }{}^{\epsilon } 
H_{\epsilon }{}^{\varepsilon \mu } R^{\alpha \beta \gamma 
\delta } \nabla_{\nu }H_{\delta \mu \tau } \nabla^{\tau 
}H_{\beta \varepsilon }{}^{\nu } + \frac{47}{24} H_{\alpha 
}{}^{\epsilon \varepsilon } H_{\gamma \epsilon }{}^{\mu } 
R^{\alpha \beta \gamma \delta } \nabla_{\tau }H_{\delta \mu 
\nu } \nabla^{\tau }H_{\beta \varepsilon }{}^{\nu }  \nn\\&&-  
\frac{13}{6} H_{\alpha \gamma }{}^{\epsilon } H_{\epsilon }{}^{
\varepsilon \mu } R^{\alpha \beta \gamma \delta } 
\nabla_{\tau }H_{\delta \mu \nu } \nabla^{\tau }H_{\beta 
\varepsilon }{}^{\nu } -  \frac{93}{128} H_{\alpha 
}{}^{\epsilon \varepsilon } H_{\gamma }{}^{\mu \nu } R^{\alpha 
\beta \gamma \delta } \nabla_{\tau }H_{\delta \epsilon 
\varepsilon } \nabla^{\tau }H_{\beta \mu \nu } \nn\\&& -  
\frac{1}{32} H_{\alpha }{}^{\epsilon \varepsilon } H_{\gamma 
\epsilon }{}^{\mu } R^{\alpha \beta \gamma \delta } 
\nabla_{\nu }H_{\delta \varepsilon \tau } \nabla^{\tau 
}H_{\beta \mu }{}^{\nu } -  \frac{263}{192} H_{\alpha 
}{}^{\epsilon \varepsilon } H_{\gamma \delta }{}^{\mu } 
R^{\alpha \beta \gamma \delta } \nabla_{\nu }H_{\epsilon 
\varepsilon \tau } \nabla^{\tau }H_{\beta \mu }{}^{\nu } \nn\\&& -  
\frac{91}{96} H_{\alpha }{}^{\epsilon \varepsilon } H_{\gamma 
\epsilon }{}^{\mu } R^{\alpha \beta \gamma \delta } 
\nabla_{\tau }H_{\delta \varepsilon \nu } \nabla^{\tau 
}H_{\beta \mu }{}^{\nu } + \frac{385}{384} H_{\alpha 
}{}^{\epsilon \varepsilon } H_{\gamma \delta }{}^{\mu } 
R^{\alpha \beta \gamma \delta } \nabla_{\tau }H_{\epsilon 
\varepsilon \nu } \nabla^{\tau }H_{\beta \mu }{}^{\nu } \nn\\&& -  
\frac{13}{64} H_{\alpha }{}^{\epsilon \varepsilon } H_{\gamma 
\epsilon \varepsilon } R^{\alpha \beta \gamma \delta } 
\nabla_{\delta }H_{\mu \nu \tau } \nabla^{\tau }H_{\beta 
}{}^{\mu \nu } + \frac{1}{16} H_{\alpha }{}^{\epsilon 
\varepsilon } H_{\gamma \epsilon \varepsilon } R^{\alpha \beta 
\gamma \delta } \nabla_{\tau }H_{\delta \mu \nu } 
\nabla^{\tau }H_{\beta }{}^{\mu \nu }  \nn\\&&-  \frac{17}{32} 
H_{\alpha }{}^{\epsilon \varepsilon } H_{\gamma \delta \epsilon 
} R^{\alpha \beta \gamma \delta } \nabla_{\tau }H_{\varepsilon 
\mu \nu } \nabla^{\tau }H_{\beta }{}^{\mu \nu } -  
\frac{263}{96} H_{\alpha }{}^{\epsilon \varepsilon } 
H_{\epsilon }{}^{\mu \nu } R^{\alpha \beta \gamma \delta } 
\nabla_{\beta }H_{\delta \nu \tau } \nabla^{\tau }H_{\gamma 
\varepsilon \mu }  \nn\\&&-  \frac{7}{32} H_{\alpha }{}^{\epsilon 
\varepsilon } H_{\epsilon \varepsilon }{}^{\mu } R^{\alpha 
\beta \gamma \delta } \nabla_{\beta }H_{\delta \nu \tau } 
\nabla^{\tau }H_{\gamma \mu }{}^{\nu } + \frac{69}{128} 
H_{\alpha }{}^{\epsilon \varepsilon } H_{\gamma }{}^{\mu \nu } 
R^{\alpha \beta \gamma \delta } \nabla_{\beta }H_{\mu \nu 
\tau } \nabla^{\tau }H_{\delta \epsilon \varepsilon } \nn\\&& + 
\frac{13}{6} H_{\alpha }{}^{\epsilon \varepsilon } H_{\gamma 
\epsilon }{}^{\mu } R^{\alpha \beta \gamma \delta } 
\nabla_{\beta }H_{\mu \nu \tau } \nabla^{\tau }H_{\delta 
\varepsilon }{}^{\nu } + \frac{39}{16} H_{\alpha \gamma 
}{}^{\epsilon } H_{\beta }{}^{\varepsilon \mu } R^{\alpha \beta 
\gamma \delta } \nabla_{\mu }H_{\epsilon \nu \tau } 
\nabla^{\tau }H_{\delta \varepsilon }{}^{\nu } \nn\\&& + \frac{1}{128} 
H_{\alpha }{}^{\epsilon \varepsilon } H_{\gamma \delta }{}^{\mu 
} R^{\alpha \beta \gamma \delta } \nabla_{\beta }H_{\mu \nu 
\tau } \nabla^{\tau }H_{\epsilon \varepsilon }{}^{\nu } -  
\frac{9}{16} H_{\alpha }{}^{\epsilon \varepsilon } H_{\gamma 
\delta }{}^{\mu } R^{\alpha \beta \gamma \delta } 
\nabla_{\beta }H_{\varepsilon \nu \tau } \nabla^{\tau 
}H_{\epsilon \mu }{}^{\nu }  \nn\\&&+ \frac{13}{16} H_{\alpha \gamma 
}{}^{\epsilon } H_{\beta \delta }{}^{\varepsilon } R^{\alpha 
\beta \gamma \delta } \nabla_{\nu }H_{\varepsilon \mu \tau } 
\nabla^{\tau }H_{\epsilon }{}^{\mu \nu } -  \frac{3}{32} 
H_{\alpha \gamma }{}^{\epsilon } H_{\beta \delta 
}{}^{\varepsilon } R^{\alpha \beta \gamma \delta } 
\nabla_{\tau }H_{\varepsilon \mu \nu } \nabla^{\tau 
}H_{\epsilon }{}^{\mu \nu }  \nn\\&&+ \frac{49}{128} H_{\alpha 
}{}^{\epsilon \varepsilon } H_{\gamma \delta \epsilon } 
R^{\alpha \beta \gamma \delta } \nabla_{\beta }H_{\mu \nu 
\tau } \nabla^{\tau }H_{\varepsilon }{}^{\mu \nu } -  
\frac{7}{192} H_{\alpha \gamma }{}^{\epsilon } H_{\beta \delta 
\epsilon } R^{\alpha \beta \gamma \delta } \nabla_{\tau 
}H_{\varepsilon \mu \nu } \nabla^{\tau }H^{\varepsilon \mu 
\nu }.\nn
\eeqa
The complete set of couplings for the \(H^4\nabla H^2 R\) structure in the Metsaev-Tseytlin scheme is presented below:
\beqa
&&[H^4\nabla H^2]_{99}=
- \frac{11}{256} H_{\alpha \beta }{}^{\delta } H^{\alpha 
\beta \gamma } H_{\epsilon }{}^{\mu \nu } H^{\epsilon 
\varepsilon \theta } \nabla_{\theta }H_{\delta \nu \tau } 
\nabla_{\mu }H_{\gamma \varepsilon }{}^{\tau }\labell{xx}\\&& -  
\frac{5087}{384} H_{\alpha }{}^{\delta \epsilon } H^{\alpha 
\beta \gamma } H_{\beta }{}^{\varepsilon \theta } H_{\delta 
}{}^{\mu \nu } \nabla_{\theta }H_{\epsilon \nu \tau } 
\nabla_{\mu }H_{\gamma \varepsilon }{}^{\tau } -  
\frac{1}{256} H_{\alpha }{}^{\delta \epsilon } H^{\alpha \beta 
\gamma } H_{\beta }{}^{\varepsilon \theta } H_{\delta 
\varepsilon }{}^{\mu } \nabla_{\theta }H_{\epsilon \nu \tau } 
\nabla_{\mu }H_{\gamma }{}^{\nu \tau } \nn\\&&-  \frac{911}{768} 
H_{\alpha }{}^{\delta \epsilon } H^{\alpha \beta \gamma } 
H_{\varepsilon }{}^{\nu \tau } H^{\varepsilon \theta \mu } 
\nabla_{\theta }H_{\beta \gamma \delta } \nabla_{\mu 
}H_{\epsilon \nu \tau } -  \frac{829}{768} H_{\alpha \beta 
}{}^{\delta } H^{\alpha \beta \gamma } H_{\gamma }{}^{\epsilon 
\varepsilon } H_{\epsilon }{}^{\theta \mu } \nabla_{\theta }H_{
\delta }{}^{\nu \tau } \nabla_{\mu }H_{\varepsilon \nu \tau }  \nn\\&&
+ \frac{2557}{384} H_{\alpha }{}^{\delta \epsilon } H^{\alpha 
\beta \gamma } H_{\beta \delta }{}^{\varepsilon } H_{\gamma 
}{}^{\theta \mu } \nabla_{\theta }H_{\epsilon }{}^{\nu \tau } 
\nabla_{\mu }H_{\varepsilon \nu \tau } + \frac{7}{6} 
H_{\alpha \beta }{}^{\delta } H^{\alpha \beta \gamma } 
H_{\gamma }{}^{\epsilon \varepsilon } H_{\delta }{}^{\theta \mu 
} \nabla_{\theta }H_{\epsilon }{}^{\nu \tau } \nabla_{\mu }H_{
\varepsilon \nu \tau } \nn\\&& + \frac{1}{128} H_{\alpha \beta 
}{}^{\delta } H^{\alpha \beta \gamma } H_{\epsilon \varepsilon 
}{}^{\mu } H^{\epsilon \varepsilon \theta } \nabla_{\delta }H_{
\gamma }{}^{\nu \tau } \nabla_{\mu }H_{\theta \nu \tau } + 
\frac{1}{256} H_{\alpha }{}^{\delta \epsilon } H^{\alpha \beta 
\gamma } H_{\beta }{}^{\varepsilon \theta } H_{\delta 
\varepsilon }{}^{\mu } \nabla_{\epsilon }H_{\gamma }{}^{\nu 
\tau } \nabla_{\mu }H_{\theta \nu \tau } \nn\\&& -  \frac{1271}{192} 
H_{\alpha }{}^{\delta \epsilon } H^{\alpha \beta \gamma } 
H_{\varepsilon }{}^{\nu \tau } H^{\varepsilon \theta \mu } 
\nabla_{\mu }H_{\gamma \epsilon \tau } \nabla_{\nu }H_{\beta 
\delta \theta } + \frac{2145}{128} H_{\alpha }{}^{\delta 
\epsilon } H^{\alpha \beta \gamma } H_{\beta }{}^{\varepsilon 
\theta } H^{\mu \nu \tau } \nabla_{\theta }H_{\epsilon 
\varepsilon \tau } \nabla_{\nu }H_{\gamma \delta \mu }  \nn\\&&-  
\frac{2315}{384} H_{\alpha }{}^{\delta \epsilon } H^{\alpha 
\beta \gamma } H_{\beta }{}^{\varepsilon \theta } H_{\delta 
}{}^{\mu \nu } \nabla_{\theta }H_{\epsilon \varepsilon \tau } 
\nabla_{\nu }H_{\gamma \mu }{}^{\tau } -  \frac{1823}{768} H_{
\alpha \beta }{}^{\delta } H^{\alpha \beta \gamma } 
H_{\epsilon }{}^{\mu \nu } H^{\epsilon \varepsilon \theta } 
\nabla_{\mu }H_{\gamma \varepsilon }{}^{\tau } \nabla_{\nu 
}H_{\delta \theta \tau } \nn\\&& -  \frac{29}{24} H_{\alpha \beta 
}{}^{\delta } H^{\alpha \beta \gamma } H_{\epsilon }{}^{\mu 
\nu } H^{\epsilon \varepsilon \theta } \nabla_{\theta 
}H_{\gamma \varepsilon }{}^{\tau } \nabla_{\nu }H_{\delta \mu 
\tau } + \frac{977}{1536} H_{\alpha \beta }{}^{\delta } 
H^{\alpha \beta \gamma } H_{\gamma }{}^{\epsilon \varepsilon } 
H^{\theta \mu \nu } \nabla_{\mu }H_{\delta \theta }{}^{\tau } 
\nabla_{\nu }H_{\epsilon \varepsilon \tau }  \nn\\&&+ \frac{1259}{96} 
H_{\alpha }{}^{\delta \epsilon } H^{\alpha \beta \gamma } 
H_{\beta }{}^{\varepsilon \theta } H_{\delta }{}^{\mu \nu } 
\nabla_{\mu }H_{\gamma \varepsilon }{}^{\tau } \nabla_{\nu 
}H_{\epsilon \theta \tau } -  \frac{575}{192} H_{\alpha 
}{}^{\delta \epsilon } H^{\alpha \beta \gamma } H_{\beta 
}{}^{\varepsilon \theta } H_{\delta }{}^{\mu \nu } 
\nabla_{\theta }H_{\gamma \varepsilon }{}^{\tau } \nabla_{\nu 
}H_{\epsilon \mu \tau }  \nn\\&&-  \frac{77}{128} H_{\alpha \beta 
}{}^{\delta } H^{\alpha \beta \gamma } H_{\gamma }{}^{\epsilon 
\varepsilon } H^{\theta \mu \nu } \nabla_{\epsilon }H_{\delta 
\theta }{}^{\tau } \nabla_{\nu }H_{\varepsilon \mu \tau } + 
\frac{73}{256} H_{\alpha }{}^{\delta \epsilon } H^{\alpha 
\beta \gamma } H_{\beta \delta }{}^{\varepsilon } H^{\theta 
\mu \nu } \nabla_{\theta }H_{\gamma \epsilon }{}^{\tau } 
\nabla_{\nu }H_{\varepsilon \mu \tau }  \nn\\&&+ \frac{7}{24} 
H_{\alpha \beta }{}^{\delta } H^{\alpha \beta \gamma } 
H_{\gamma }{}^{\epsilon \varepsilon } H^{\theta \mu \nu } 
\nabla_{\theta }H_{\delta \epsilon }{}^{\tau } \nabla_{\nu 
}H_{\varepsilon \mu \tau } + \frac{1123}{768} H_{\alpha \beta 
}{}^{\delta } H^{\alpha \beta \gamma } H_{\epsilon }{}^{\mu 
\nu } H^{\epsilon \varepsilon \theta } \nabla_{\delta 
}H_{\gamma \varepsilon }{}^{\tau } \nabla_{\nu }H_{\theta \mu 
\tau } \nn\\&& -  \frac{1817}{384} H_{\alpha }{}^{\delta \epsilon } H^{
\alpha \beta \gamma } H_{\beta }{}^{\varepsilon \theta } 
H_{\delta }{}^{\mu \nu } \nabla_{\epsilon }H_{\gamma 
\varepsilon }{}^{\tau } \nabla_{\nu }H_{\theta \mu \tau } -  
\frac{1877}{384} H_{\alpha }{}^{\delta \epsilon } H^{\alpha 
\beta \gamma } H_{\beta }{}^{\varepsilon \theta } H_{\delta 
}{}^{\mu \nu } \nabla_{\varepsilon }H_{\gamma \epsilon 
}{}^{\tau } \nabla_{\nu }H_{\theta \mu \tau } \nn\\&& -  
\frac{499}{256} H_{\alpha }{}^{\delta \epsilon } H^{\alpha 
\beta \gamma } H_{\beta }{}^{\varepsilon \theta } H_{\gamma 
}{}^{\mu \nu } \nabla_{\varepsilon }H_{\delta \epsilon 
}{}^{\tau } \nabla_{\nu }H_{\theta \mu \tau } + \frac{1}{16} 
H_{\alpha }{}^{\delta \epsilon } H^{\alpha \beta \gamma } 
H_{\varepsilon }{}^{\nu \tau } H^{\varepsilon \theta \mu } 
\nabla_{\nu }H_{\beta \delta \theta } \nabla_{\tau }H_{\gamma 
\epsilon \mu }  \nn\\&&-  \frac{2057}{384} H_{\alpha }{}^{\delta 
\epsilon } H^{\alpha \beta \gamma } H_{\varepsilon }{}^{\nu 
\tau } H^{\varepsilon \theta \mu } \nabla_{\mu }H_{\beta 
\delta \theta } \nabla_{\tau }H_{\gamma \epsilon \nu } -  
\frac{895}{1536} H_{\alpha }{}^{\delta \epsilon } H^{\alpha 
\beta \gamma } H_{\beta }{}^{\varepsilon \theta } H^{\mu \nu 
\tau } \nabla_{\theta }H_{\delta \epsilon \varepsilon } 
\nabla_{\tau }H_{\gamma \mu \nu }  \nn\\&&-  \frac{203}{384} 
H_{\alpha }{}^{\delta \epsilon } H^{\alpha \beta \gamma } 
H_{\varepsilon }{}^{\nu \tau } H^{\varepsilon \theta \mu } 
\nabla_{\nu }H_{\beta \gamma \theta } \nabla_{\tau }H_{\delta 
\epsilon \mu } + \frac{509}{384} H_{\alpha }{}^{\delta 
\epsilon } H^{\alpha \beta \gamma } H_{\varepsilon }{}^{\nu 
\tau } H^{\varepsilon \theta \mu } \nabla_{\mu }H_{\beta 
\gamma \theta } \nabla_{\tau }H_{\delta \epsilon \nu }  \nn\\&&+ 
\frac{919}{768} H_{\alpha \beta }{}^{\delta } H^{\alpha \beta 
\gamma } H^{\epsilon \varepsilon \theta } H^{\mu \nu \tau } 
\nabla_{\varepsilon }H_{\gamma \epsilon \mu } \nabla_{\tau 
}H_{\delta \theta \nu } -  \frac{1217}{768} H_{\alpha \beta 
}{}^{\delta } H^{\alpha \beta \gamma } H^{\epsilon \varepsilon 
\theta } H^{\mu \nu \tau } \nabla_{\mu }H_{\gamma \epsilon 
\varepsilon } \nabla_{\tau }H_{\delta \theta \nu } \nn\\&& + 
\frac{1}{256} H_{\alpha \beta }{}^{\delta } H^{\alpha \beta 
\gamma } H^{\epsilon \varepsilon \theta } H^{\mu \nu \tau } 
\nabla_{\theta }H_{\gamma \epsilon \varepsilon } \nabla_{\tau 
}H_{\delta \mu \nu } + \frac{461}{384} H_{\alpha }{}^{\delta 
\epsilon } H^{\alpha \beta \gamma } H_{\beta }{}^{\varepsilon 
\theta } H^{\mu \nu \tau } \nabla_{\delta }H_{\gamma \mu \nu 
} \nabla_{\tau }H_{\epsilon \varepsilon \theta }  \nn\\&&-  
\frac{5519}{768} H_{\alpha }{}^{\delta \epsilon } H^{\alpha 
\beta \gamma } H_{\beta }{}^{\varepsilon \theta } H^{\mu \nu 
\tau } \nabla_{\nu }H_{\gamma \delta \mu } \nabla_{\tau 
}H_{\epsilon \varepsilon \theta } + \frac{1835}{384} H_{\alpha 
}{}^{\delta \epsilon } H^{\alpha \beta \gamma } H_{\beta 
}{}^{\varepsilon \theta } H^{\mu \nu \tau } 
\nabla_{\varepsilon }H_{\gamma \delta \mu } \nabla_{\tau 
}H_{\epsilon \theta \nu } \nn\\&& + \frac{35}{8} H_{\alpha }{}^{\delta 
\epsilon } H^{\alpha \beta \gamma } H_{\varepsilon }{}^{\nu 
\tau } H^{\varepsilon \theta \mu } \nabla_{\delta }H_{\beta 
\gamma \theta } \nabla_{\tau }H_{\epsilon \mu \nu } -  
\frac{1}{128} H_{\alpha }{}^{\delta \epsilon } H^{\alpha \beta 
\gamma } H_{\varepsilon }{}^{\nu \tau } H^{\varepsilon \theta 
\mu } \nabla_{\theta }H_{\beta \gamma \delta } \nabla_{\tau 
}H_{\epsilon \mu \nu } \nn\\&& -  \frac{1171}{1536} H_{\alpha \beta 
}{}^{\delta } H^{\alpha \beta \gamma } H^{\epsilon \varepsilon 
\theta } H^{\mu \nu \tau } \nabla_{\delta }H_{\gamma \epsilon 
\mu } \nabla_{\tau }H_{\varepsilon \theta \nu } -  
\frac{3847}{384} H_{\alpha }{}^{\delta \epsilon } H^{\alpha 
\beta \gamma } H_{\beta }{}^{\varepsilon \theta } H^{\mu \nu 
\tau } \nabla_{\epsilon }H_{\gamma \delta \mu } \nabla_{\tau 
}H_{\varepsilon \theta \nu }  \nn\\&&+ \frac{67}{128} H_{\alpha 
}{}^{\delta \epsilon } H^{\alpha \beta \gamma } H_{\beta 
}{}^{\varepsilon \theta } H^{\mu \nu \tau } \nabla_{\mu 
}H_{\gamma \delta \epsilon } \nabla_{\tau }H_{\varepsilon 
\theta \nu } -  \frac{121}{384} H_{\alpha \beta }{}^{\delta } 
H^{\alpha \beta \gamma } H_{\gamma }{}^{\epsilon \varepsilon } 
H^{\theta \mu \nu } \nabla_{\theta }H_{\delta \epsilon 
}{}^{\tau } \nabla_{\tau }H_{\varepsilon \mu \nu }  \nn\\&&+ 
\frac{1}{64} H_{\alpha \beta }{}^{\delta } H^{\alpha \beta 
\gamma } H^{\epsilon \varepsilon \theta } H^{\mu \nu \tau } 
\nabla_{\delta }H_{\gamma \epsilon \varepsilon } \nabla_{\tau 
}H_{\theta \mu \nu } + \frac{95}{384} H_{\alpha \beta 
}{}^{\delta } H^{\alpha \beta \gamma } H_{\epsilon }{}^{\mu 
\nu } H^{\epsilon \varepsilon \theta } \nabla_{\delta 
}H_{\gamma \varepsilon }{}^{\tau } \nabla_{\tau }H_{\theta \mu 
\nu } \nn\\&& + \frac{859}{768} H_{\alpha }{}^{\delta \epsilon } 
H^{\alpha \beta \gamma } H_{\beta }{}^{\varepsilon \theta } H^{
\mu \nu \tau } \nabla_{\epsilon }H_{\gamma \delta \varepsilon 
} \nabla_{\tau }H_{\theta \mu \nu } -  \frac{767}{384} 
H_{\alpha }{}^{\delta \epsilon } H^{\alpha \beta \gamma } 
H_{\beta }{}^{\varepsilon \theta } H_{\delta }{}^{\mu \nu } 
\nabla_{\epsilon }H_{\gamma \varepsilon }{}^{\tau } 
\nabla_{\tau }H_{\theta \mu \nu }  \nn\\&&+ \frac{41}{96} H_{\alpha 
}{}^{\delta \epsilon } H^{\alpha \beta \gamma } H_{\beta 
}{}^{\varepsilon \theta } H^{\mu \nu \tau } 
\nabla_{\varepsilon }H_{\gamma \delta \epsilon } \nabla_{\tau 
}H_{\theta \mu \nu } + \frac{655}{96} H_{\alpha }{}^{\delta 
\epsilon } H^{\alpha \beta \gamma } H_{\beta }{}^{\varepsilon 
\theta } H_{\delta }{}^{\mu \nu } \nabla_{\varepsilon 
}H_{\gamma \epsilon }{}^{\tau } \nabla_{\tau }H_{\theta \mu 
\nu } \nn\\&& -  \frac{1}{64} H_{\alpha \beta }{}^{\delta } H^{\alpha 
\beta \gamma } H_{\gamma }{}^{\epsilon \varepsilon } H^{\theta 
\mu \nu } \nabla_{\varepsilon }H_{\delta \epsilon }{}^{\tau } 
\nabla_{\tau }H_{\theta \mu \nu } + \frac{245}{768} H_{\alpha 
\beta }{}^{\delta } H^{\alpha \beta \gamma } H_{\gamma 
}{}^{\epsilon \varepsilon } H_{\epsilon }{}^{\theta \mu } 
\nabla_{\varepsilon }H_{\delta }{}^{\nu \tau } \nabla_{\tau 
}H_{\theta \mu \nu }  \nn\\&&+ \frac{2557}{192} H_{\alpha }{}^{\delta 
\epsilon } H^{\alpha \beta \gamma } H_{\beta }{}^{\varepsilon 
\theta } H_{\delta }{}^{\mu \nu } \nabla_{\nu }H_{\theta \mu 
\tau } \nabla^{\tau }H_{\gamma \epsilon \varepsilon } -  
\frac{1289}{192} H_{\alpha }{}^{\delta \epsilon } H^{\alpha 
\beta \gamma } H_{\beta }{}^{\varepsilon \theta } H_{\delta 
}{}^{\mu \nu } \nabla_{\tau }H_{\theta \mu \nu } 
\nabla^{\tau }H_{\gamma \epsilon \varepsilon } \nn\\&& + 
\frac{81}{256} H_{\alpha }{}^{\delta \epsilon } H^{\alpha 
\beta \gamma } H_{\beta \delta }{}^{\varepsilon } H^{\theta 
\mu \nu } \nabla_{\nu }H_{\varepsilon \mu \tau } 
\nabla^{\tau }H_{\gamma \epsilon \theta } -  \frac{103}{512} 
H_{\alpha }{}^{\delta \epsilon } H^{\alpha \beta \gamma } 
H_{\beta \delta }{}^{\varepsilon } H^{\theta \mu \nu } 
\nabla_{\tau }H_{\varepsilon \mu \nu } \nabla^{\tau 
}H_{\gamma \epsilon \theta }  \nn\\&&-  \frac{1}{256} H_{\alpha 
}{}^{\delta \epsilon } H^{\alpha \beta \gamma } H_{\beta 
}{}^{\varepsilon \theta } H_{\delta \varepsilon }{}^{\mu } 
\nabla_{\nu }H_{\theta \mu \tau } \nabla^{\tau }H_{\gamma 
\epsilon }{}^{\nu } + \frac{5}{256} H_{\alpha }{}^{\delta 
\epsilon } H^{\alpha \beta \gamma } H_{\beta }{}^{\varepsilon 
\theta } H_{\delta \varepsilon }{}^{\mu } \nabla_{\tau 
}H_{\theta \mu \nu } \nabla^{\tau }H_{\gamma \epsilon 
}{}^{\nu } \nn\\&& -  \frac{349}{768} H_{\alpha \beta }{}^{\delta } H^{
\alpha \beta \gamma } H_{\epsilon }{}^{\mu \nu } H^{\epsilon 
\varepsilon \theta } \nabla_{\nu }H_{\delta \mu \tau } 
\nabla^{\tau }H_{\gamma \varepsilon \theta } + \frac{277}{384} 
H_{\alpha }{}^{\delta \epsilon } H^{\alpha \beta \gamma } 
H_{\beta }{}^{\varepsilon \theta } H_{\delta }{}^{\mu \nu } 
\nabla_{\nu }H_{\epsilon \mu \tau } \nabla^{\tau }H_{\gamma 
\varepsilon \theta }  \nn\\&&-  \frac{1}{256} H_{\alpha \beta 
}{}^{\delta } H^{\alpha \beta \gamma } H_{\epsilon }{}^{\mu 
\nu } H^{\epsilon \varepsilon \theta } \nabla_{\tau }H_{\delta 
\mu \nu } \nabla^{\tau }H_{\gamma \varepsilon \theta } + 
\frac{295}{768} H_{\alpha }{}^{\delta \epsilon } H^{\alpha 
\beta \gamma } H_{\beta }{}^{\varepsilon \theta } H_{\delta 
}{}^{\mu \nu } \nabla_{\tau }H_{\epsilon \mu \nu } 
\nabla^{\tau }H_{\gamma \varepsilon \theta }  \nn\\&&+ 
\frac{2545}{192} H_{\alpha }{}^{\delta \epsilon } H^{\alpha 
\beta \gamma } H_{\beta }{}^{\varepsilon \theta } H_{\delta 
}{}^{\mu \nu } \nabla_{\theta }H_{\epsilon \nu \tau } 
\nabla^{\tau }H_{\gamma \varepsilon \mu } + \frac{1105}{128} 
H_{\alpha \beta }{}^{\delta } H^{\alpha \beta \gamma } 
H_{\epsilon }{}^{\mu \nu } H^{\epsilon \varepsilon \theta } 
\nabla_{\nu }H_{\delta \theta \tau } \nabla^{\tau }H_{\gamma 
\varepsilon \mu } \nn\\&& -  \frac{643}{192} H_{\alpha \beta 
}{}^{\delta } H^{\alpha \beta \gamma } H_{\epsilon }{}^{\mu 
\nu } H^{\epsilon \varepsilon \theta } \nabla_{\tau }H_{\delta 
\theta \nu } \nabla^{\tau }H_{\gamma \varepsilon \mu } -  
\frac{3}{128} H_{\alpha }{}^{\delta \epsilon } H^{\alpha \beta 
\gamma } H_{\beta }{}^{\varepsilon \theta } H_{\delta }{}^{\mu 
\nu } \nabla_{\tau }H_{\epsilon \theta \nu } \nabla^{\tau 
}H_{\gamma \varepsilon \mu } \nn\\&& -  \frac{305}{64} H_{\alpha 
}{}^{\delta \epsilon } H^{\alpha \beta \gamma } H_{\beta 
\delta }{}^{\varepsilon } H^{\theta \mu \nu } 
\nabla_{\varepsilon }H_{\epsilon \nu \tau } \nabla^{\tau 
}H_{\gamma \theta \mu } + \frac{1213}{1536} H_{\alpha \beta 
}{}^{\delta } H^{\alpha \beta \gamma } H_{\epsilon \varepsilon 
}{}^{\mu } H^{\epsilon \varepsilon \theta } \nabla_{\nu 
}H_{\delta \mu \tau } \nabla^{\tau }H_{\gamma \theta }{}^{\nu 
}  \nn\\&&-  \frac{1205}{1536} H_{\alpha \beta }{}^{\delta } H^{\alpha 
\beta \gamma } H_{\epsilon \varepsilon }{}^{\mu } H^{\epsilon 
\varepsilon \theta } \nabla_{\tau }H_{\delta \mu \nu } 
\nabla^{\tau }H_{\gamma \theta }{}^{\nu } + \frac{5453}{768} 
H_{\alpha }{}^{\delta \epsilon } H^{\alpha \beta \gamma } 
H_{\beta }{}^{\varepsilon \theta } H_{\delta }{}^{\mu \nu } 
\nabla_{\theta }H_{\epsilon \varepsilon \tau } \nabla^{\tau 
}H_{\gamma \mu \nu }  \nn\\&&-  \frac{1053}{512} H_{\alpha }{}^{\delta 
\epsilon } H^{\alpha \beta \gamma } H_{\beta }{}^{\varepsilon 
\theta } H_{\delta }{}^{\mu \nu } \nabla_{\tau }H_{\epsilon 
\varepsilon \theta } \nabla^{\tau }H_{\gamma \mu \nu } -  
\frac{325}{384} H_{\alpha }{}^{\delta \epsilon } H^{\alpha 
\beta \gamma } H_{\beta }{}^{\varepsilon \theta } H_{\gamma 
}{}^{\mu \nu } \nabla_{\nu }H_{\theta \mu \tau } 
\nabla^{\tau }H_{\delta \epsilon \varepsilon } \nn\\&& + 
\frac{623}{1536} H_{\alpha }{}^{\delta \epsilon } H^{\alpha 
\beta \gamma } H_{\beta }{}^{\varepsilon \theta } H_{\gamma 
}{}^{\mu \nu } \nabla_{\tau }H_{\theta \mu \nu } 
\nabla^{\tau }H_{\delta \epsilon \varepsilon } -  
\frac{7}{768} H_{\alpha \beta }{}^{\delta } H^{\alpha \beta 
\gamma } H_{\gamma }{}^{\epsilon \varepsilon } H^{\theta \mu 
\nu } \nabla_{\tau }H_{\theta \mu \nu } \nabla^{\tau 
}H_{\delta \epsilon \varepsilon } \nn\\&& -  \frac{35}{48} H_{\alpha 
\beta }{}^{\delta } H^{\alpha \beta \gamma } H_{\gamma 
}{}^{\epsilon \varepsilon } H^{\theta \mu \nu } \nabla_{\nu 
}H_{\varepsilon \mu \tau } \nabla^{\tau }H_{\delta \epsilon 
\theta } + \frac{175}{768} H_{\alpha \beta }{}^{\delta } 
H^{\alpha \beta \gamma } H_{\gamma }{}^{\epsilon \varepsilon } 
H^{\theta \mu \nu } \nabla_{\tau }H_{\varepsilon \mu \nu } 
\nabla^{\tau }H_{\delta \epsilon \theta }  \nn\\&&-  \frac{253}{768} 
H_{\alpha \beta }{}^{\delta } H^{\alpha \beta \gamma } 
H_{\gamma }{}^{\epsilon \varepsilon } H_{\epsilon }{}^{\theta 
\mu } \nabla_{\nu }H_{\theta \mu \tau } \nabla^{\tau 
}H_{\delta \varepsilon }{}^{\nu } + \frac{13}{24} H_{\alpha 
\beta }{}^{\delta } H^{\alpha \beta \gamma } H_{\gamma 
}{}^{\epsilon \varepsilon } H_{\epsilon }{}^{\theta \mu } 
\nabla_{\tau }H_{\theta \mu \nu } \nabla^{\tau }H_{\delta 
\varepsilon }{}^{\nu } \nn\\&& -  \frac{139}{1536} H_{\alpha \beta 
}{}^{\delta } H^{\alpha \beta \gamma } H_{\gamma }{}^{\epsilon 
\varepsilon } H^{\theta \mu \nu } \nabla_{\nu }H_{\epsilon 
\varepsilon \tau } \nabla^{\tau }H_{\delta \theta \mu } + 
\frac{221}{1536} H_{\alpha \beta }{}^{\delta } H^{\alpha \beta 
\gamma } H_{\gamma }{}^{\epsilon \varepsilon } H^{\theta \mu 
\nu } \nabla_{\tau }H_{\epsilon \varepsilon \nu } 
\nabla^{\tau }H_{\delta \theta \mu }  \nn\\&&-  \frac{95}{48} 
H_{\alpha \beta }{}^{\delta } H^{\alpha \beta \gamma } 
H_{\gamma }{}^{\epsilon \varepsilon } H_{\epsilon }{}^{\theta 
\mu } \nabla_{\mu }H_{\varepsilon \nu \tau } \nabla^{\tau 
}H_{\delta \theta }{}^{\nu } + \frac{97}{384} H_{\alpha \beta 
}{}^{\delta } H^{\alpha \beta \gamma } H_{\gamma }{}^{\epsilon 
\varepsilon } H_{\epsilon }{}^{\theta \mu } \nabla_{\nu 
}H_{\varepsilon \mu \tau } \nabla^{\tau }H_{\delta \theta 
}{}^{\nu } \nn\\&& -  \frac{85}{384} H_{\alpha \beta }{}^{\delta } 
H^{\alpha \beta \gamma } H_{\gamma }{}^{\epsilon \varepsilon } 
H_{\epsilon }{}^{\theta \mu } \nabla_{\tau }H_{\varepsilon \mu 
\nu } \nabla^{\tau }H_{\delta \theta }{}^{\nu } + 
\frac{157}{768} H_{\alpha \beta }{}^{\delta } H^{\alpha \beta 
\gamma } H_{\gamma }{}^{\epsilon \varepsilon } H_{\epsilon 
\varepsilon }{}^{\theta } \nabla_{\nu }H_{\theta \mu \tau } 
\nabla^{\tau }H_{\delta }{}^{\mu \nu } \nn\\&& + \frac{1}{128} 
H_{\alpha \beta }{}^{\delta } H^{\alpha \beta \gamma } 
H_{\gamma }{}^{\epsilon \varepsilon } H_{\epsilon \varepsilon 
}{}^{\theta } \nabla_{\tau }H_{\theta \mu \nu } \nabla^{\tau 
}H_{\delta }{}^{\mu \nu } + \frac{85}{3072} H_{\alpha \beta 
}{}^{\delta } H^{\alpha \beta \gamma } H_{\gamma }{}^{\epsilon 
\varepsilon } H^{\theta \mu \nu } \nabla_{\delta }H_{\mu \nu 
\tau } \nabla^{\tau }H_{\epsilon \varepsilon \theta }  \nn\\&&-  
\frac{1127}{512} H_{\alpha }{}^{\delta \epsilon } H^{\alpha 
\beta \gamma } H_{\beta \delta }{}^{\varepsilon } H_{\gamma 
}{}^{\theta \mu } \nabla_{\nu }H_{\theta \mu \tau } 
\nabla^{\tau }H_{\epsilon \varepsilon }{}^{\nu } -  
\frac{1609}{3072} H_{\alpha \beta }{}^{\delta } H^{\alpha 
\beta \gamma } H_{\gamma }{}^{\epsilon \varepsilon } H_{\delta 
}{}^{\theta \mu } \nabla_{\nu }H_{\theta \mu \tau } 
\nabla^{\tau }H_{\epsilon \varepsilon }{}^{\nu } \nn\\&& + 
\frac{601}{256} H_{\alpha }{}^{\delta \epsilon } H^{\alpha 
\beta \gamma } H_{\beta \delta }{}^{\varepsilon } H_{\gamma 
}{}^{\theta \mu } \nabla_{\tau }H_{\theta \mu \nu } 
\nabla^{\tau }H_{\epsilon \varepsilon }{}^{\nu } + 
\frac{185}{384} H_{\alpha \beta }{}^{\delta } H^{\alpha \beta 
\gamma } H_{\gamma }{}^{\epsilon \varepsilon } H_{\delta 
}{}^{\theta \mu } \nabla_{\tau }H_{\theta \mu \nu } 
\nabla^{\tau }H_{\epsilon \varepsilon }{}^{\nu }  \nn\\&&+ 
\frac{151}{384} H_{\alpha \beta }{}^{\delta } H^{\alpha \beta 
\gamma } H_{\gamma }{}^{\epsilon \varepsilon } H^{\theta \mu 
\nu } \nabla_{\delta }H_{\varepsilon \nu \tau } \nabla^{\tau 
}H_{\epsilon \theta \mu } + \frac{637}{48} H_{\alpha 
}{}^{\delta \epsilon } H^{\alpha \beta \gamma } H_{\beta 
\delta }{}^{\varepsilon } H_{\gamma }{}^{\theta \mu } 
\nabla_{\mu }H_{\varepsilon \nu \tau } \nabla^{\tau 
}H_{\epsilon \theta }{}^{\nu }  \nn\\&&-  \frac{1}{64} H_{\alpha 
}{}^{\delta \epsilon } H^{\alpha \beta \gamma } H_{\beta 
\delta }{}^{\varepsilon } H_{\gamma }{}^{\theta \mu } 
\nabla_{\nu }H_{\varepsilon \mu \tau } \nabla^{\tau 
}H_{\epsilon \theta }{}^{\nu } + \frac{1337}{1536} H_{\alpha 
\beta }{}^{\delta } H^{\alpha \beta \gamma } H_{\gamma 
}{}^{\epsilon \varepsilon } H_{\delta }{}^{\theta \mu } 
\nabla_{\nu }H_{\varepsilon \mu \tau } \nabla^{\tau 
}H_{\epsilon \theta }{}^{\nu }  \nn\\&&+ \frac{3}{64} H_{\alpha 
}{}^{\delta \epsilon } H^{\alpha \beta \gamma } H_{\beta 
\delta }{}^{\varepsilon } H_{\gamma }{}^{\theta \mu } 
\nabla_{\tau }H_{\varepsilon \mu \nu } \nabla^{\tau 
}H_{\epsilon \theta }{}^{\nu } -  \frac{1349}{1536} H_{\alpha 
\beta }{}^{\delta } H^{\alpha \beta \gamma } H_{\gamma 
}{}^{\epsilon \varepsilon } H_{\delta }{}^{\theta \mu } 
\nabla_{\tau }H_{\varepsilon \mu \nu } \nabla^{\tau 
}H_{\epsilon \theta }{}^{\nu } \nn\\&& -  \frac{1921}{384} H_{\alpha 
\beta }{}^{\delta } H^{\alpha \beta \gamma } H_{\gamma 
}{}^{\epsilon \varepsilon } H_{\epsilon }{}^{\theta \mu } 
\nabla_{\delta }H_{\mu \nu \tau } \nabla^{\tau 
}H_{\varepsilon \theta }{}^{\nu } -  \frac{637}{96} H_{\alpha 
}{}^{\delta \epsilon } H^{\alpha \beta \gamma } H_{\beta 
\delta }{}^{\varepsilon } H_{\gamma \epsilon }{}^{\theta } 
\nabla_{\nu }H_{\theta \mu \tau } \nabla^{\tau 
}H_{\varepsilon }{}^{\mu \nu }  \nn\\&&-  \frac{485}{192} H_{\alpha 
\beta }{}^{\delta } H^{\alpha \beta \gamma } H_{\gamma 
}{}^{\epsilon \varepsilon } H_{\delta \epsilon }{}^{\theta } 
\nabla_{\nu }H_{\theta \mu \tau } \nabla^{\tau 
}H_{\varepsilon }{}^{\mu \nu } + \frac{1271}{384} H_{\alpha 
}{}^{\delta \epsilon } H^{\alpha \beta \gamma } H_{\beta 
\delta }{}^{\varepsilon } H_{\gamma \epsilon }{}^{\theta } 
\nabla_{\tau }H_{\theta \mu \nu } \nabla^{\tau 
}H_{\varepsilon }{}^{\mu \nu } \nn\\&& + \frac{229}{192} H_{\alpha 
\beta }{}^{\delta } H^{\alpha \beta \gamma } H_{\gamma 
}{}^{\epsilon \varepsilon } H_{\delta \epsilon }{}^{\theta } 
\nabla_{\tau }H_{\theta \mu \nu } \nabla^{\tau 
}H_{\varepsilon }{}^{\mu \nu } + \frac{7}{1536} H_{\alpha 
\beta }{}^{\delta } H^{\alpha \beta \gamma } H_{\gamma 
}{}^{\epsilon \varepsilon } H_{\delta \epsilon \varepsilon } 
\nabla_{\tau }H_{\theta \mu \nu } \nabla^{\tau }H^{\theta 
\mu \nu }.\nn
\eeqa


\end{document}